\documentclass[structabstract]{aa}  
                                   \usepackage{graphicx}
\usepackage{txfonts}
\usepackage{amssymb}
\usepackage{natbib}
\usepackage{rotating}
\usepackage{fixltx2e}
\bibpunct{(}{)}{;}{a}{}{,}

\begin{document}
   \title{APEX-CHAMP$^+$ high-$J$ CO observations of \\low--mass young stellar objects: }
   \subtitle{III. NGC~1333~IRAS~4A/4B envelope, outflow and UV heating}
   \author{Umut A. Y{\i}ld{\i}z\inst{1}
          \and Lars E. Kristensen\inst{1}
          \and Ewine F. van Dishoeck\inst{1,2}
          \and Arnaud Belloche\inst{3}
          \and Tim A. van Kempen\inst{1,4}
          \and \\Michiel R. Hogerheijde\inst{1}
          \and Rolf G{\"u}sten\inst{3}
          \and Nienke van der Marel\inst{1}
          }

          \institute{$^1$ Leiden Observatory, Leiden University, P.O. Box 9513, 2300 RA Leiden, The Netherlands\\
          $^2$ Max-Planck Institut f\"ur Extraterrestrische Physik (MPE), Giessenbachstr.\ 1, 85748 Garching, Germany \\
          $^3$ Max Planck Institut f\"ur Radioastronomie, Auf dem H\"ugel 69, D-53121, Bonn, Germany\\
          $^4$ Joint ALMA Offices, Av. Alonso de Cordova 3107, Vitacura, Santiago, Chile \\
          \email{yildiz@strw.leidenuniv.nl}
             }
   \date{Accepted: 05/03/2012}
   \titlerunning{NGC~1333~IRAS~4A/4B envelope, outflow and UV heating}

\def\placeFigureIrasFourABCOsixfivespectraandoutflowINSET{
\begin{figure}[!t]
\begin{center}
\includegraphics[scale=0.45]{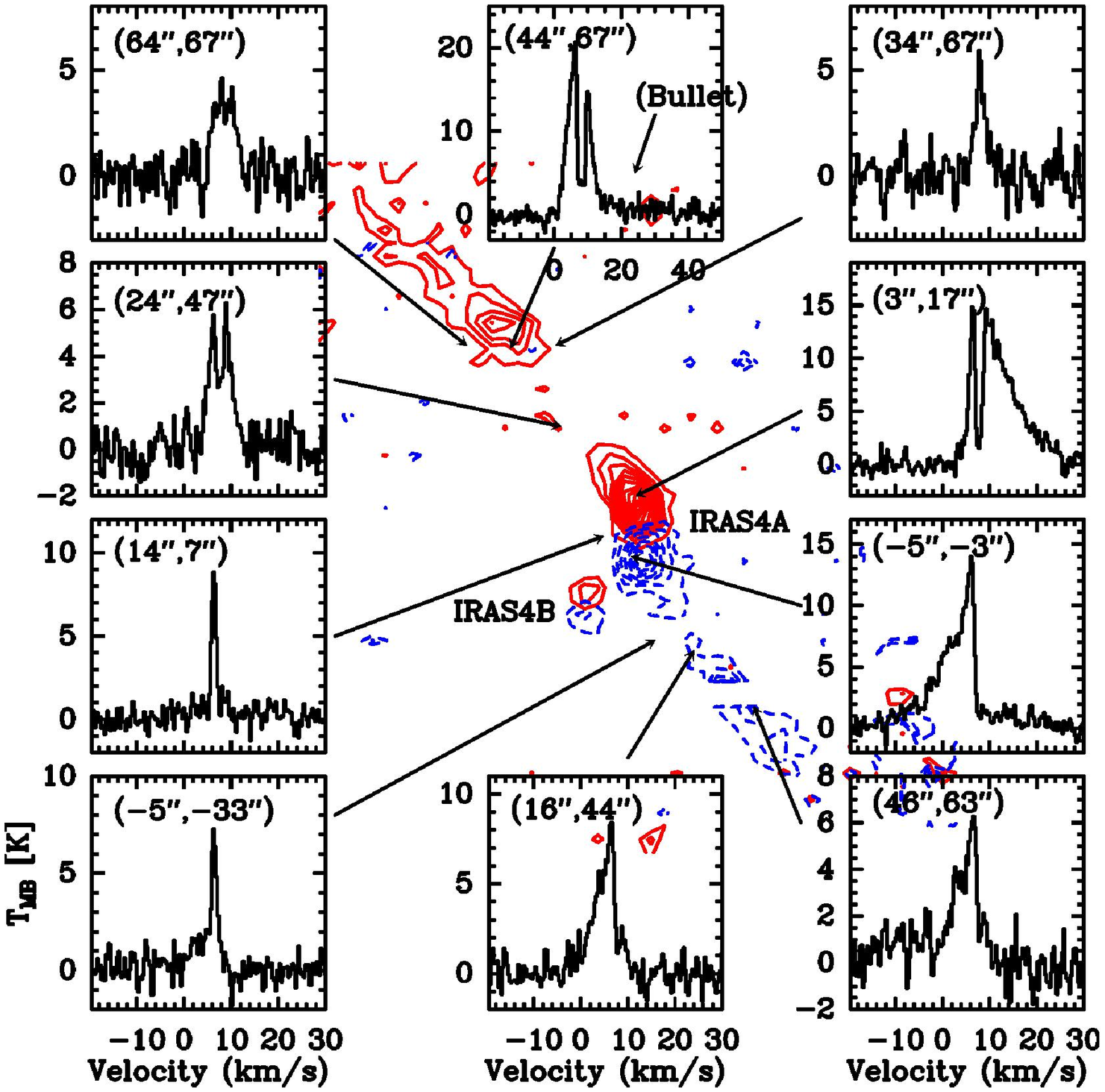} \end{center}
\caption{Gallery of \mbox{$^{12}$CO~6--5} spectra from 10 different locations. Spectra of the IRAS~4A and 4B central positions are shown in Fig.~\ref{fig:iras4abspectra}.
  The arrows indicate the exact locations of the
  corresponding spectra with respect to the outflow lobes and each
  spectrum is given with the offset from IRAS~4A. Note the mix of
  narrow ($<$2 km s$^{-1}$) and medium (10--15 km s$^{-1}$) profiles together with the broad lines (25--30 km s$^{-1}$) at the outflowing positions close to the center of IRAS~4A. Also note that the velocity scale of the (44$\arcsec$,67$\arcsec$) panel is different to emphasize the weak ``bullet'' emission (see text). The vertical scale is for $T_{\rm mb}$. The contours are \mbox{$^{12}$CO~6--5} emission where the levels start from 3$\sigma$ (15~K~km~s$^{-1}$) with an increasing step size of 2$\sigma$ (10~K~km~s$^{-1}$). The blue and red velocity ranges are selected from -20 to 2.7 and from 10.5 to 30~km~s$^{-1}$, respectively.
  }
\label{fig:co65insets}
\end{figure}
}

\def\placeFigureIrasFourABCOsixfivespectraandoutflow{
\begin{figure*}[!th]
\begin{center}
\includegraphics[scale=0.455]{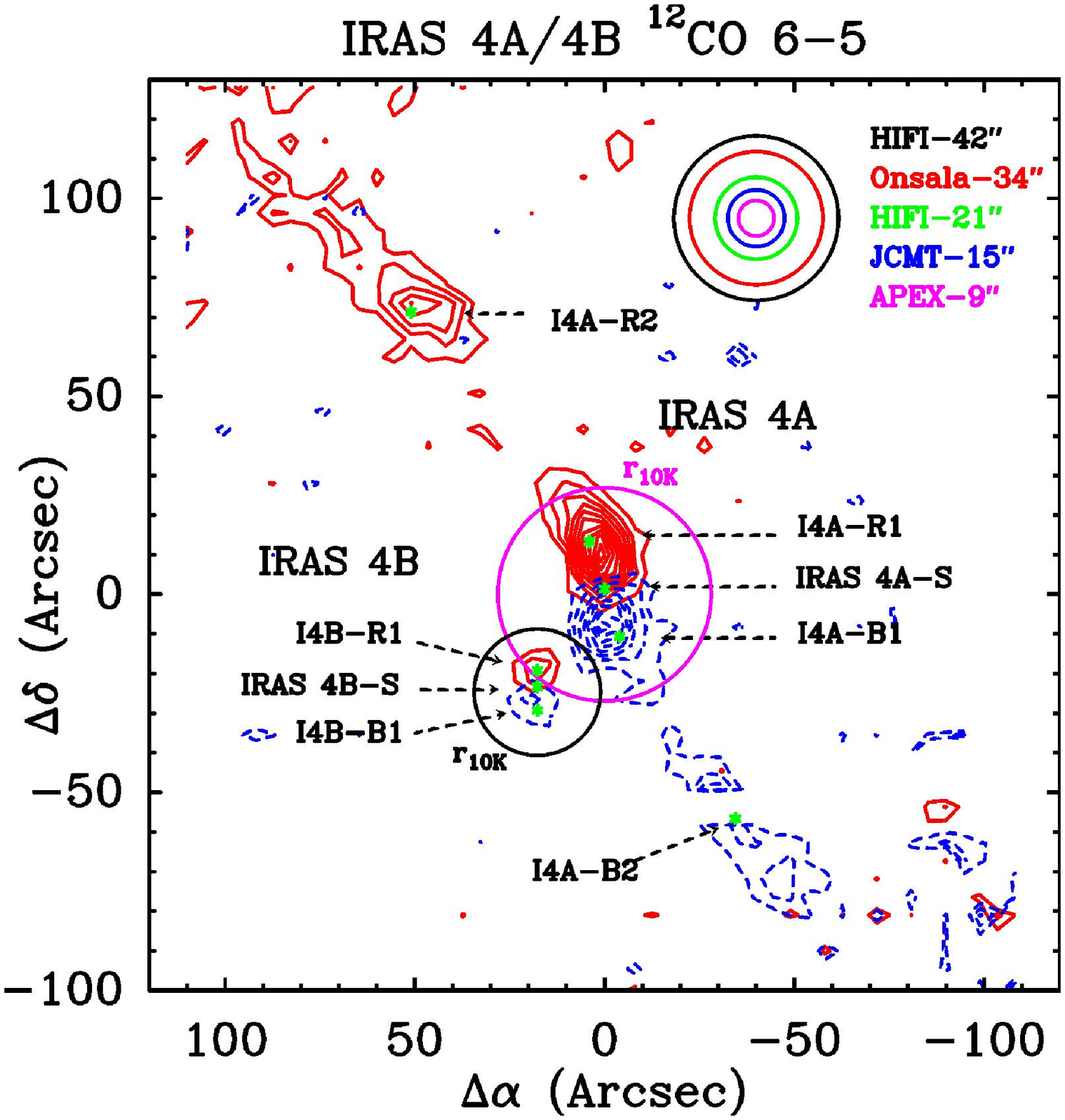} \includegraphics[scale=0.33]{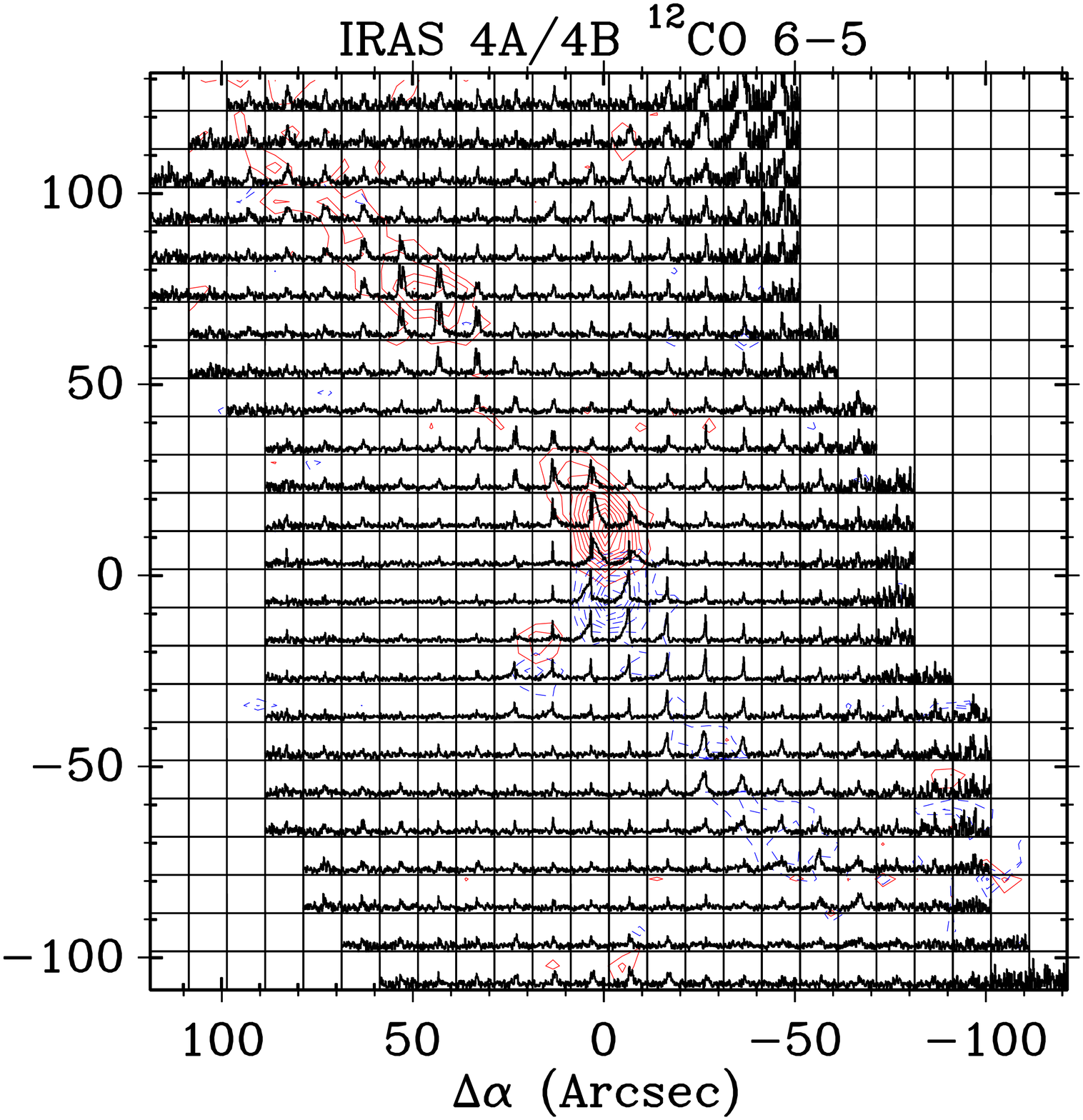} \end{center}
\caption{({\it Right:}) \mbox{$^{12}$CO~6--5} spectral map of IRAS~4A and 4B over the
  240$\arcsec\times$240$\arcsec$ mapping area. Individual spectra are
  shown on the $T_{\rm MB}$ scale from --2 to 12 K and velocity scale
  from --20 to 30 \mbox{km~s$^{-1}$}. The outflows of IRAS~4A and 4B
  are overplotted over the entire spectral map. The map is centered on
  IRAS~4A. The contour levels start from 3$\sigma$ (15~K~km~s$^{-1}$)
  with an increasing step size of 2$\sigma$ (10~K~km~s$^{-1}$). ({\it Left:}) Envelopes of
  IRAS~4A and 4B at the 10~K radius are shown together with the
  beam-sizes compared. See Fig. 1 caption and Sect. 3.2.1 for the details and the positions at which deep spectra are obtained.}
\label{fig:co65map}
\end{figure*}
}

\def\placeFigureTemperatureMaps{
\begin{figure*}[h]
    \centering
    \includegraphics[scale=0.63]{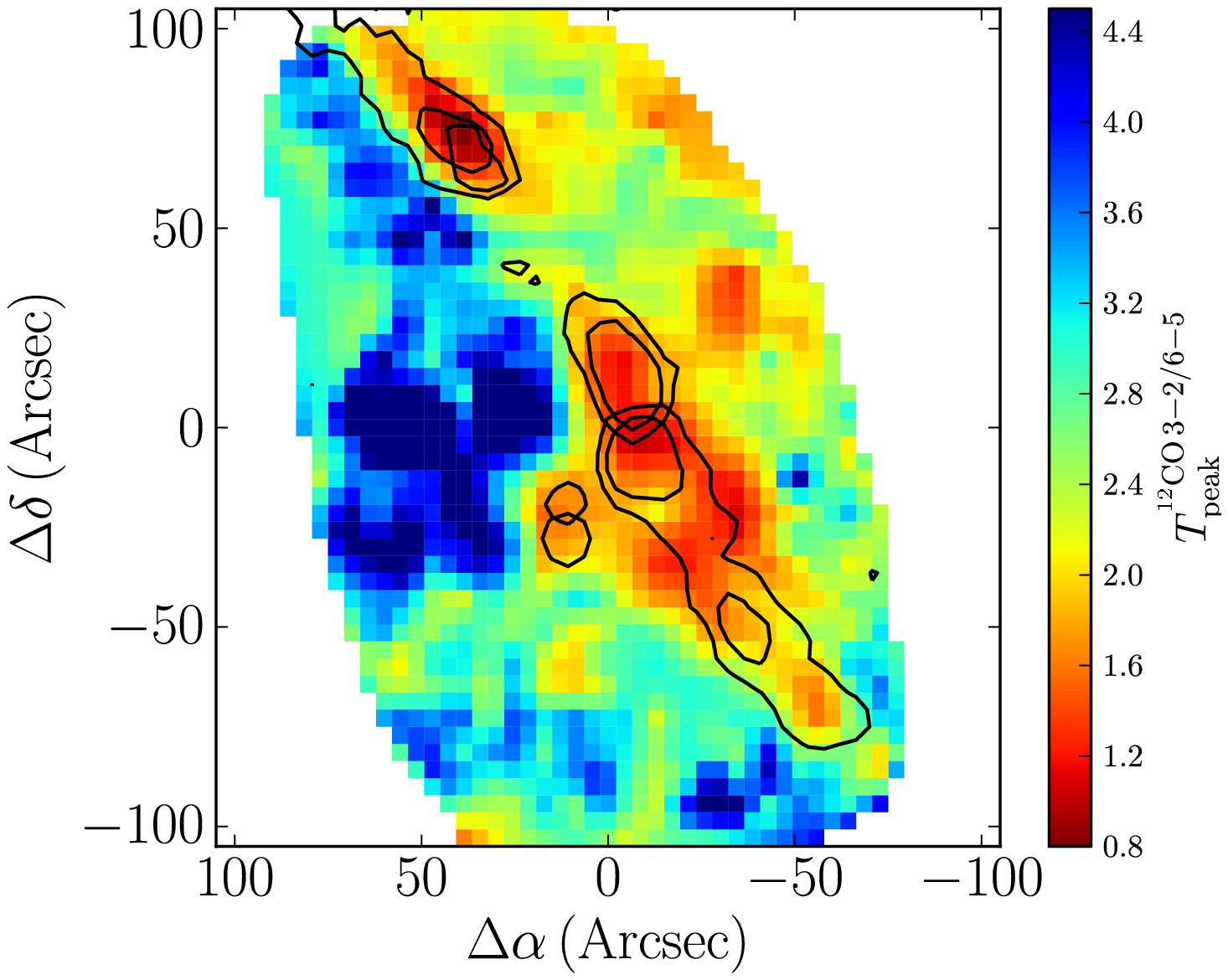}     \includegraphics[scale=0.63]{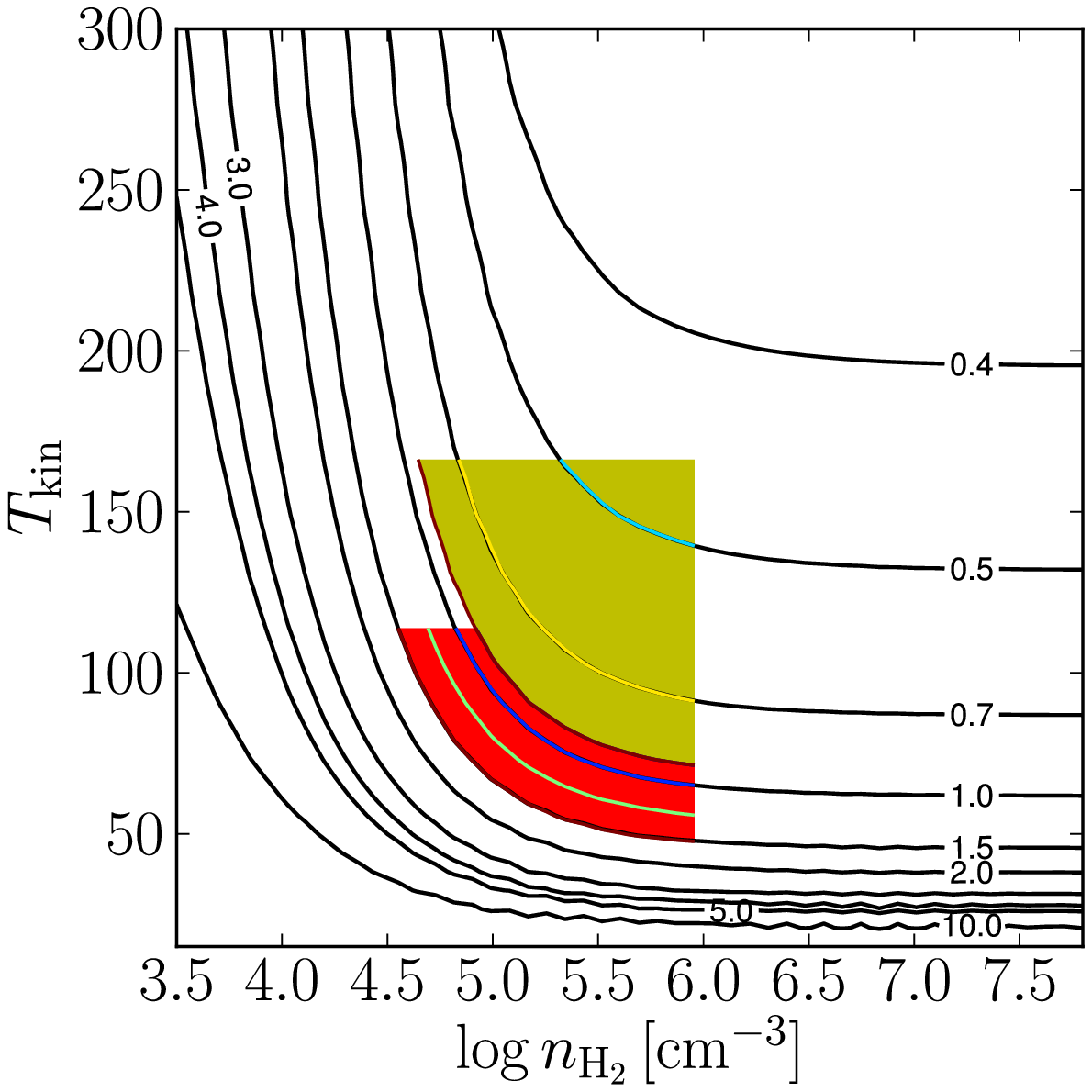}     \caption{\small {\it Left:} Map of peak intensity ratios of the
      $^{12}$CO~3--2 /6--5 lines.  {\it Right:} Model of the CO~3--2 /
      CO~6--5 line intensity ratio as function of temperature and
      density. The red region represents the observed range for
      IRAS~4A, the yellow range for IRAS 4B. CO column
      density is taken to be 10$^{17}$ cm$^{-2}$ with a line width of
      10 km s$^{-1}$, these conditions are chosen because they are representative
 of the observed CO~6--5 flux and line width.
     The coloured lines give the range of densities
      within the 20$\arcsec$ beam for the two sources based on the
      models of \citet{Kristensen12subm}. In the relevant density
      range, lower ratios imply higher temperatures.}
    \label{fig:12co32-co65mapratio}
    \label{fig:radexmodels}
\end{figure*}
}

\def\placeFigureDeltaVMaps{
\begin{figure*}[h]
    \centering
    \includegraphics[scale=0.42]{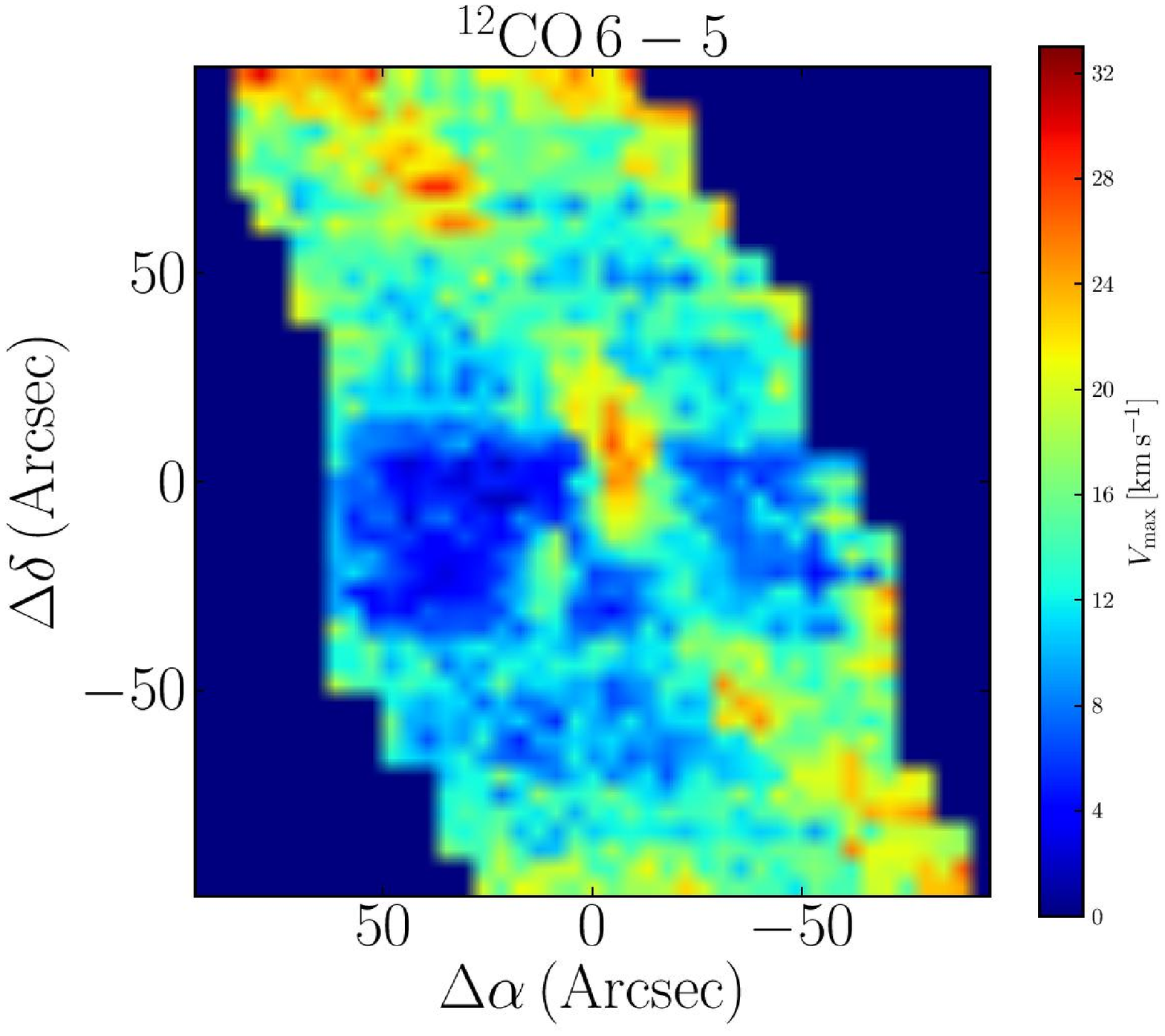}         \hspace*{0.5cm}
    \includegraphics[scale=0.42]{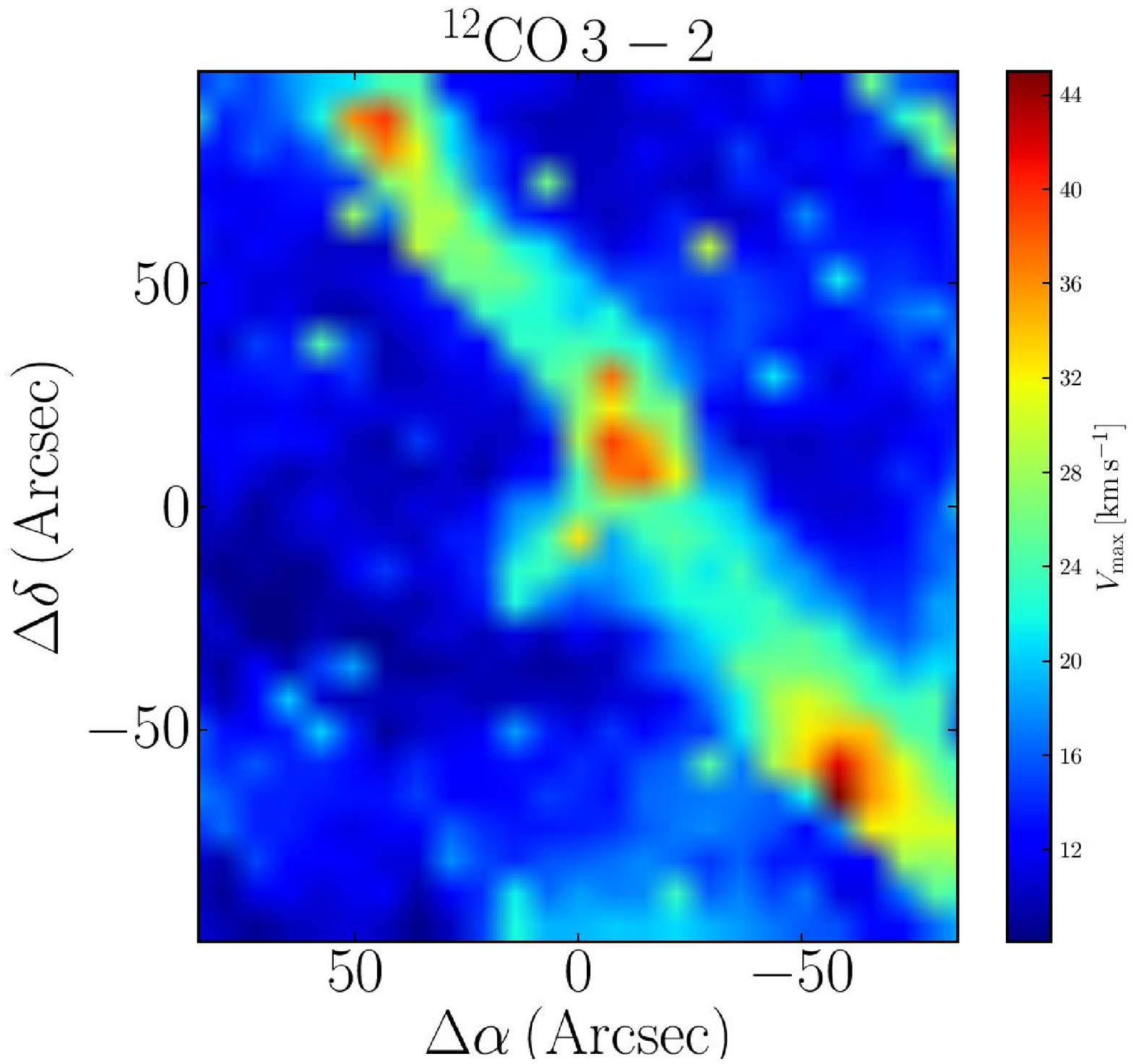}     \caption{\small  Maps of $V_{\rm max}$ obtained from full width at zero intensity (FWZI) at each position in both CO~6--5 ({\it left}) and CO~3--2 ({\it right}) maps.}
    \label{fig:Vmaxmaps}
\end{figure*}
}

\def\placeFigureIrasFourABThirteenCOSpectraAndOutflow{
\begin{figure*}[!th]
\begin{center}
\includegraphics[scale=0.18]{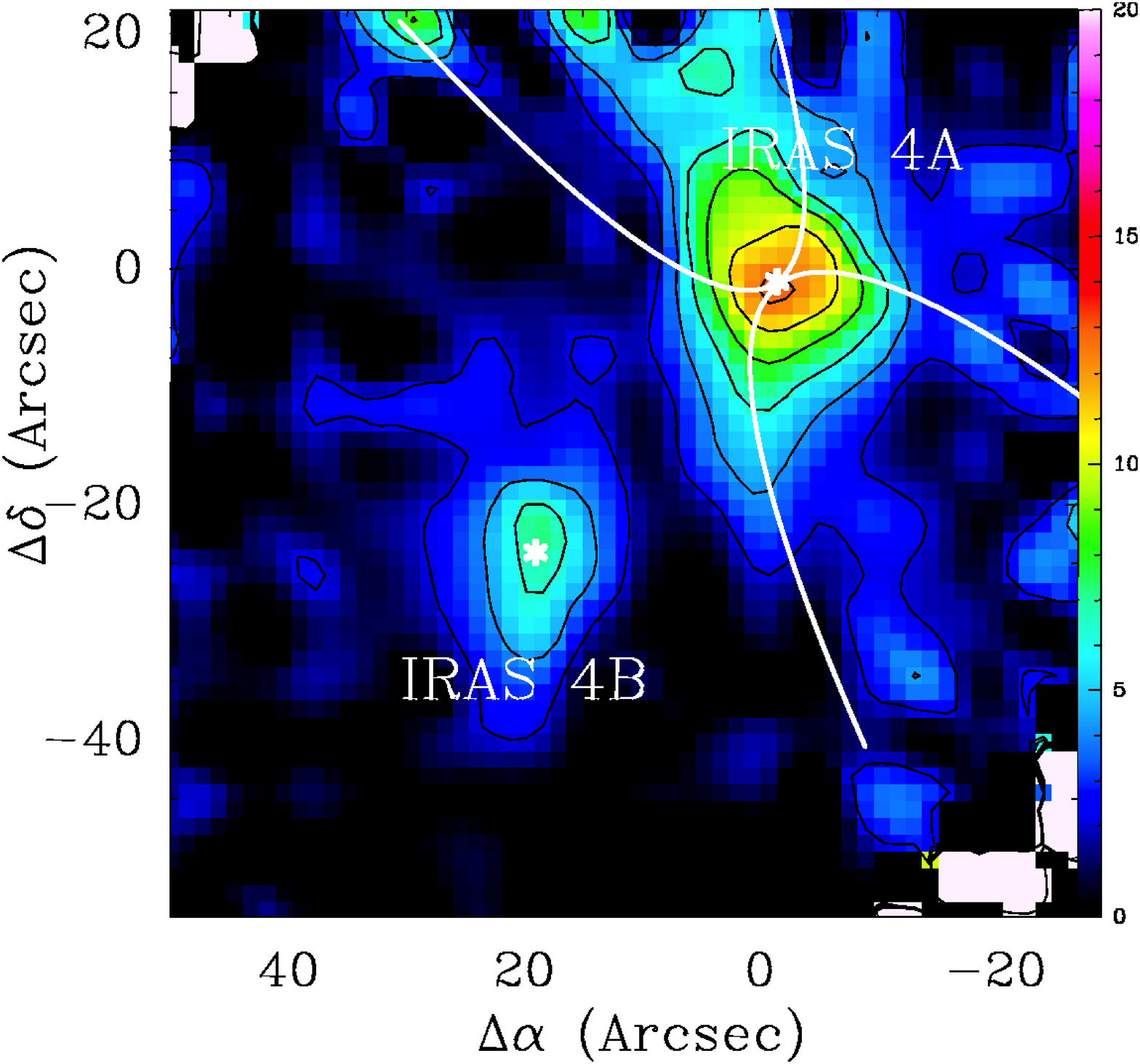}         \hspace*{1.2cm}
\includegraphics[scale=0.35]{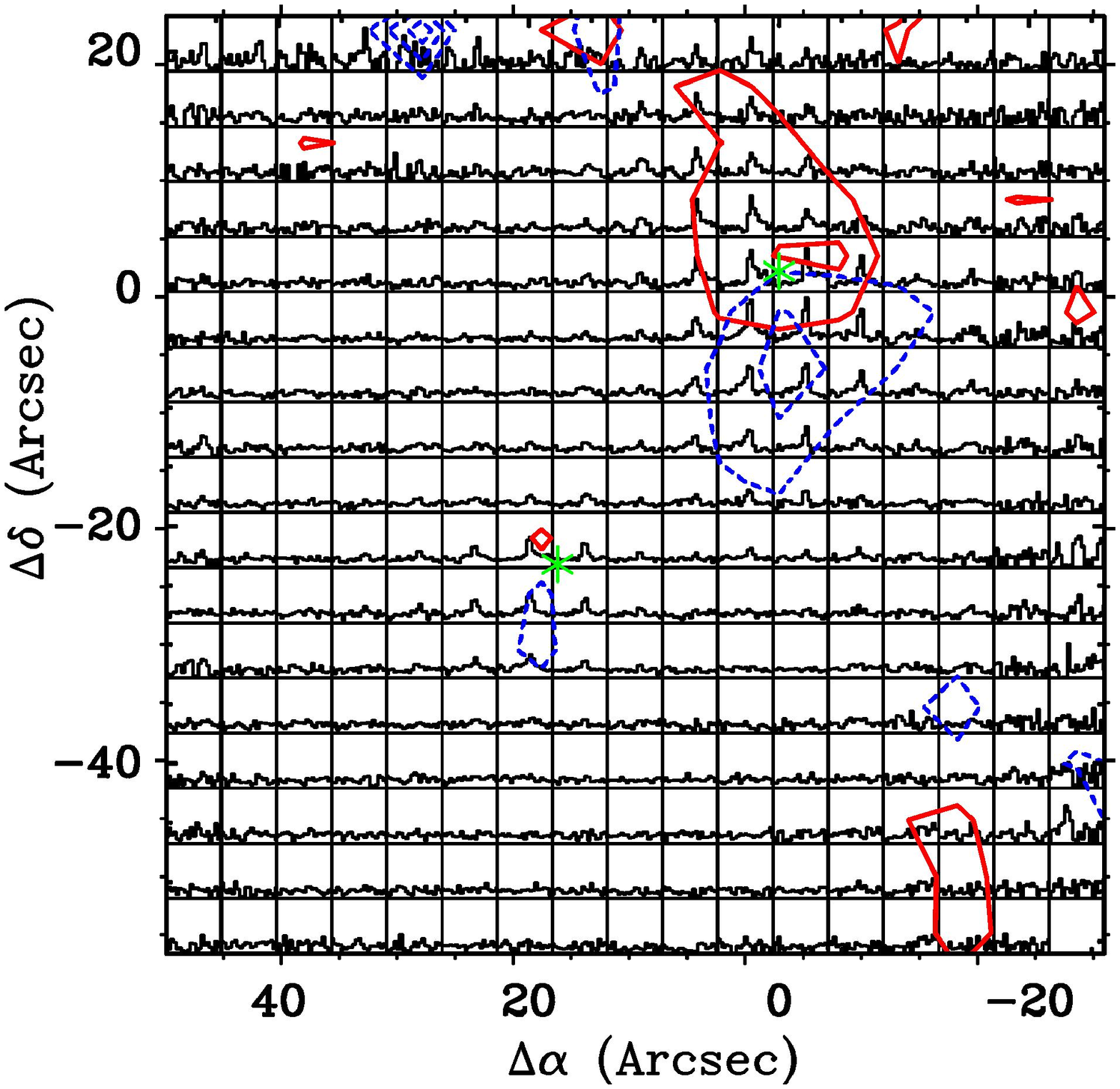} \end{center}
\caption{\mbox{$^{13}$CO~6--5} maps of IRAS~4A (0,0) and IRAS~4B
  (22.5,-22.8). \textit{Left}: Integrated intensity map of IRAS~4A and
  4B in a 80$\arcsec\times$80$\arcsec$ area. The
  white elliptical biconical shape delineate the outflow cones as discussed in
  Sect.~\ref{sec:analysisUV}. \textit{Right}: Blue
  and redshifted outflows seen in the $^{13}$CO 6-5 line profiles
  overplotted on a spectral map of the same region. Individual spectra
  are shown on a $T_{\rm MB}$ scale from --1 to 4~K and the velocity
  scale runs from --5 to 15 \mbox{km~s$^{-1}$}. The contour levels for both figures are 3$\sigma$,
  6$\sigma$, 9$\sigma$, ... where $\sigma$=0.6~K.}
\label{fig:iras4ab13co65}
\end{figure*}
}

\def\placeFigureIRASThereD{
\begin{figure}[!ht]
    \centering
    \includegraphics[scale=0.12]{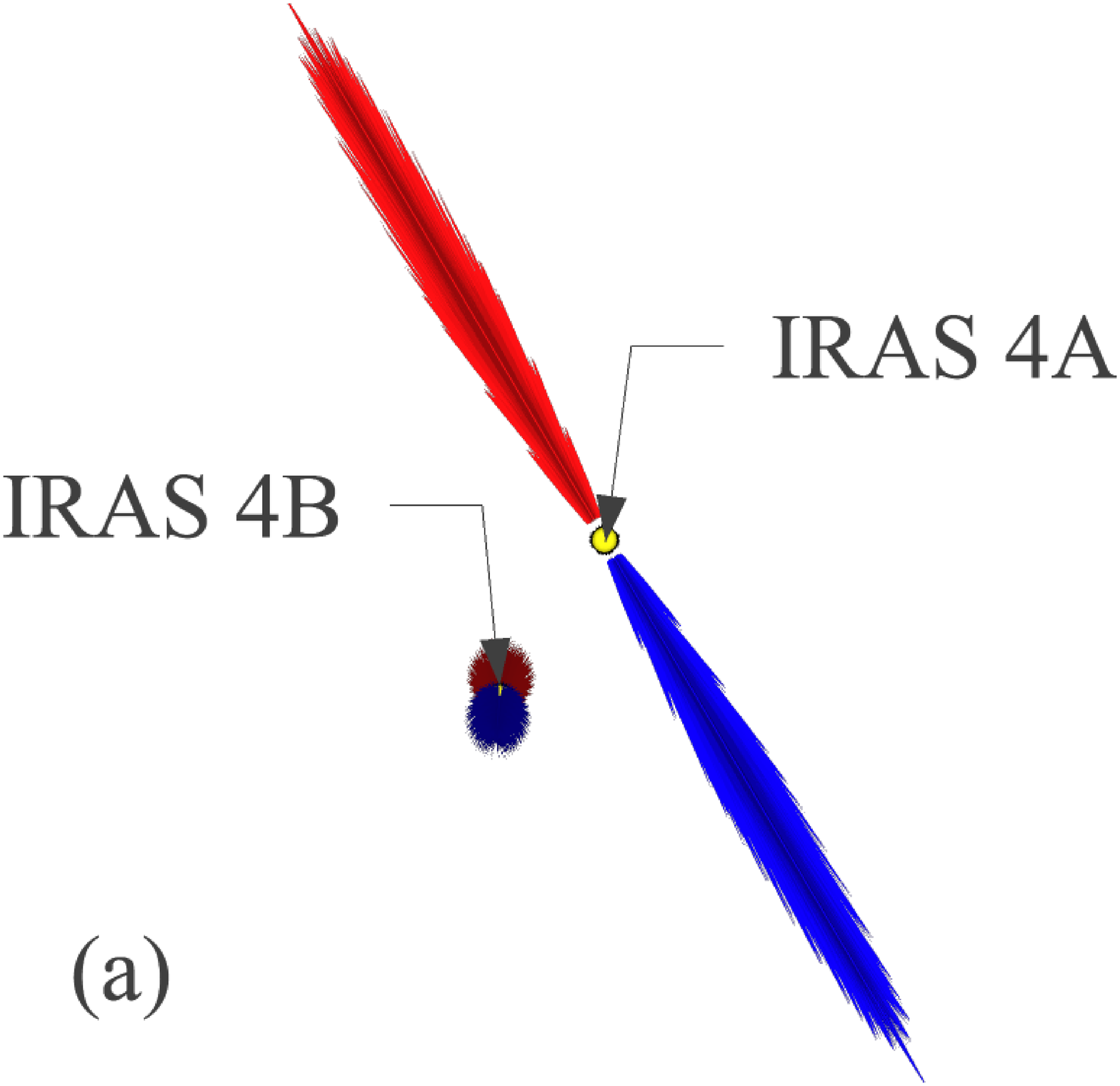}     \includegraphics[scale=0.12]{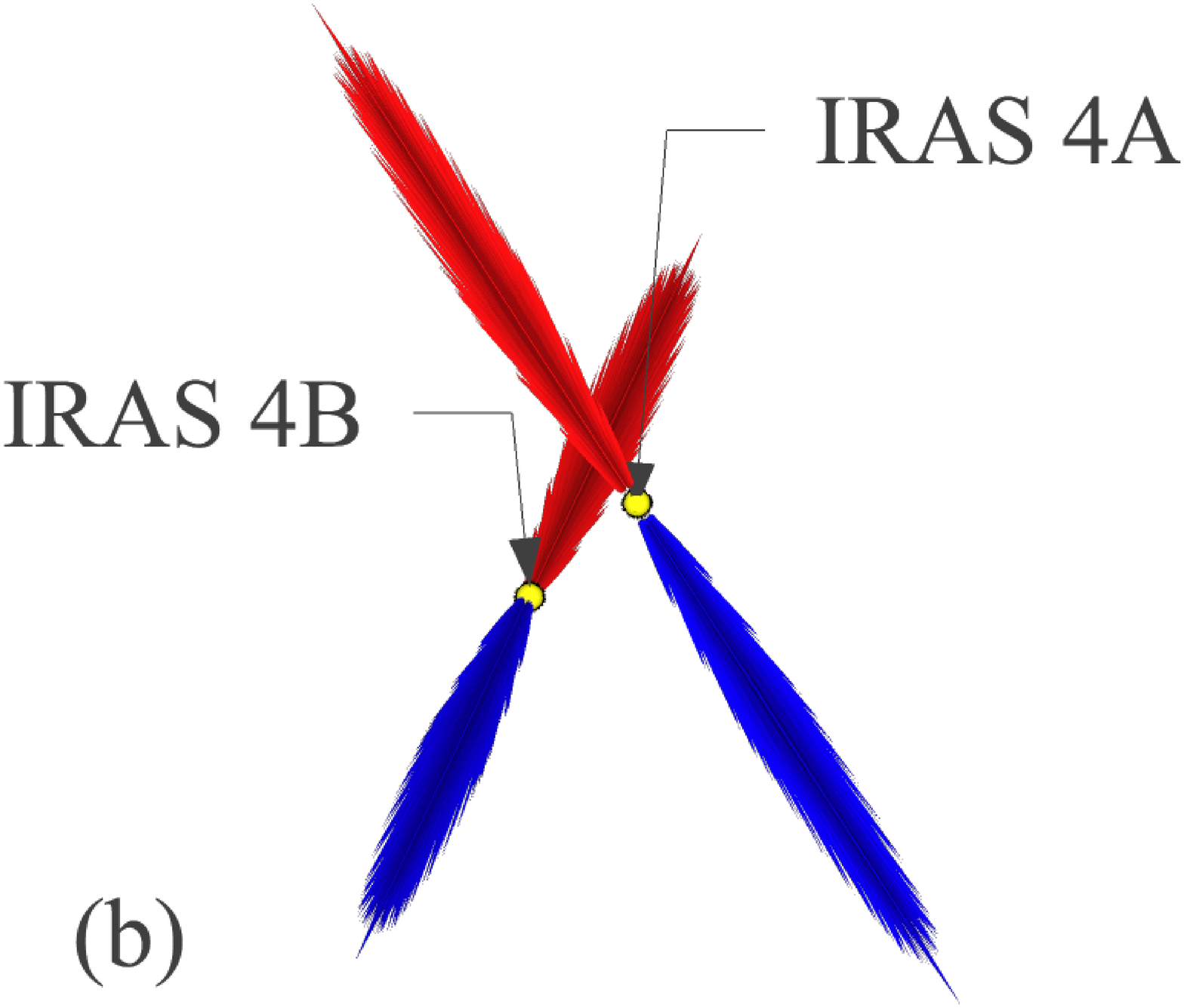}     \includegraphics[scale=0.12]{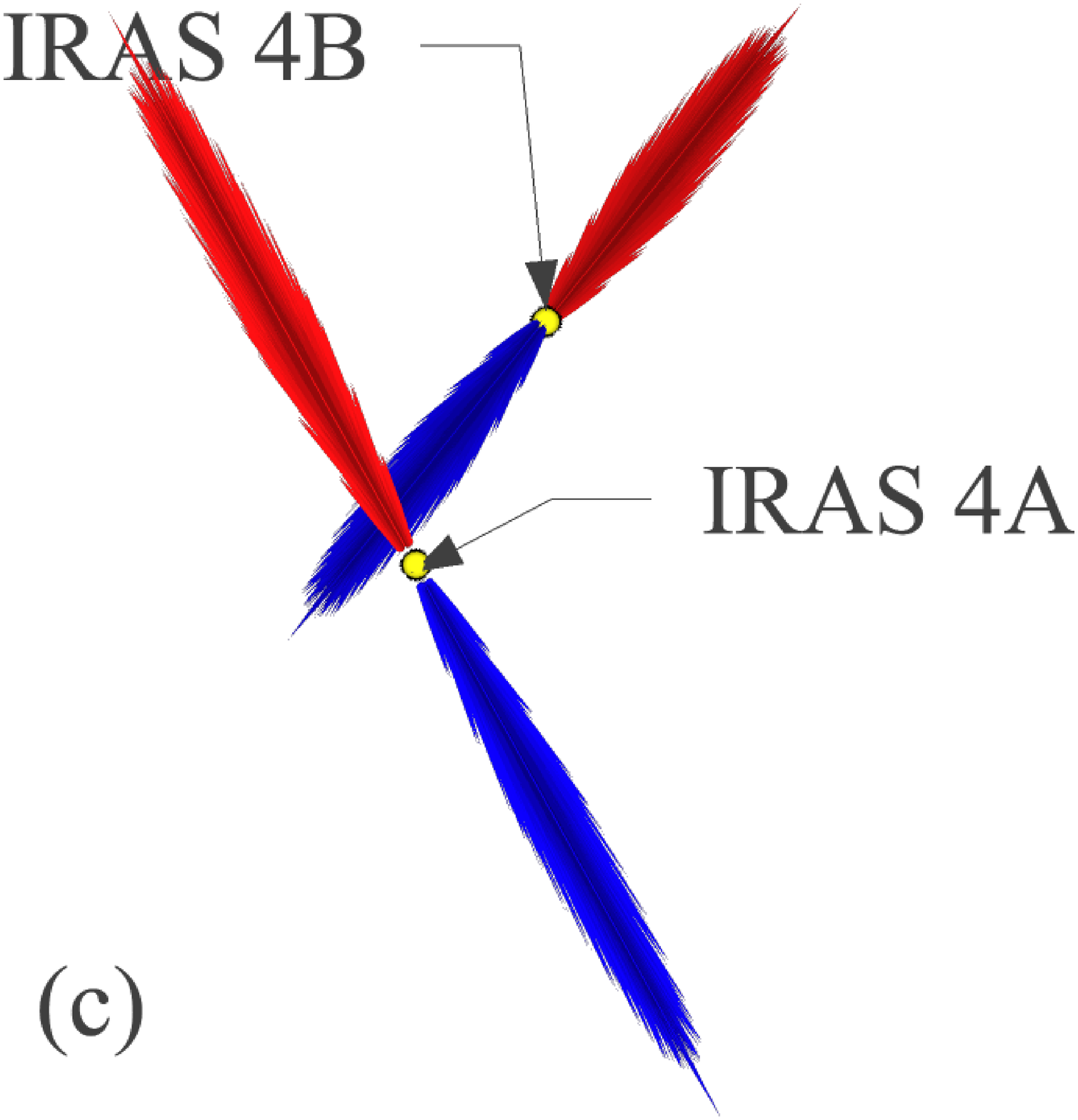}     \includegraphics[scale=0.12]{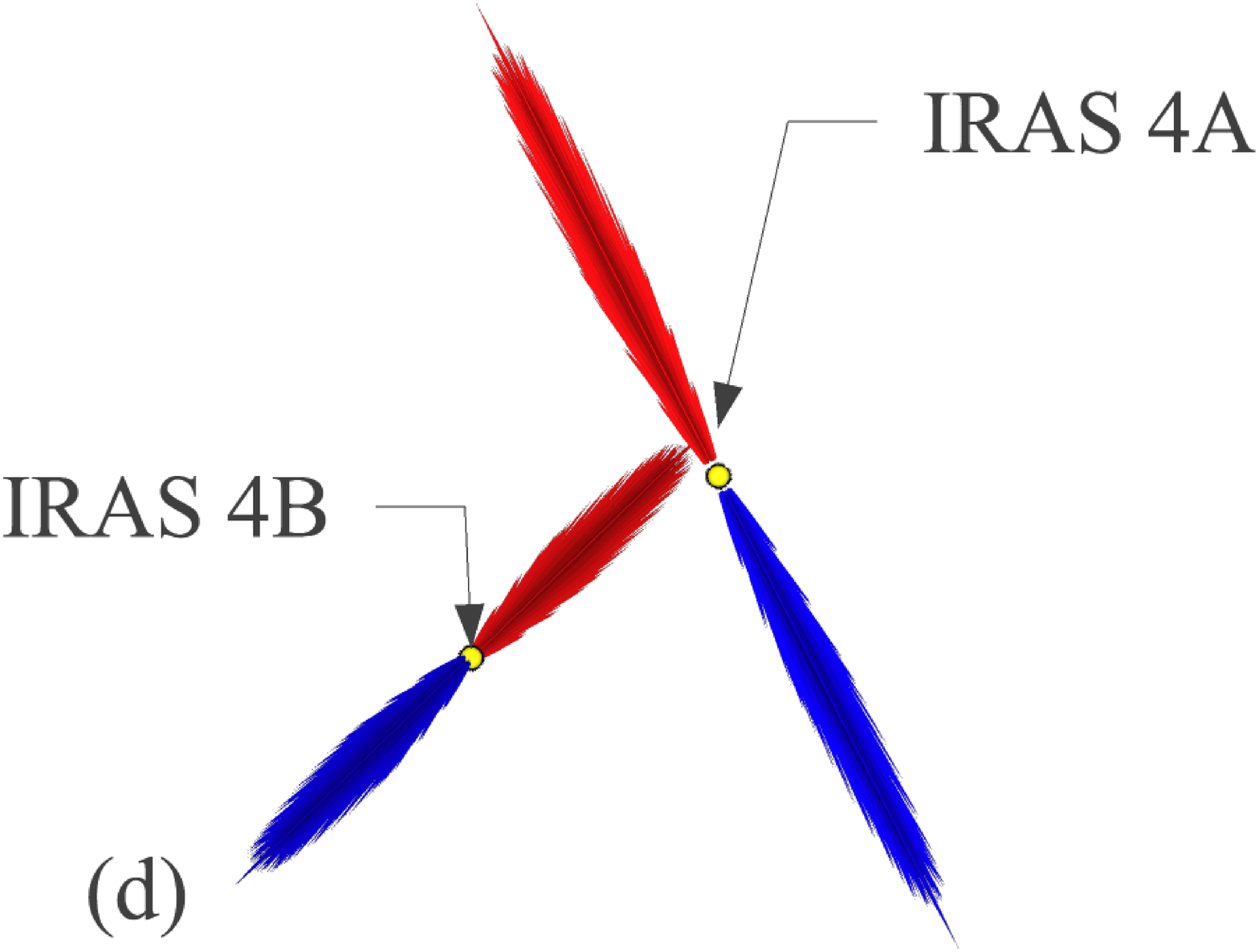}     \caption{\small Four different projection scenarios of the IRAS 4A and 4B outflows are presented, assuming the two outflows have similar physical extents. These scenarios are treated by keeping the position of IRAS~4A fixed and rotating the plane of the sky $\sim$60$\degr$ for better comprehension the difference of IRAS~4B in (b)-(d).
    ({\it a}) shows the geometry as projected to the plane of the sky,
    ({\it b}) the protostars are at the same distance and very close to each other so that their envelopes overlap, 
    ({\it c}) IRAS~4A is in front of IRAS~4B,
    ({\it d}) IRAS~4B is in front of IRAS~4A. 
In the latter two scenarios the envelopes may be sufficiently distant to each other and may not overlap.}
    \label{fig:iras4ab3D}
\end{figure}
}

\def\placeFigureCOladder{
\begin{figure}[!tb]
    \centering
    \includegraphics[scale=0.8]{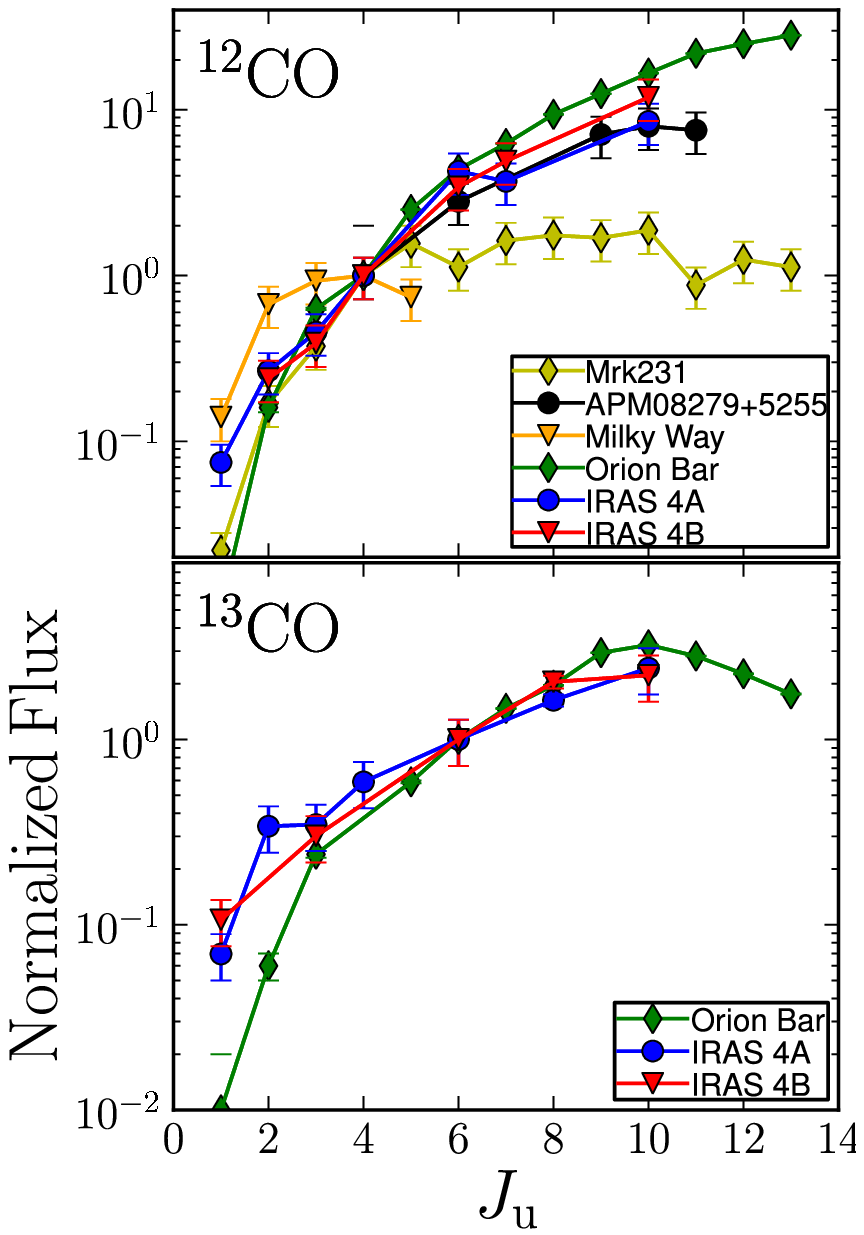}     \caption{\small CO line fluxes for the observed transitions. The
      \mbox{$^{12}$CO} and \mbox{$^{13}$CO} lines are normalized
      relative to the $J$=4--3 and $J$=6--5 lines,
      respectively. Observations of the Milky Way \citep{Wright1991},
      the dense Orion Bar PDR \citep{Habart10}, the ultraluminous
      galaxy Mrk231 \citep{vanderWerf10} and the high redshift quasar
      APM08279+5255 \citep{Weiss2007} are compared. In IRAS~4A
      and 4B, the available maps are convolved to 20$\arcsec$ in order to
      compare similar spatial regions.}
    \label{fig:COladder}
\end{figure}
}

\def\placeFigureRotDiags{
\begin{figure}[tb]
    \centering
    \includegraphics[scale=0.7]{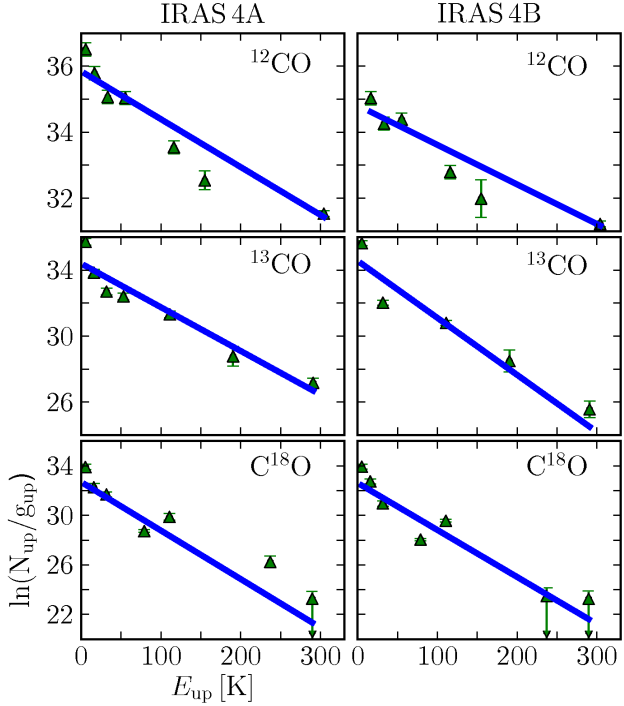}     \caption{\small Rotation diagrams measured for the CO and
      isotopolog lines at the source positions of IRAS 4A and 4B. The
      column density of each observation in each rotational level
      divided by the statistical weight is plotted against the
      excitation energy of the level. The fitted line shows the
      Boltzmann distribution of the rotational populations. Derived
      values of the rotation temperatures are presented in
      Table~\ref{tbl:rotdiaglist}.}
    \label{fig:RotationDiagrams}
\end{figure}
}

\def\placeFigureIrasFourACOThreetoTwoandSixtoFiveRatios{
\begin{figure}[!t]
    \centering
    \includegraphics[scale=0.125]{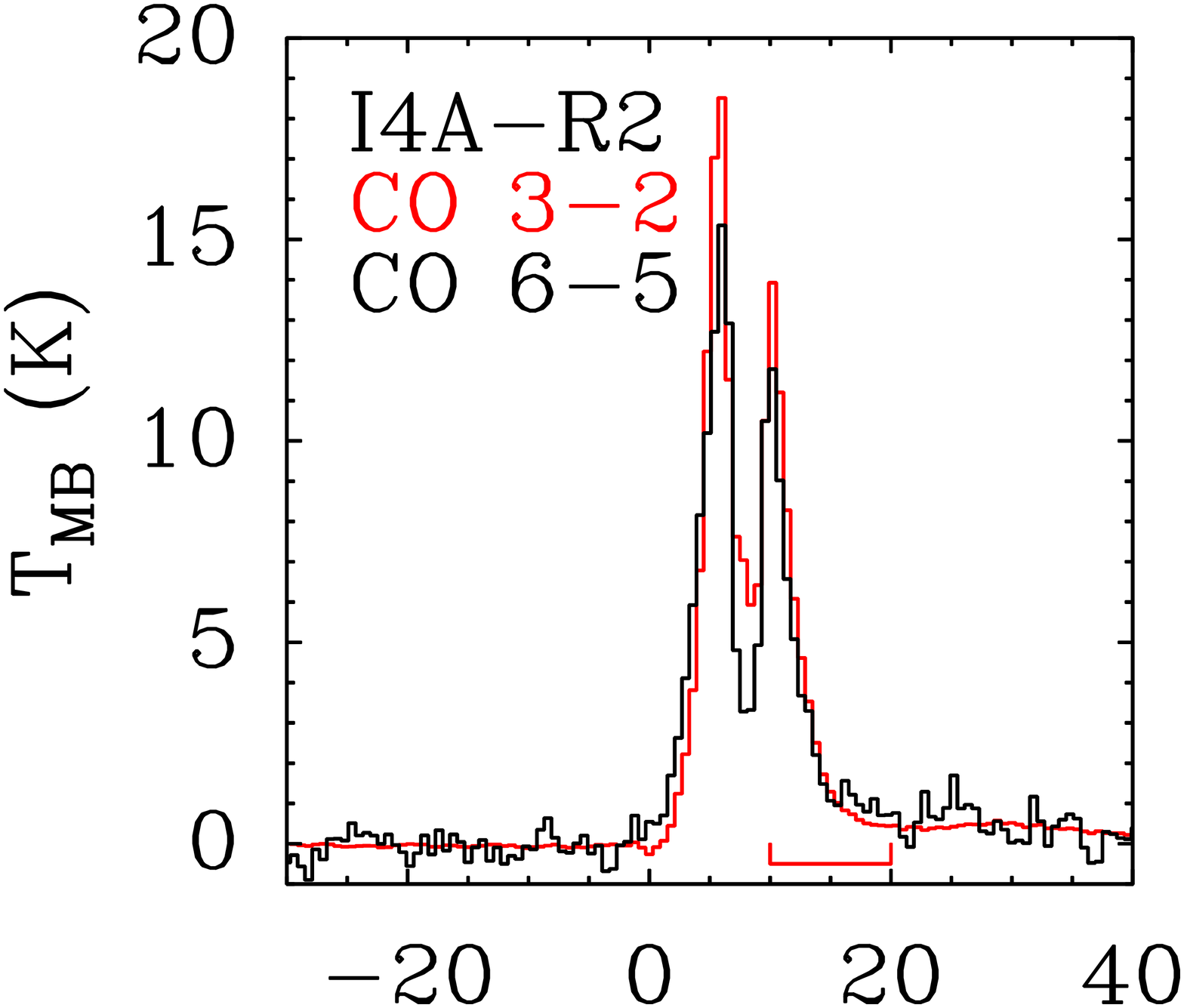}     \hspace*{2.3cm} 
    \includegraphics[scale=0.125]{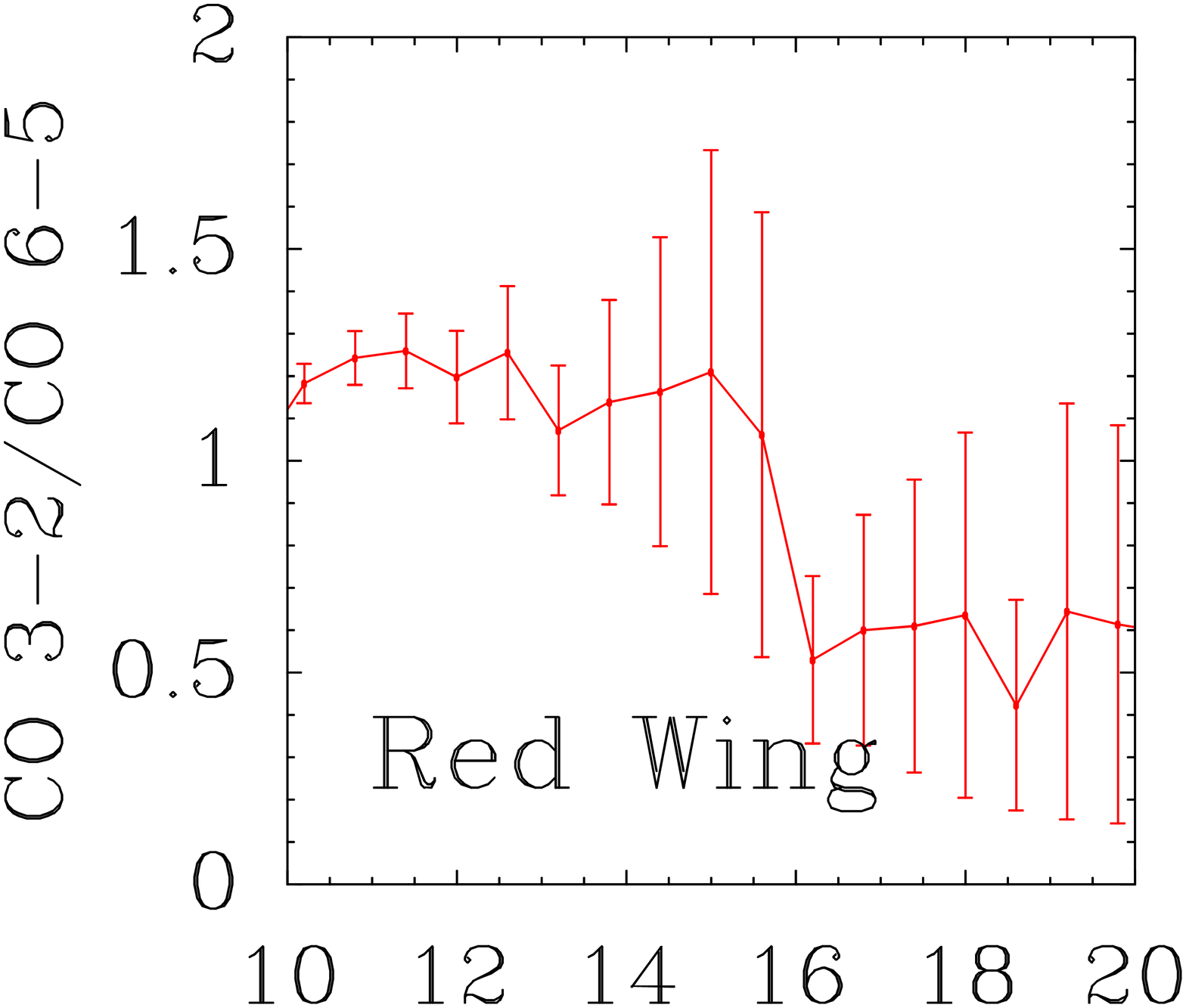}\\     \includegraphics[scale=0.125]{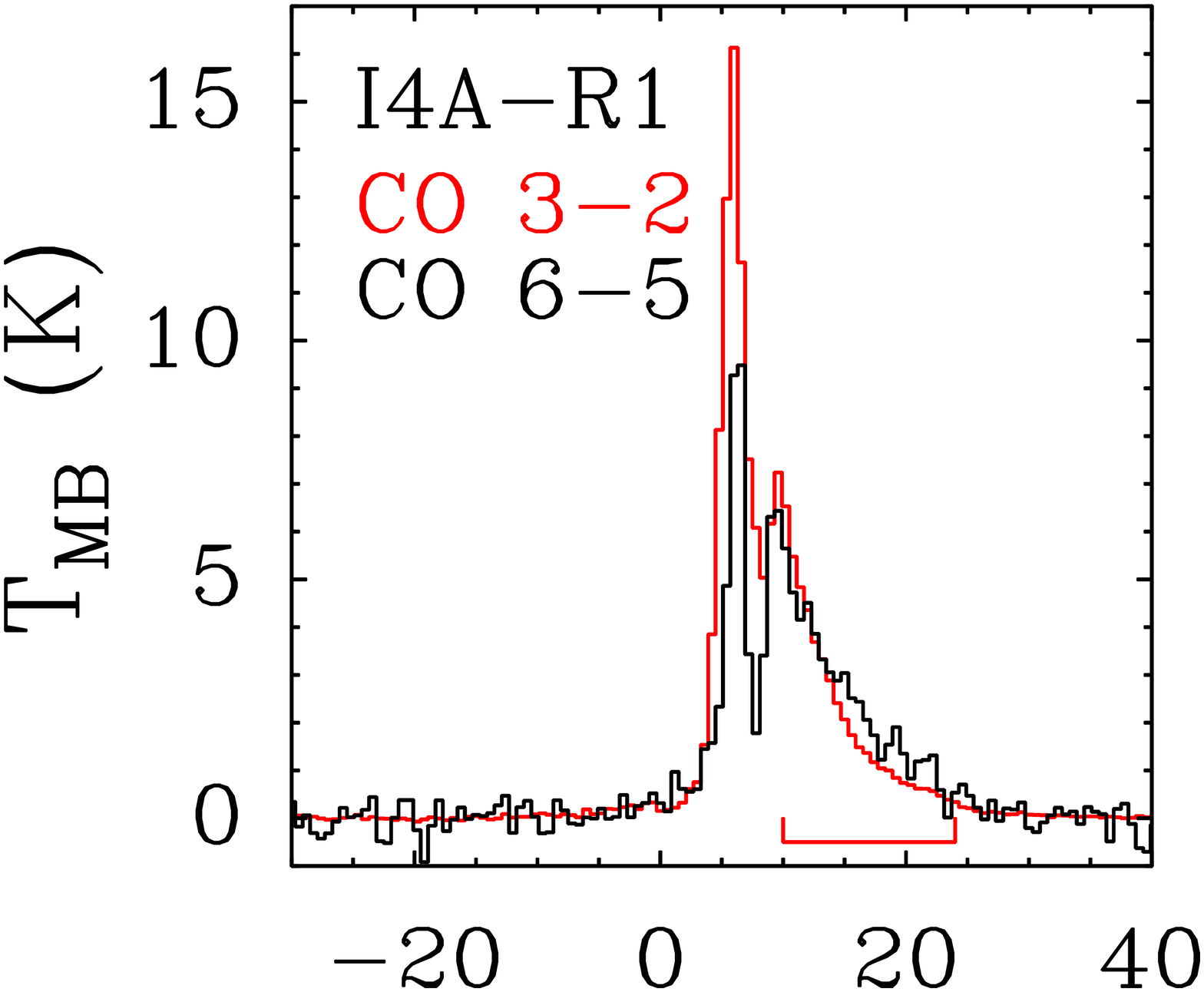}     \hspace*{2.3cm} 
    \includegraphics[scale=0.125]{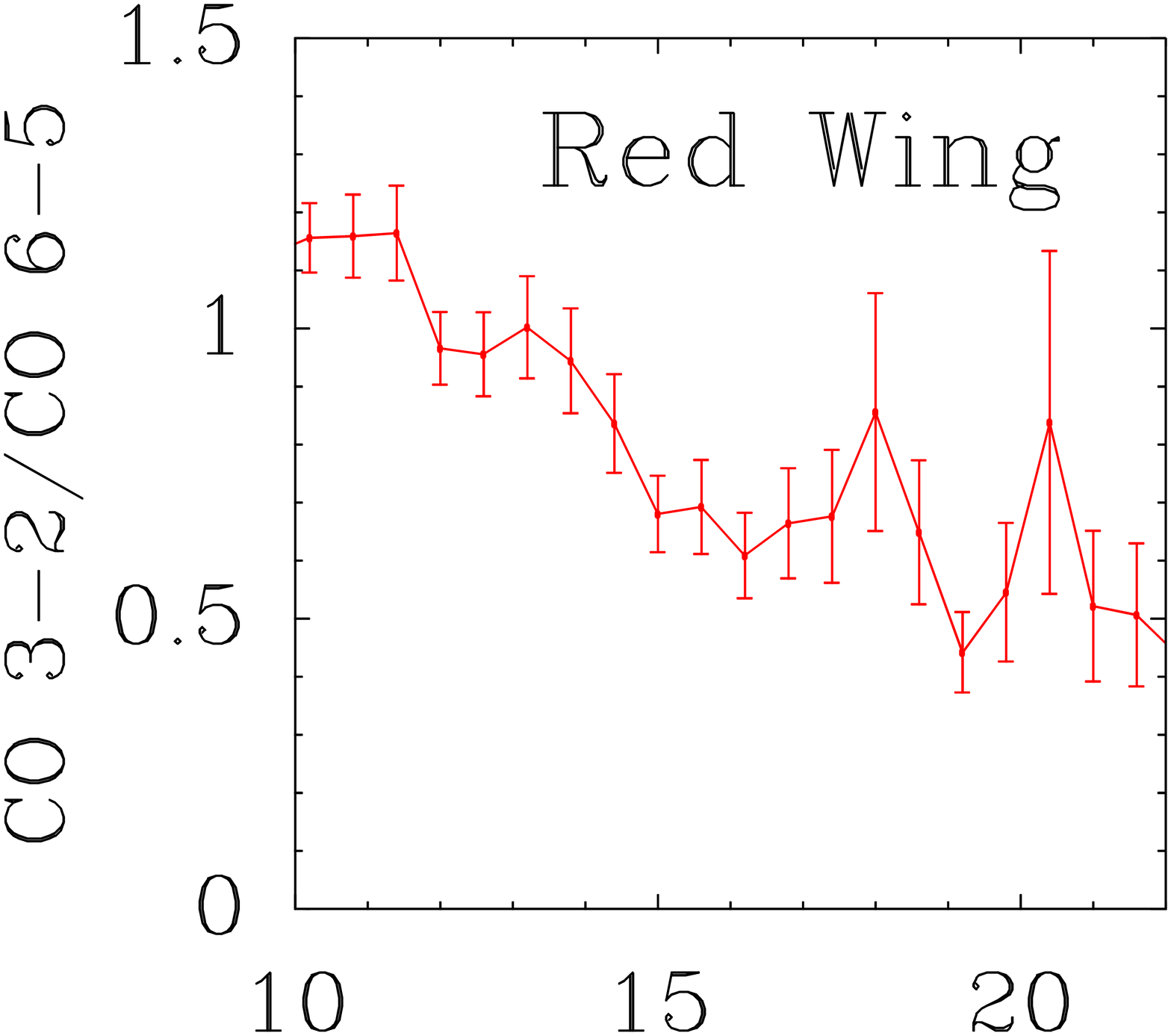}\\     \includegraphics[scale=0.125]{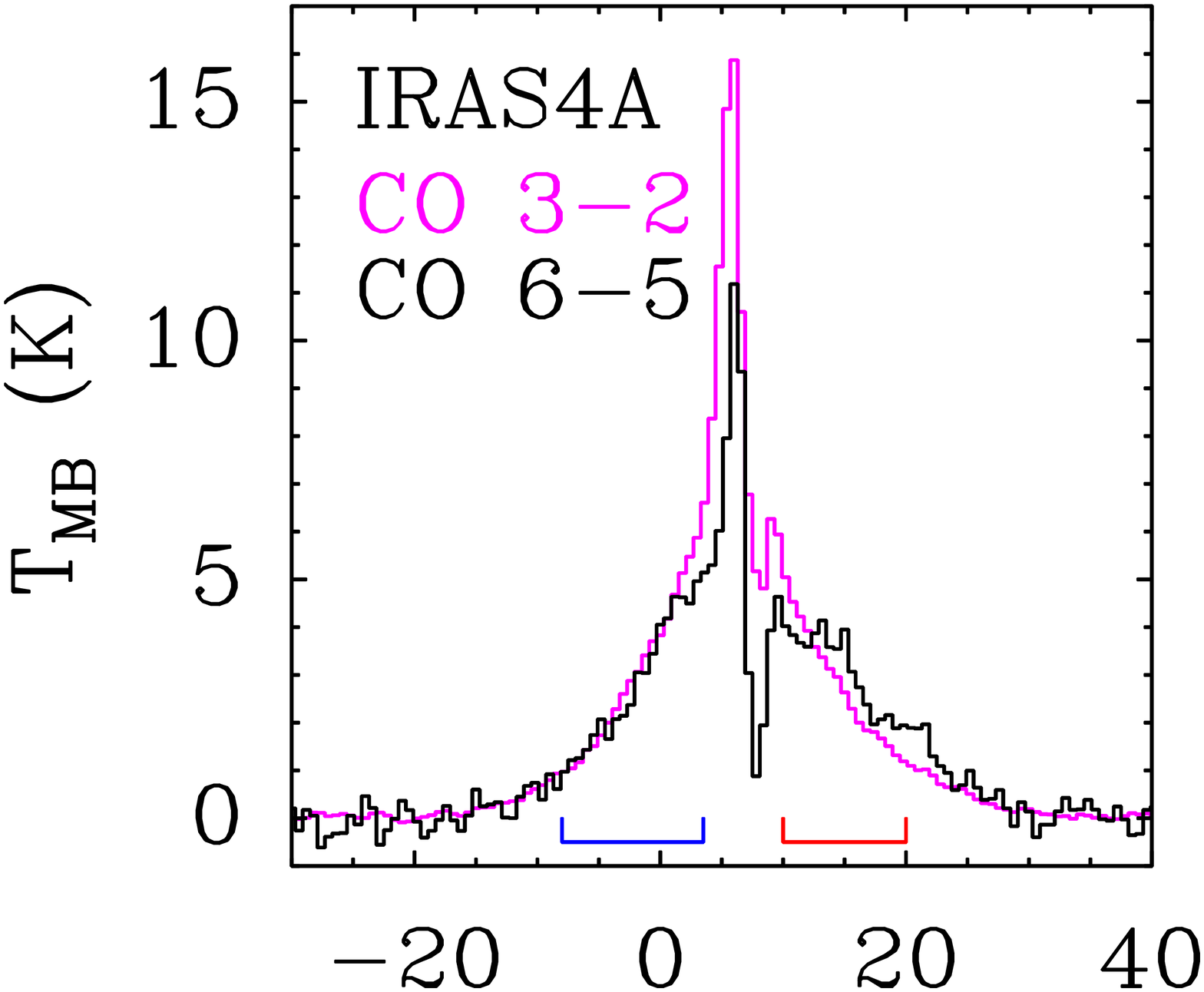}     \includegraphics[scale=0.125]{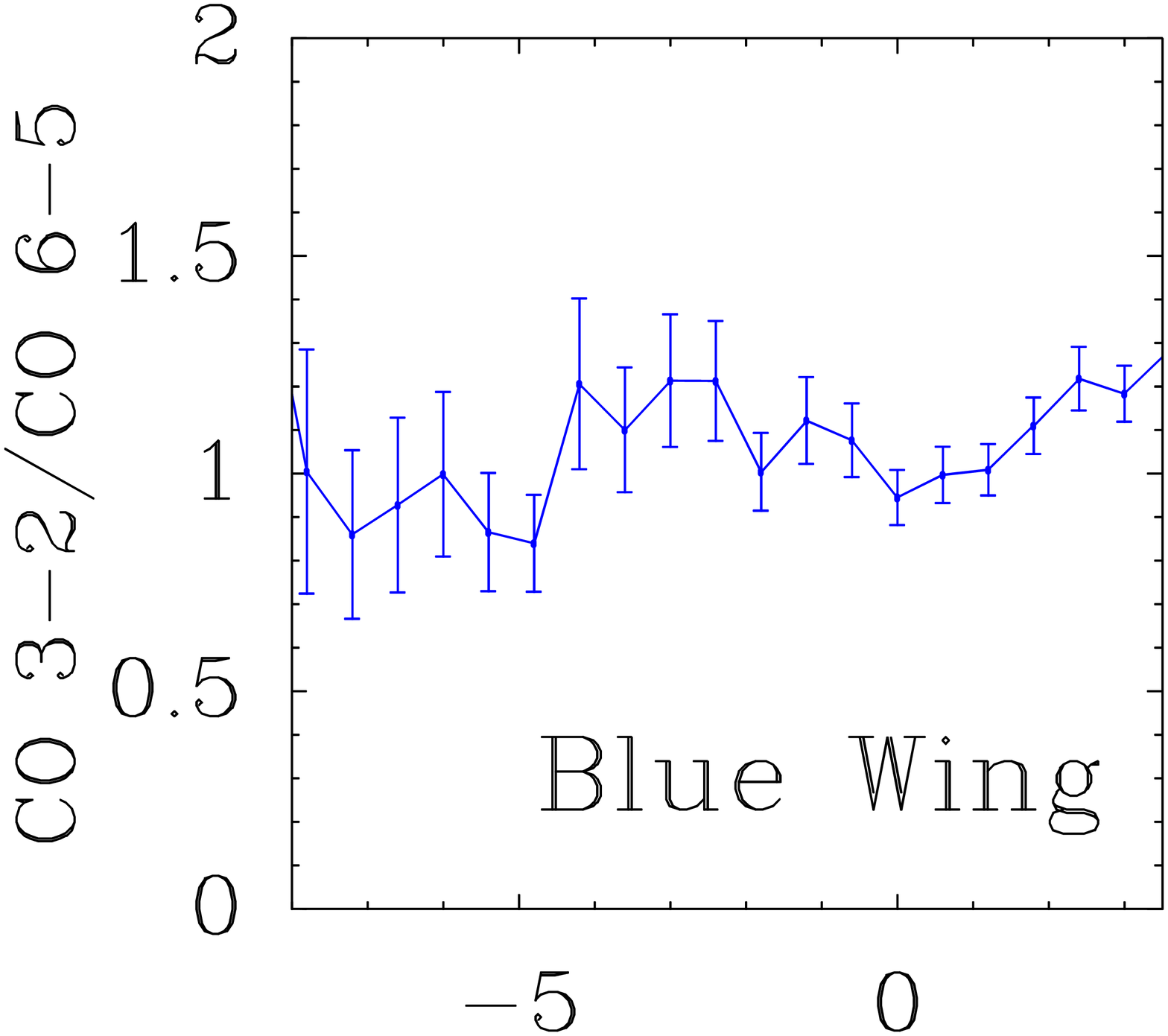}     \includegraphics[scale=0.125]{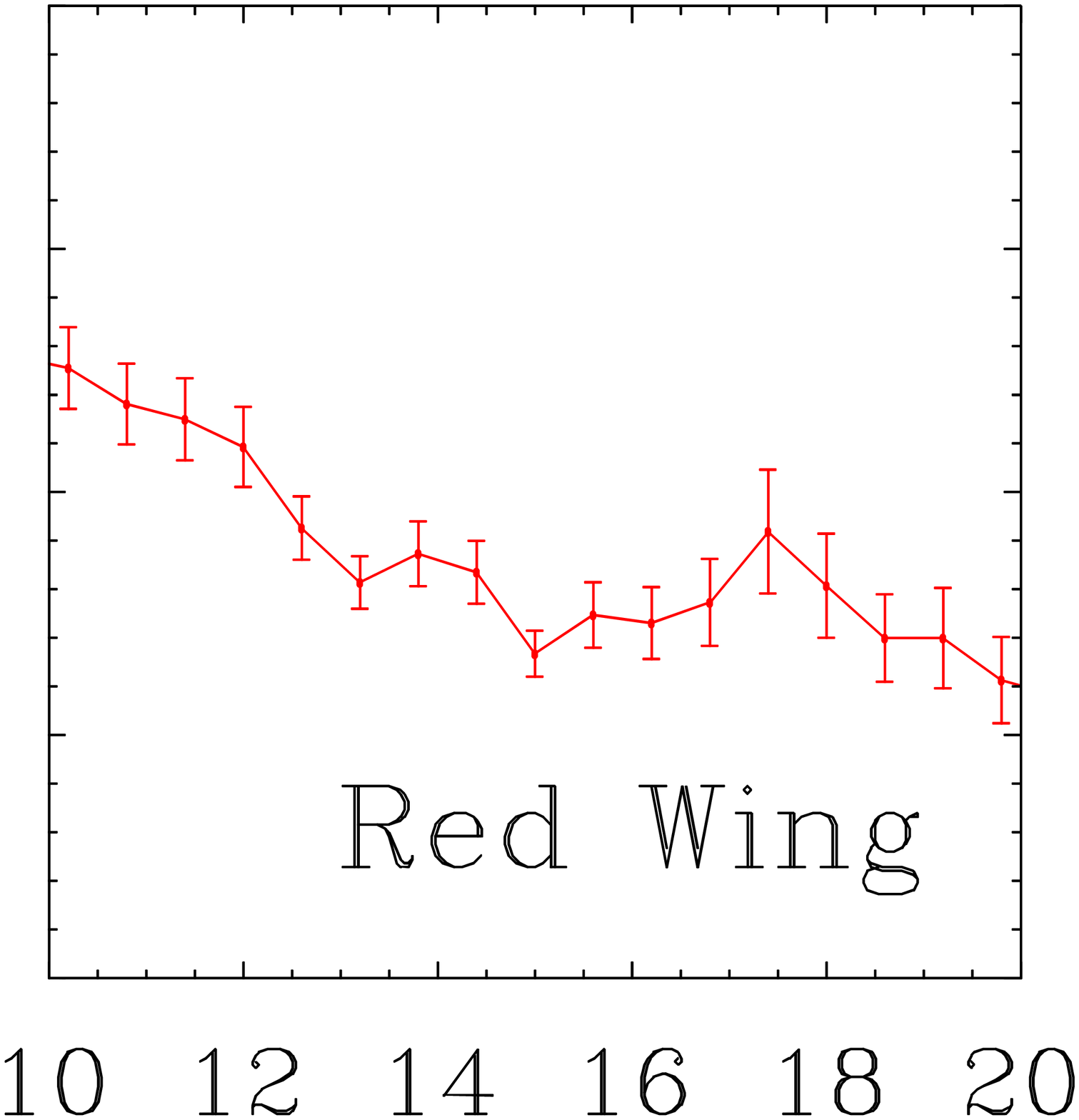}\\     \includegraphics[scale=0.125]{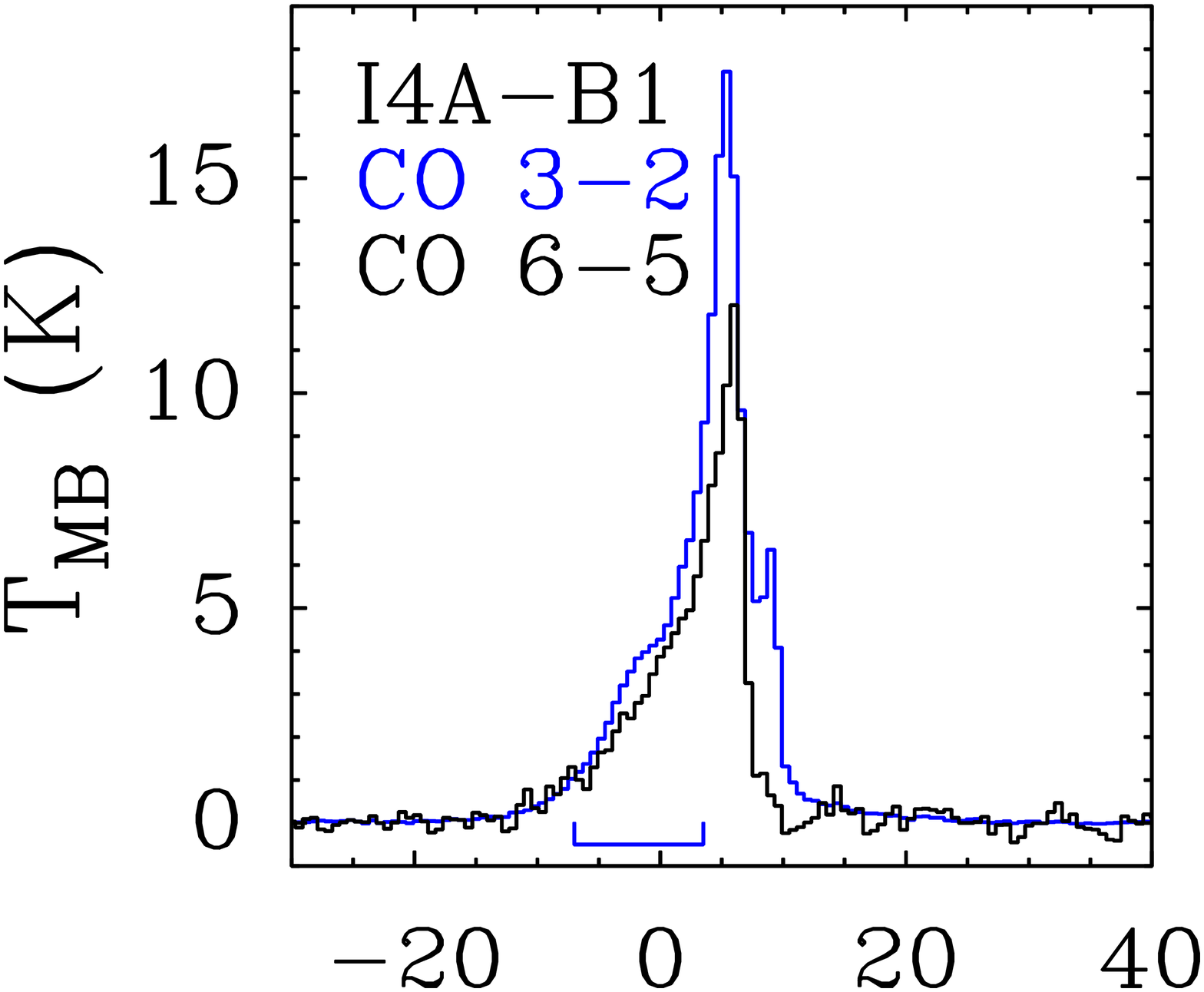}     \includegraphics[scale=0.125]{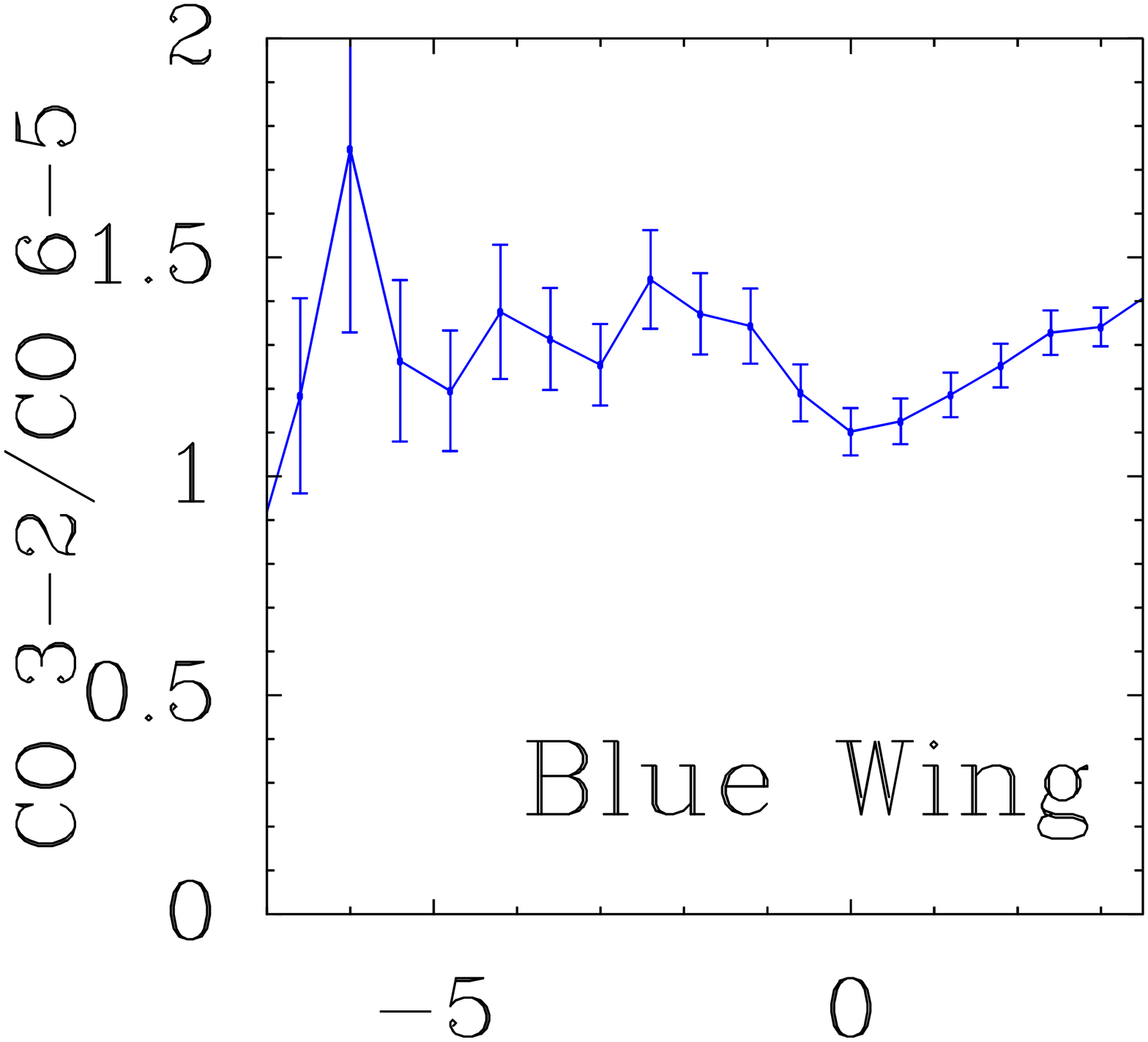}     \hspace*{2.5cm} \\
    \includegraphics[scale=0.125]{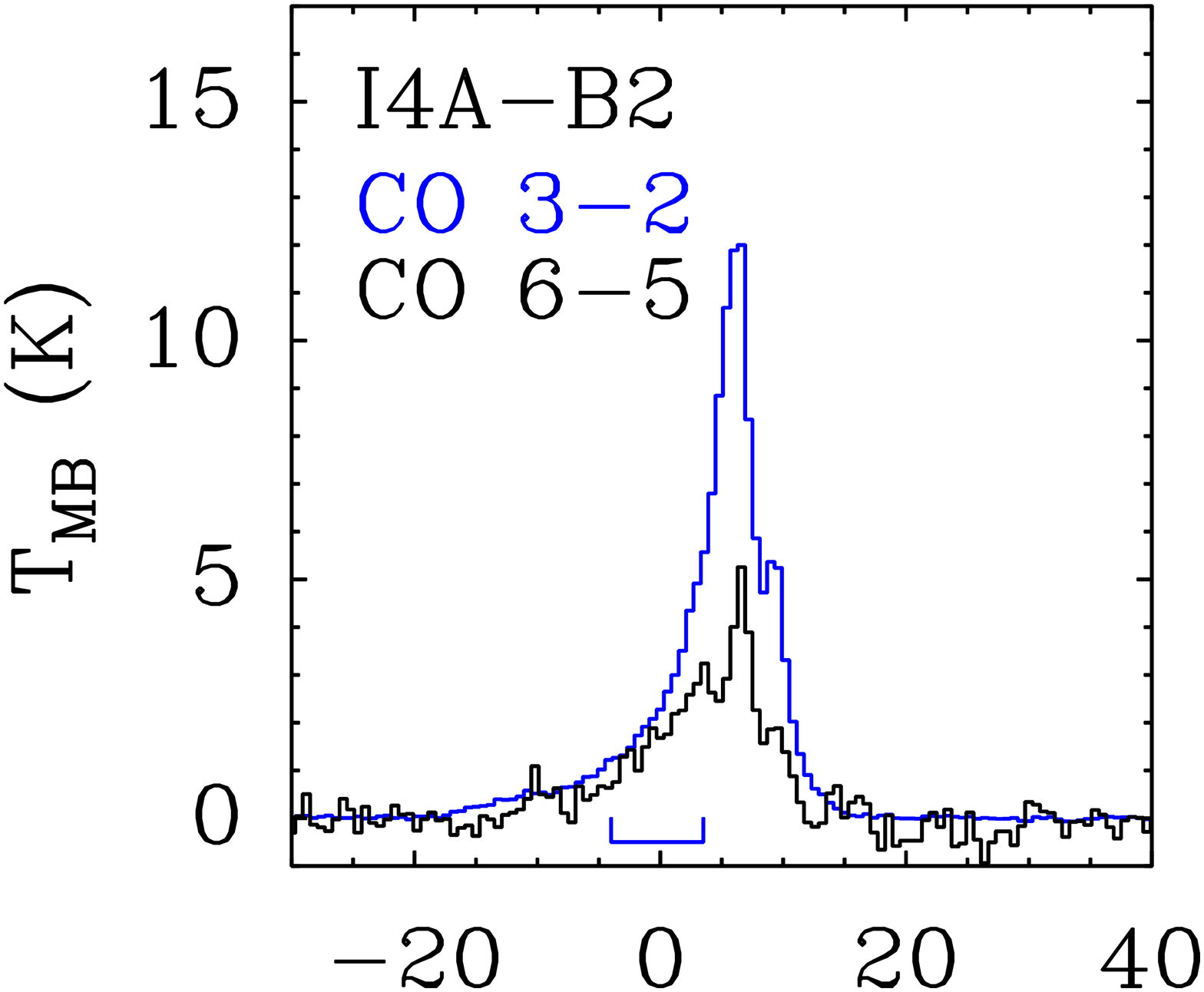}     \includegraphics[scale=0.125]{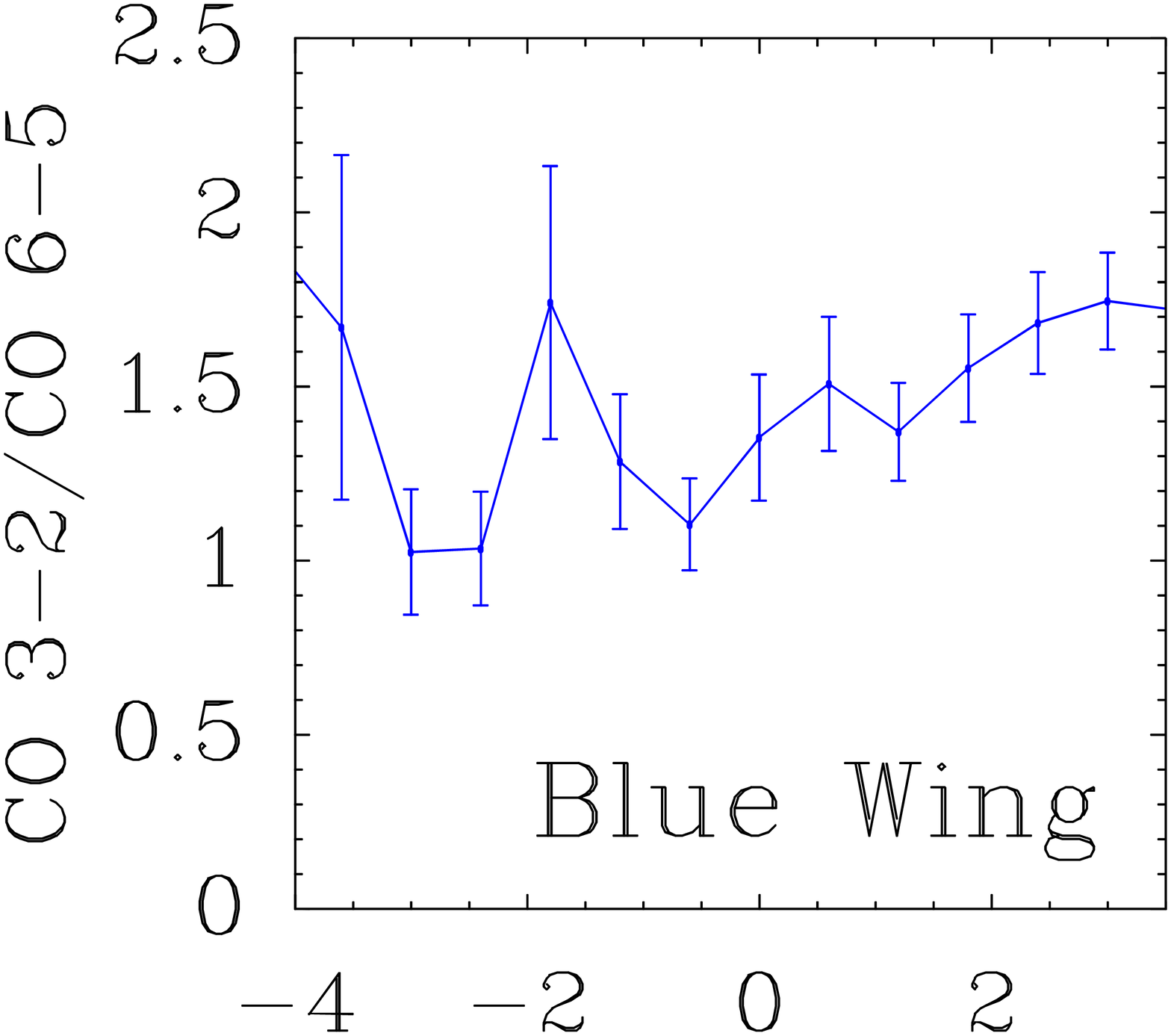}     \hspace*{2.5cm} \\
    \centering
    \includegraphics[scale=0.125]{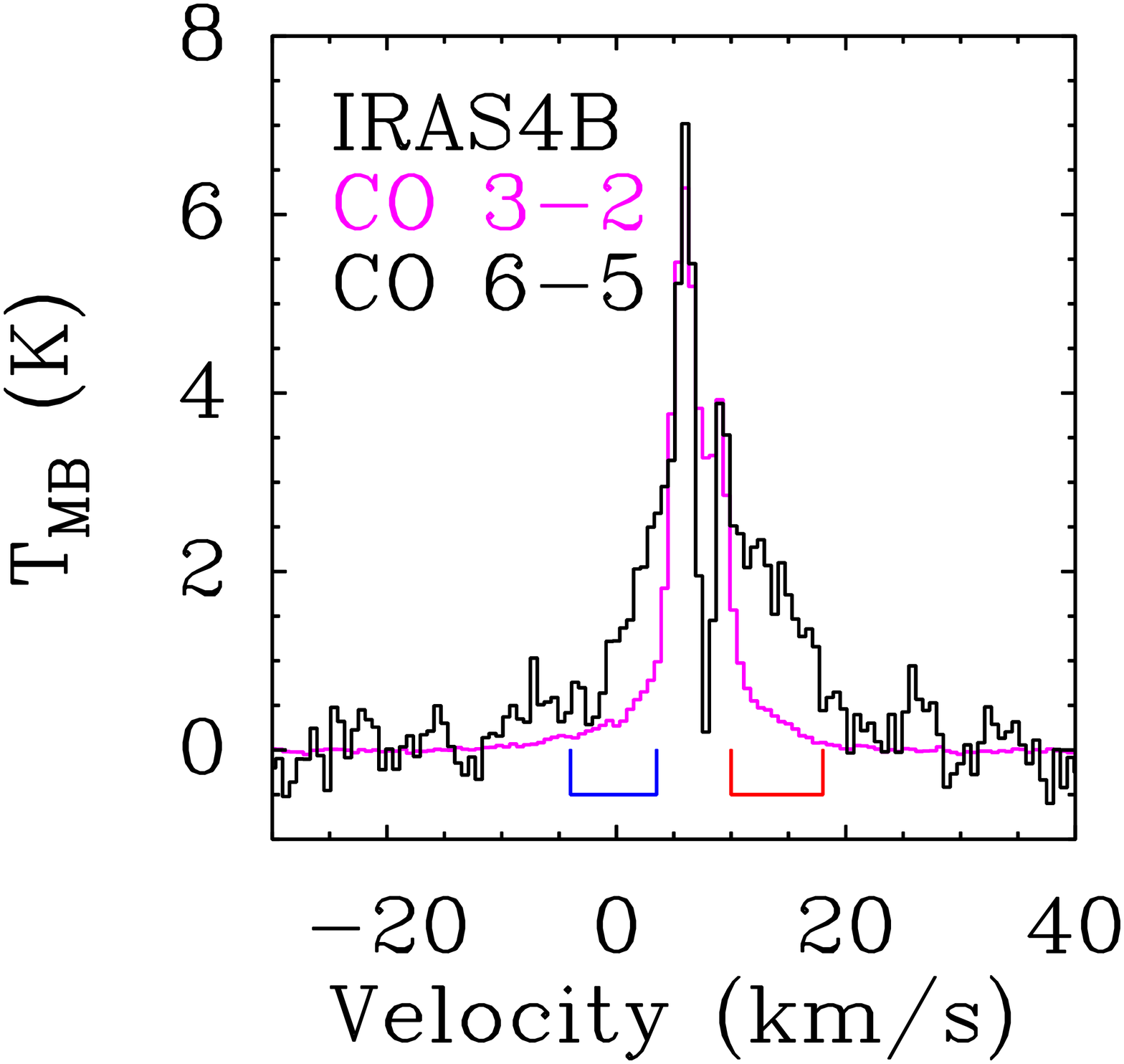}     \includegraphics[scale=0.125]{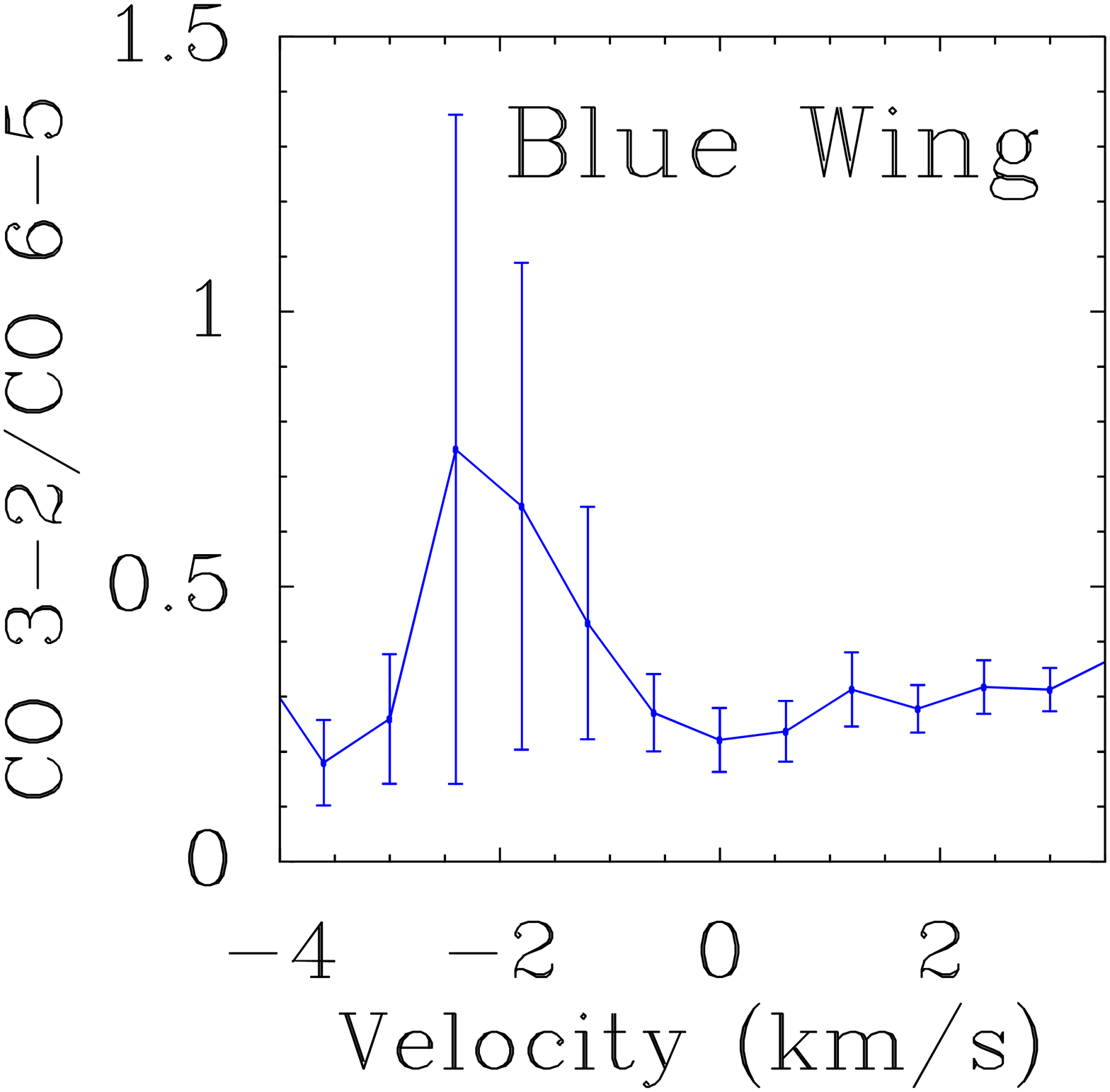}     \includegraphics[scale=0.125]{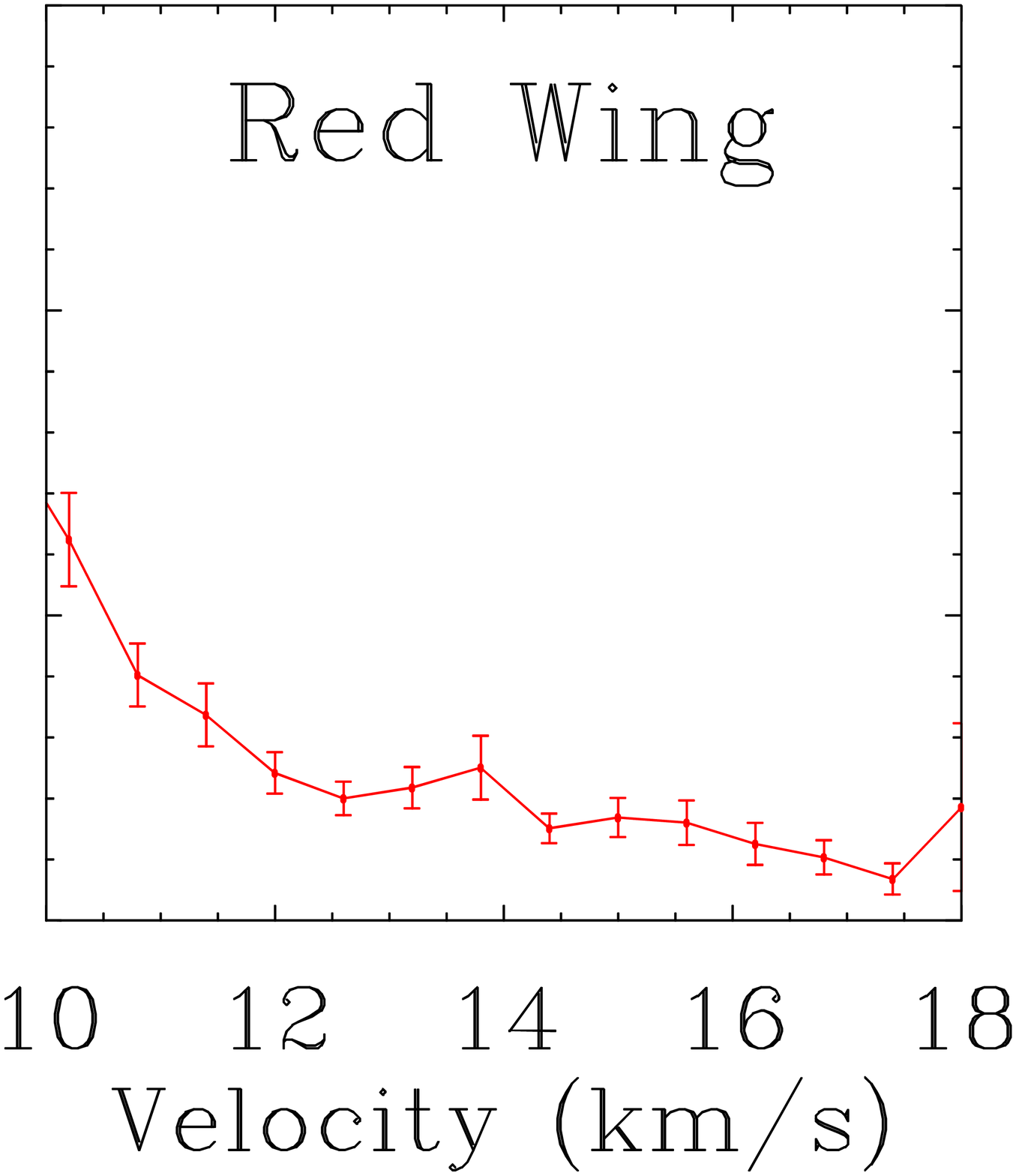}\\     \caption{\small Ratio of the $T_{\rm MB}$ temperatures of $^{12}$CO~3--2/$^{12}$CO~6--5. From top to bottom; left hand column shows IRAS~4A red outflow knots I4A-R2, I4A-R1, central source, blue outflow knots I4A-B1, I4A-B2 and IRAS~4B central source positions. Coordinates of these positions are given in Table \ref{tbl:overviewobs}. The spectra are binned to 0.6 km s$^{-1}$. The blue and red masks under the spectra in the left column show the range is used for the ratio calculations. Right hand column shows the ratios of these transitions.}
    \label{fig:i4a_co3-2_6-5ratios}
\end{figure}
}

\def\placeFigureIrasFourABlueRedOutflowsOpticalDepthRatioSou{
\begin{figure*}[tb]
    \centering
    \includegraphics[scale=0.20]{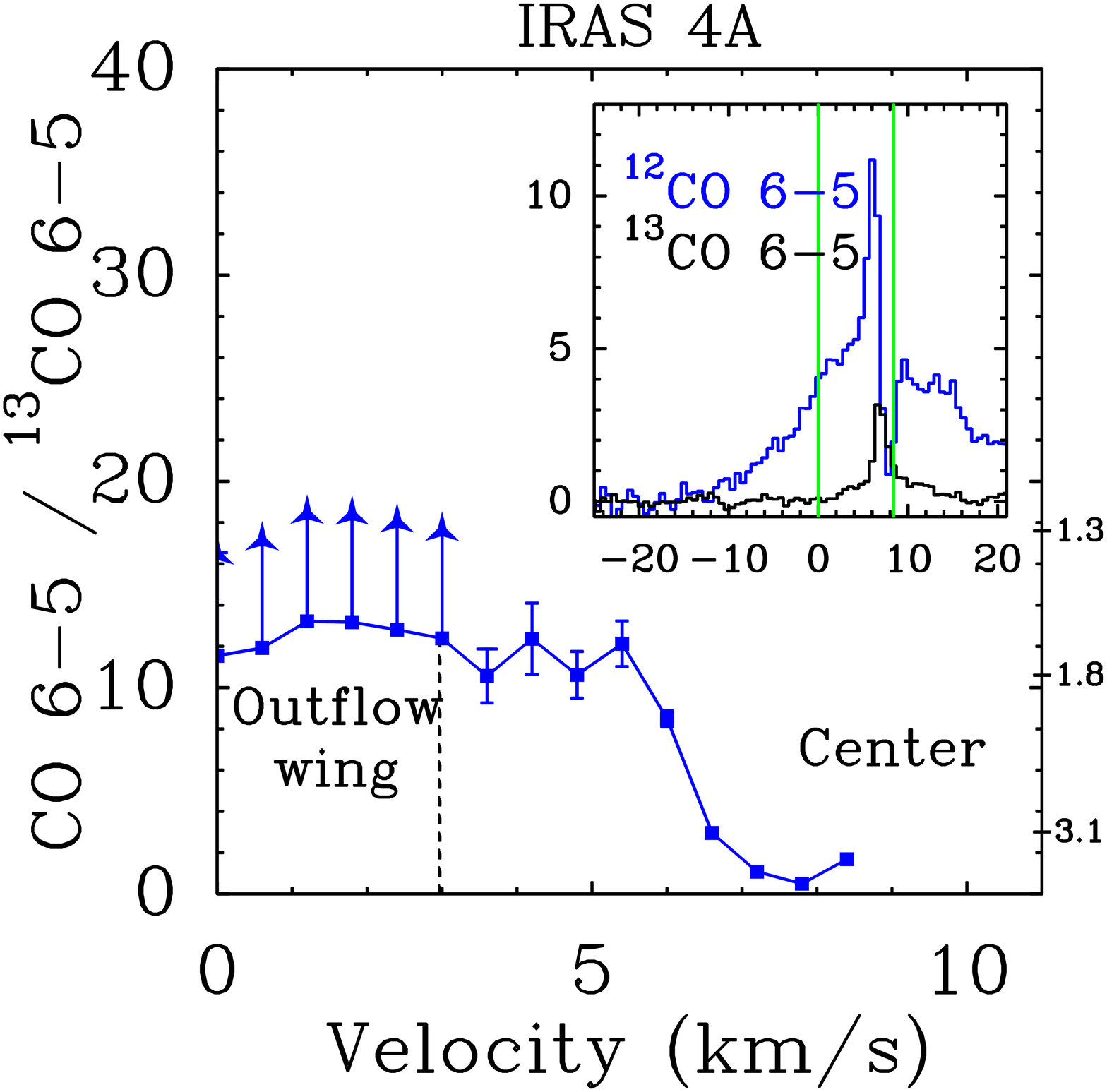}     \hspace*{0.5cm}
    \includegraphics[scale=0.20]{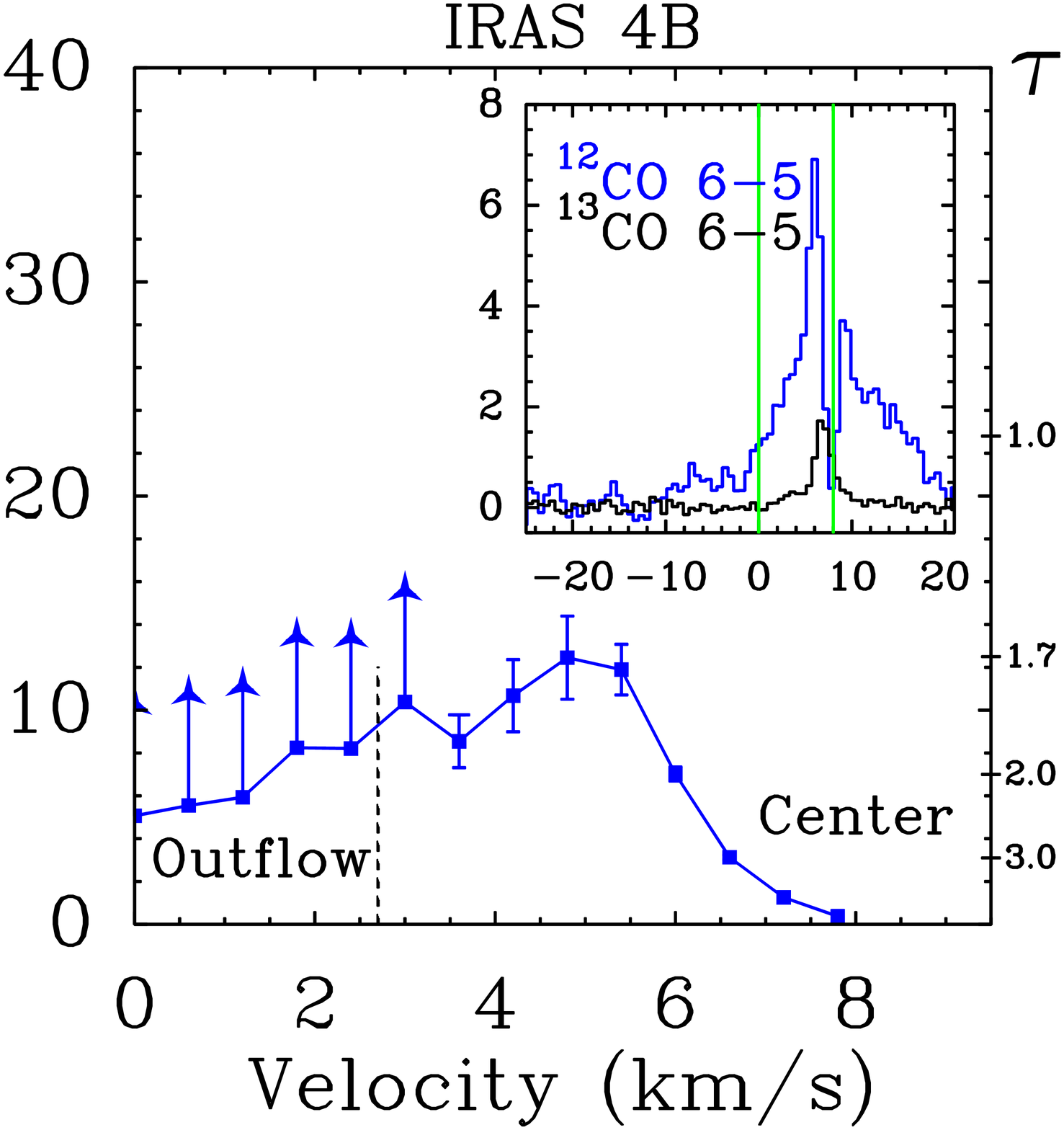}     \hspace*{0.5cm}
    \includegraphics[scale=0.20]{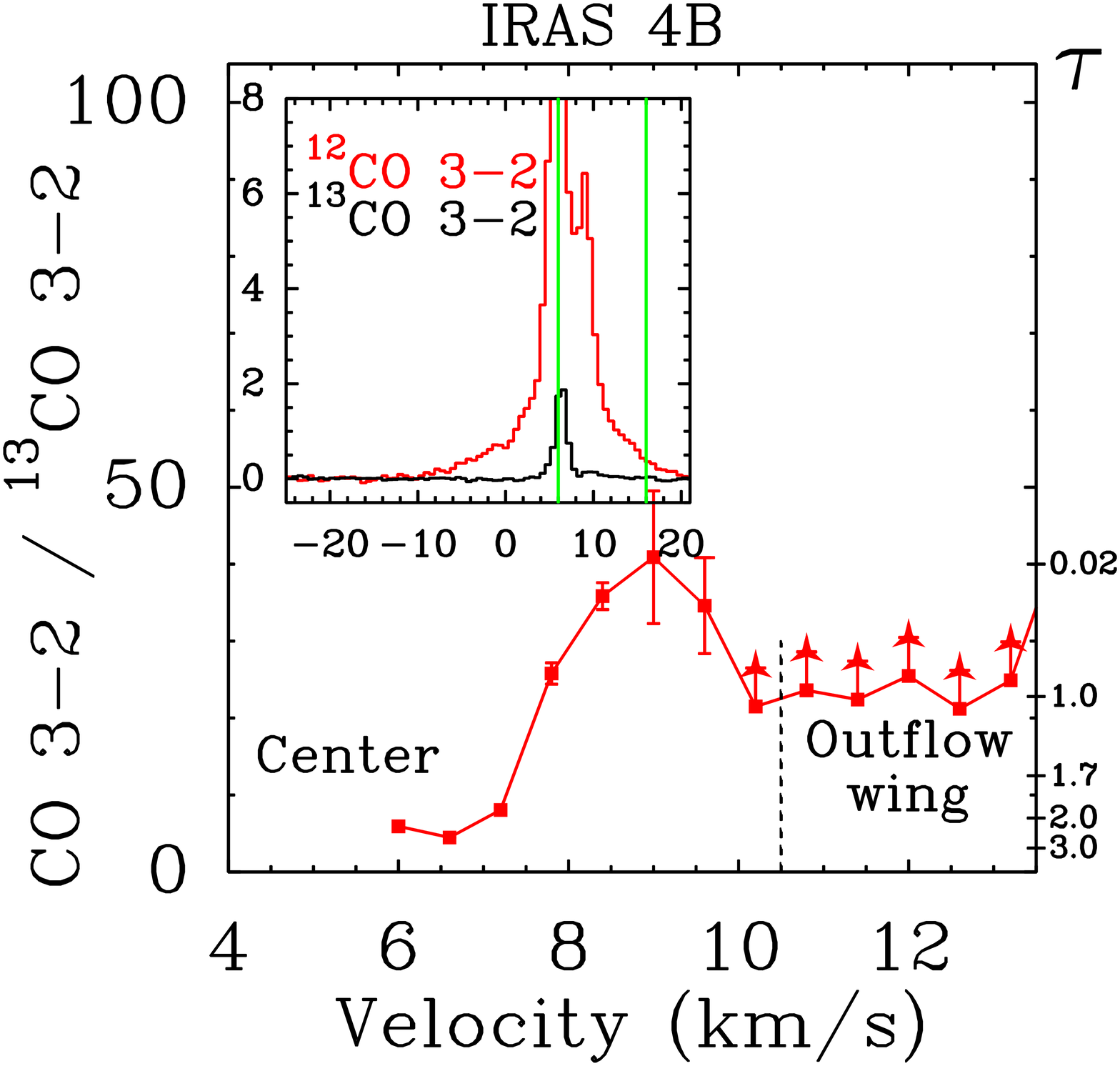}     \caption{\small Ratio of $T_{\rm MB}$ 
      $^{12}$CO~6--5/$^{13}$CO~6--5 at the IRAS~4A and IRAS~4B source
      positions and $^{12}$CO~3--2/$^{13}$CO~3--2 at the IRAS~4B in left, 
      middle and right figures, respectively. The insets
      display the corresponding spectra and the green lines show the
      limits of the velocities over which these ratios are taken.
      The resulting optical depths of $^{12}$CO as a function of velocity are shown on the
      right-hand axes. The spectra are binned to 0.6~km~s$^{-1}$.}
    \label{fig:i4a_sou_coratios}
\end{figure*}
}

\def\placeFigureIrasFourABlueRedOutflowsOpticalDepthRatioOutflows{
\begin{figure*}[htb]
    \centering
    \includegraphics[scale=0.20]{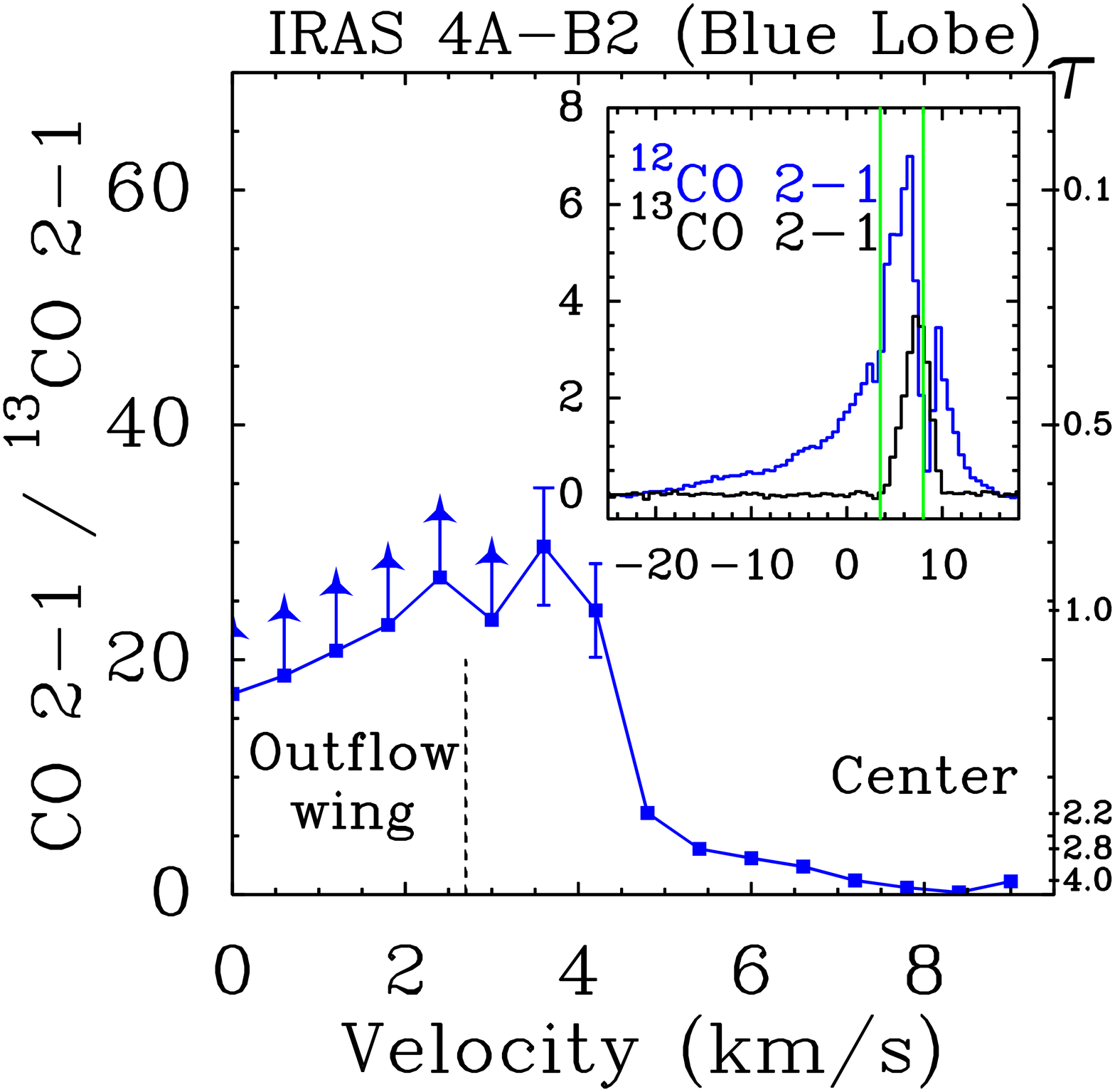}     \hspace*{0.5cm}
    \includegraphics[scale=0.20]{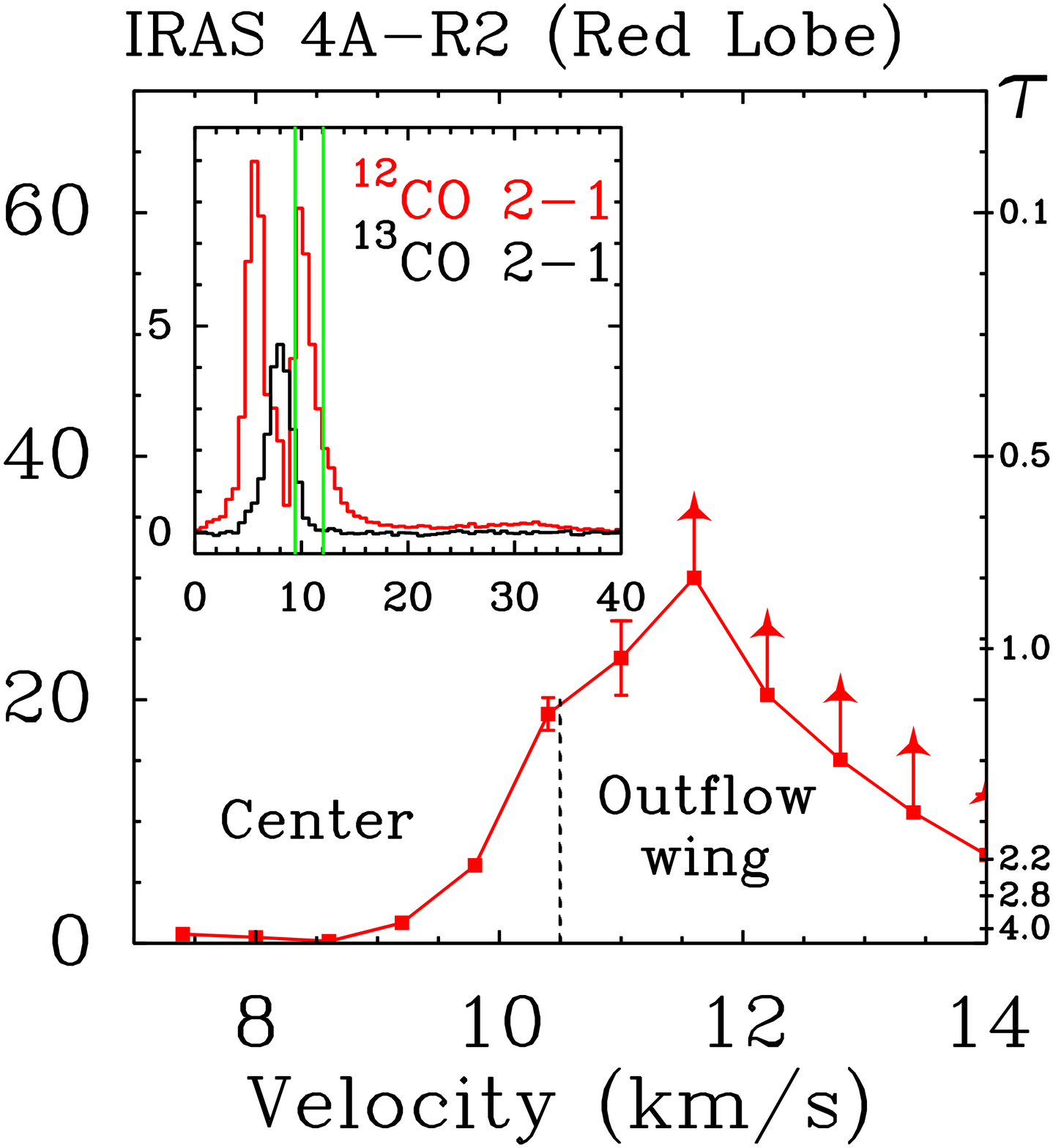}     \caption{\small Ratio of $T_{\rm MB}$ 
      $^{12}$CO 2--1/$^{13}$CO 2--1 at the I4A-B2 (left)
      and I4A-R2 (right) outflow positions. The insets display the corresponding
      spectra and the green lines show the limits of the velocities over which
      these ratios are taken.
    The resulting optical depths of $^{12}$CO as a function of velocity are shown on the right-hand axes. The spectra are binned to 0.6~km~s$^{-1}$.}
    \label{fig:i4a_b2r2_coratios}
\end{figure*}
}

\def\placeFigureEnvelopePlot{
\begin{figure}[tb]
    \centering
    \includegraphics[scale=0.18]{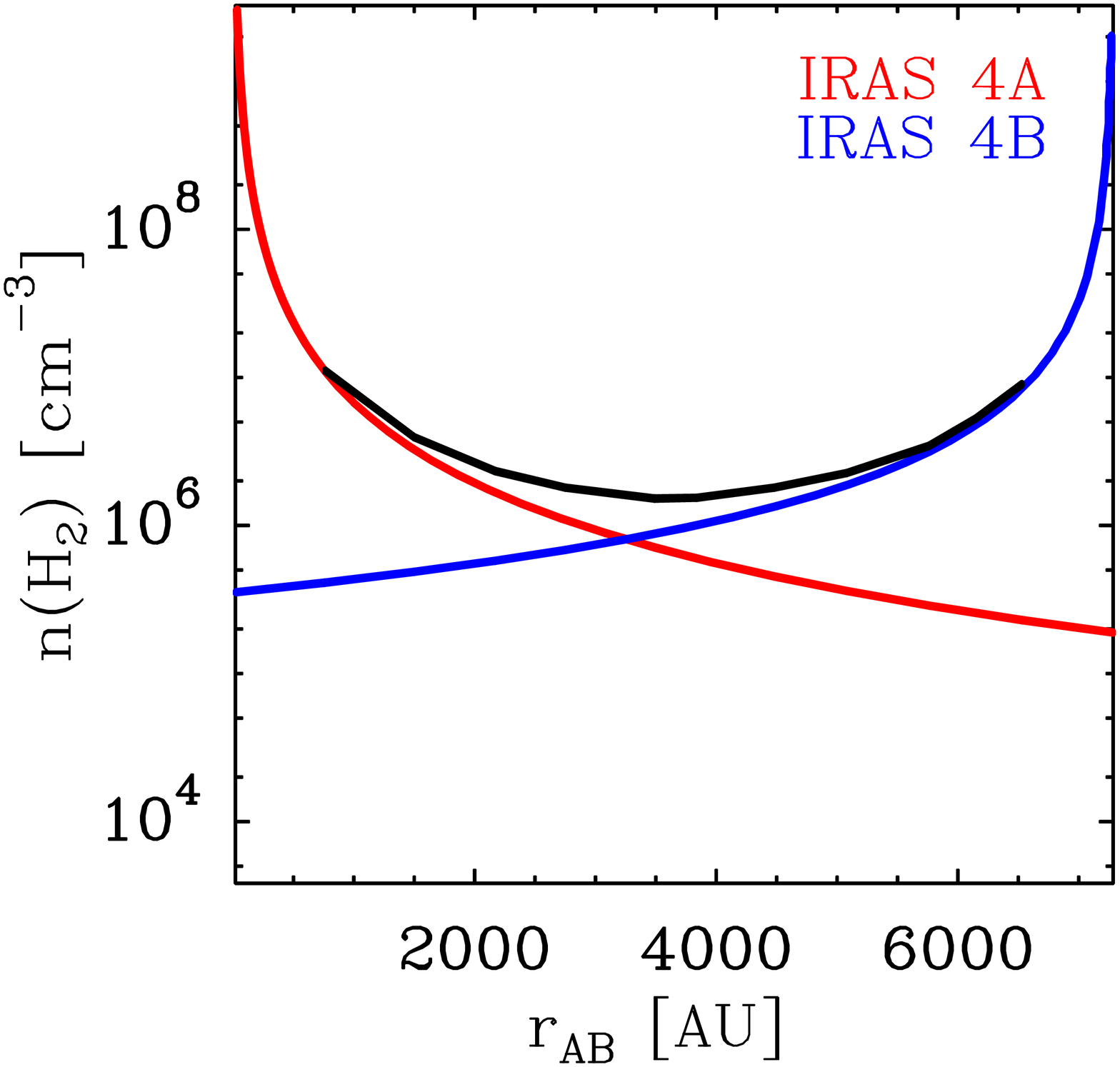}     \includegraphics[scale=0.18]{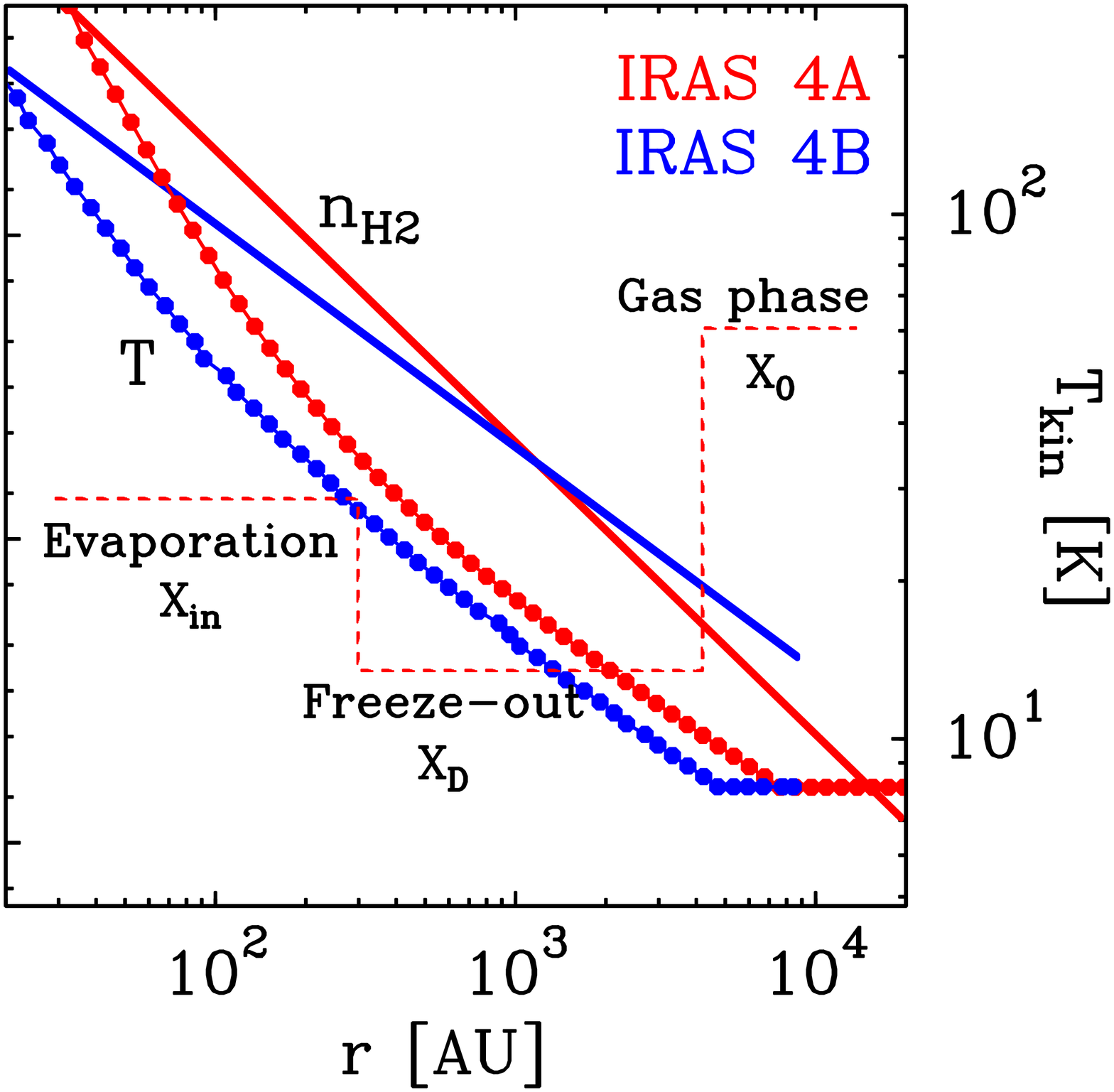}     \caption{\small Power-law density profiles discussed for two
      scenarios in Section 4.3. In the left panel, the IRAS~4A
      position is taken as the reference and $r_{\rm AB}$ indicates
      the 4A-4B distance. In the right-hand panel the individual
      envelope profiles are shown. This panel includes a typical
      drop-abundance profile, with an outer abundance $X_0$, a
      freeze-out abundance $X_D$, and an inner abundance $X_{\rm
        in}$. In an anti-jump profile, the evaporation jump in the
      inner envelope is lacking.}
    \label{fig:envelope_plot}
\end{figure}
}

\def\placeFigureAntiJumpAllChi{
\begin{figure}[tb]
    \centering
    \includegraphics[scale=0.43]{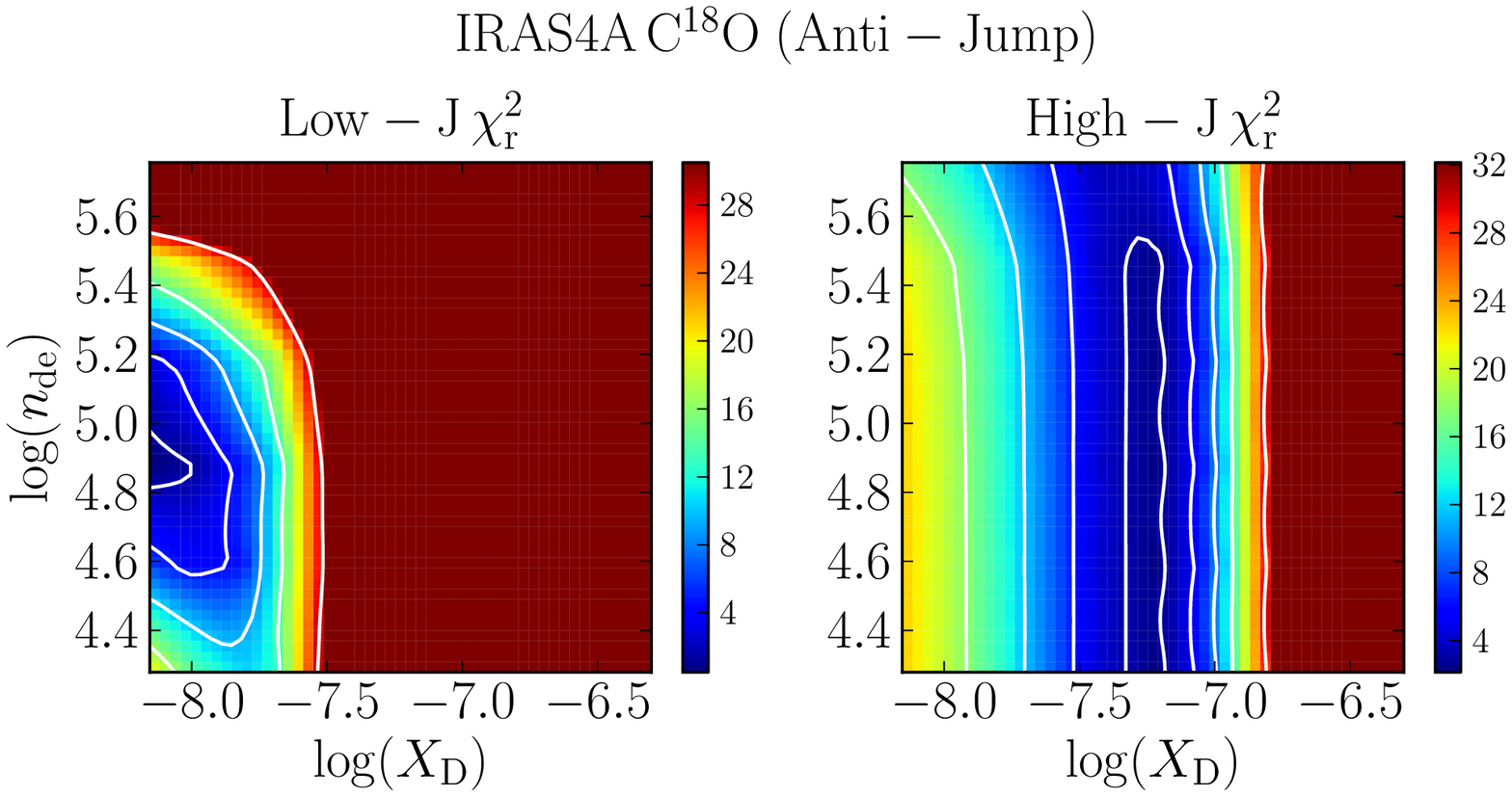}     \caption{\small Reduced $\chi^{2}$ plots for the anti-jump abundance profile in IRAS~4A for the C$^{18}$O lines in which the freeze-out abundance $X_D$ and depletion density $n_{\rm de}$ are varied. The left panel take the \mbox{low-$J$} lines 
      C$^{18}$O~1--0, 2--1, 3--2 into account, whereas the right panels use the \mbox{high-$J$} C$^{18}$O~5--4, 6--5, 9--8 and 10--9 lines. The
      contours are plotted for the 1$\sigma$, 2$\sigma$, 3$\sigma$, 4$\sigma$,
      5$\sigma$ confidence levels.}
    \label{fig:antijumpallchi4ab}
\end{figure}
}

\def\placeFigureDropAllChiXDXin{
\begin{figure}[tb]
    \centering
    \includegraphics[scale=0.36]{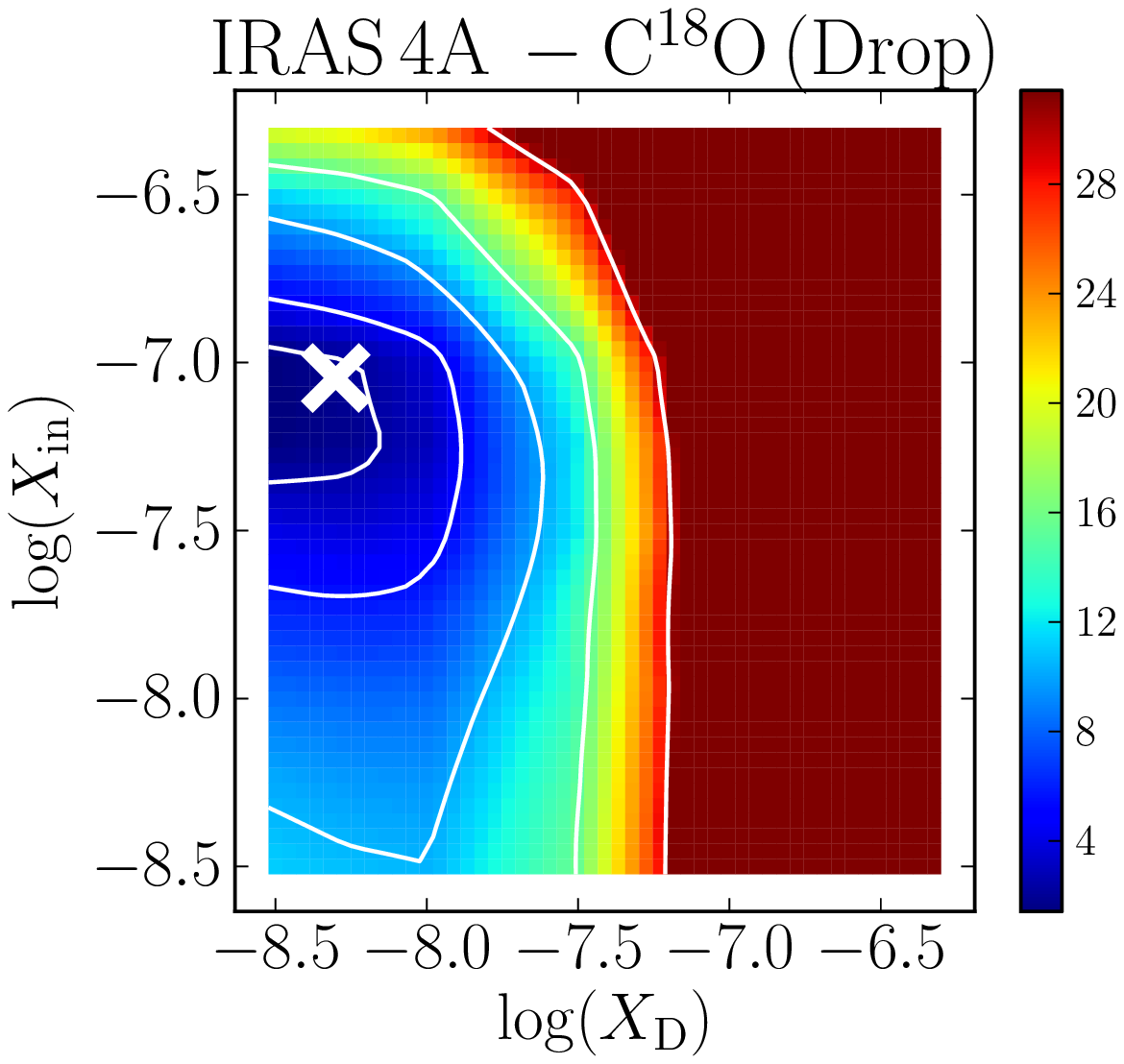}     \includegraphics[scale=0.36]{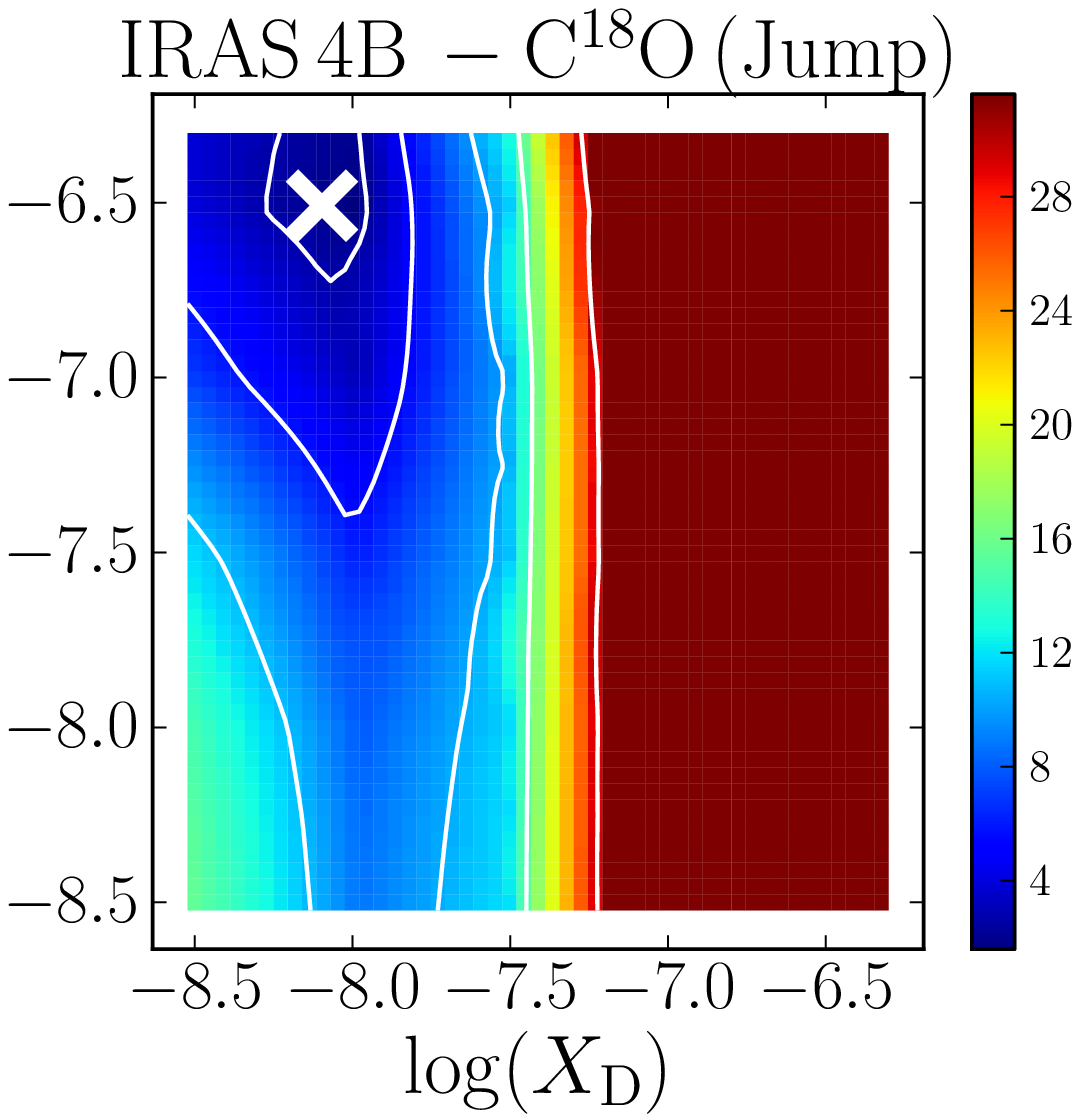}     \caption{\small Reduced $\chi^{2}$ plots for the drop and jump abundance
      profile for the C$^{18}$O lines in IRAS~4A and 4B, respectively. 
    The freeze-out abundance $X_{\rm D}$ and inner abundance $X_{\rm
        in}$ are varied. All lines are taken into account except $J$=1--0 and 5--4 due to comparatively larger beam sizes. 
       The contours are plotted at
      1$\sigma$, 2$\sigma$, 3$\sigma$, 4$\sigma$ confidence levels and
      white crosses show the best-fit values.}
    \label{fig:dropallXDXin}
\end{figure}
}

\def\placeFigureBestfitXR{
\begin{figure}[tb]
    \centering
    \includegraphics[scale=0.46]{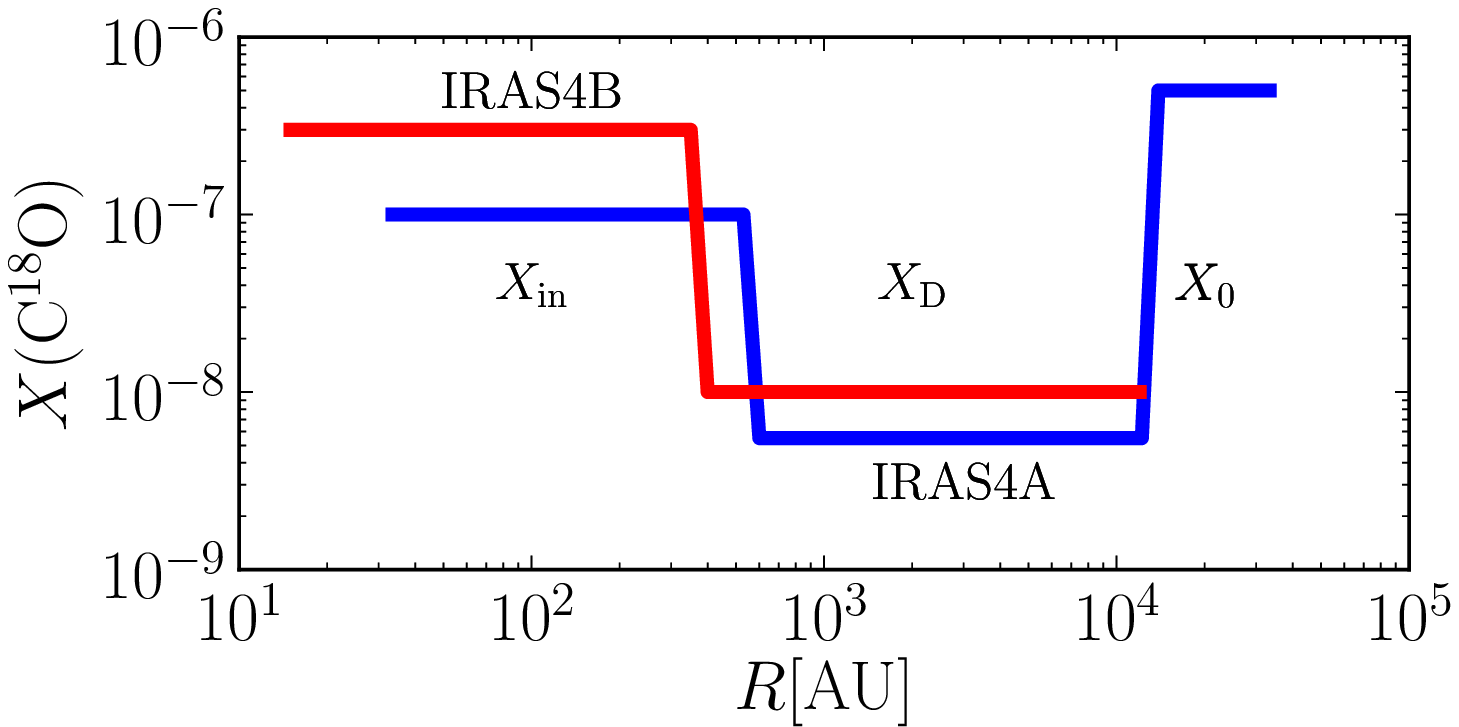}     \caption{\small Schematic diagram showing the best-fit abundance profiles for IRAS~4A (blue) and IRAS~4B (red).}
    \label{fig:bestfit_X_vs_R}
\end{figure}
}

\def\placeFigureDropOverplots{
\begin{figure}[tb]
    \centering
    \includegraphics[scale=0.24]{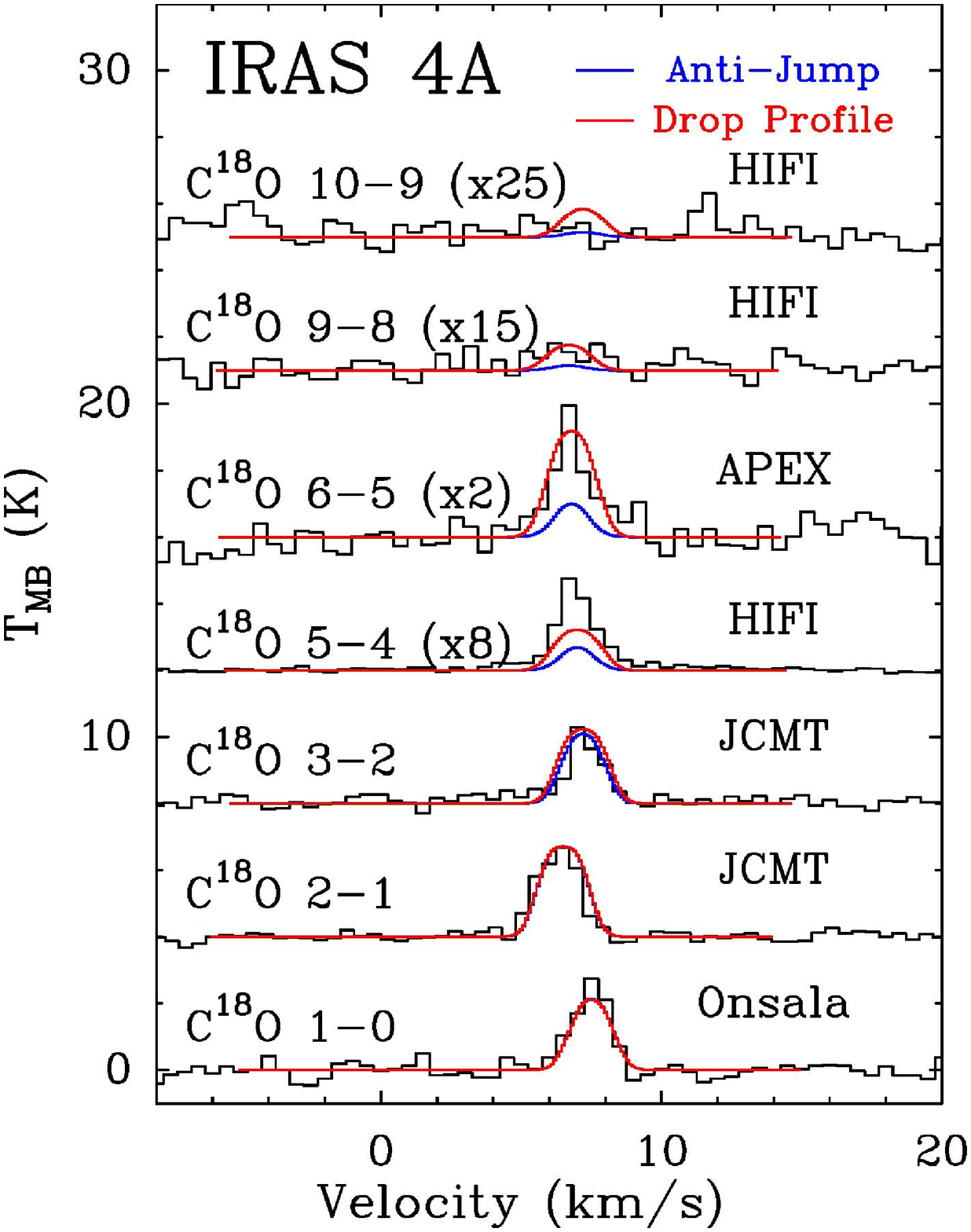}     \includegraphics[scale=0.211]{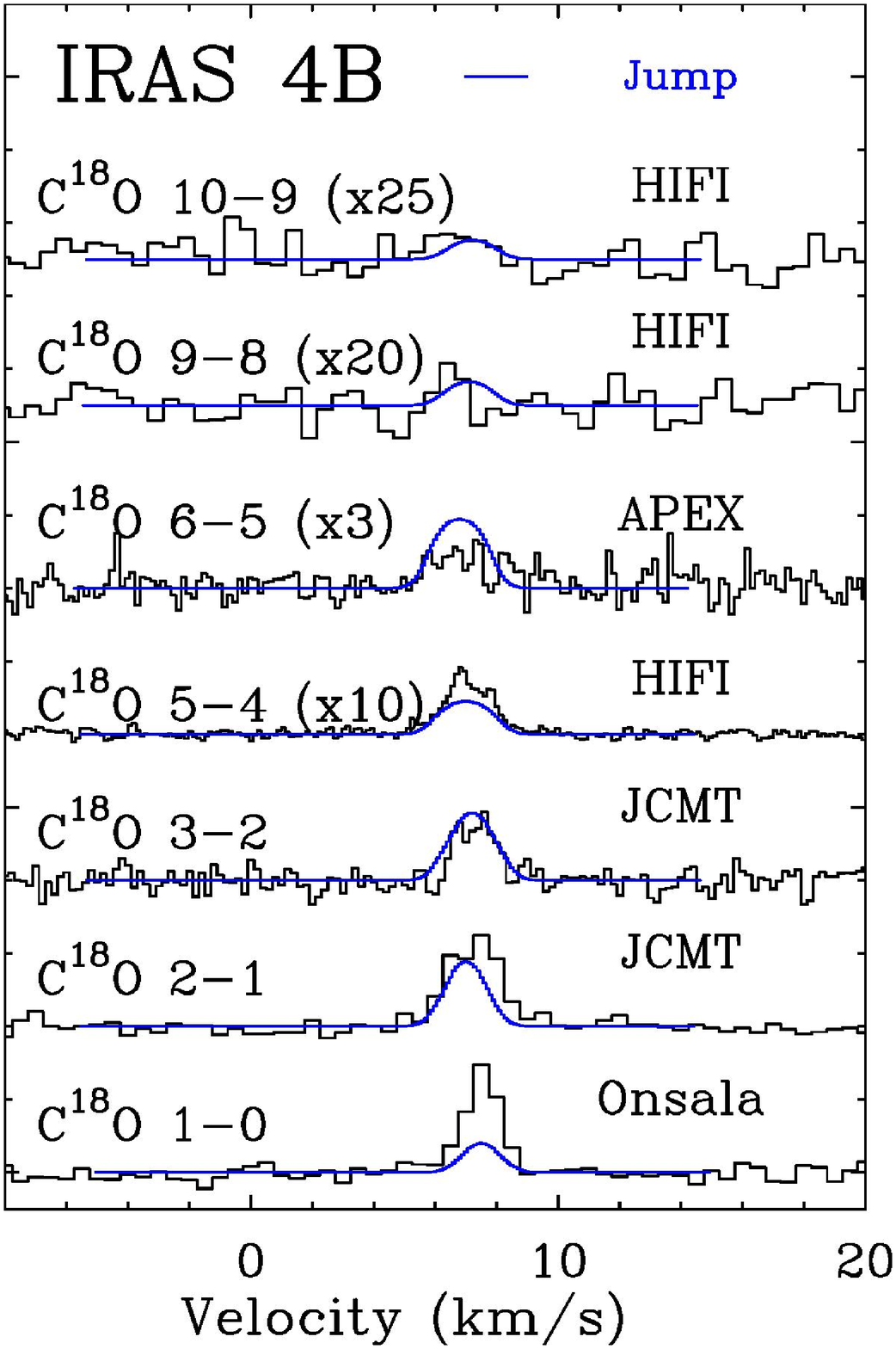}     \caption{\small {\it Left:} Line profiles obtained with the best-fit anti-jump
      (blue) and drop abundance (red) envelope models overplotted on the
      observed C$^{18}$O lines in IRAS~4A. {\it Right:} Similar best-fit jump abundance profile for IRAS~4B.
      See Table~\ref{tbl:abundancetable} for best-fit parameters.}
    \label{fig:dropoverplot4ab}
\end{figure}
}

\def\placeFigureDropThirteenCO{
\begin{figure}[tb]
    \centering
    \includegraphics[scale=0.4]{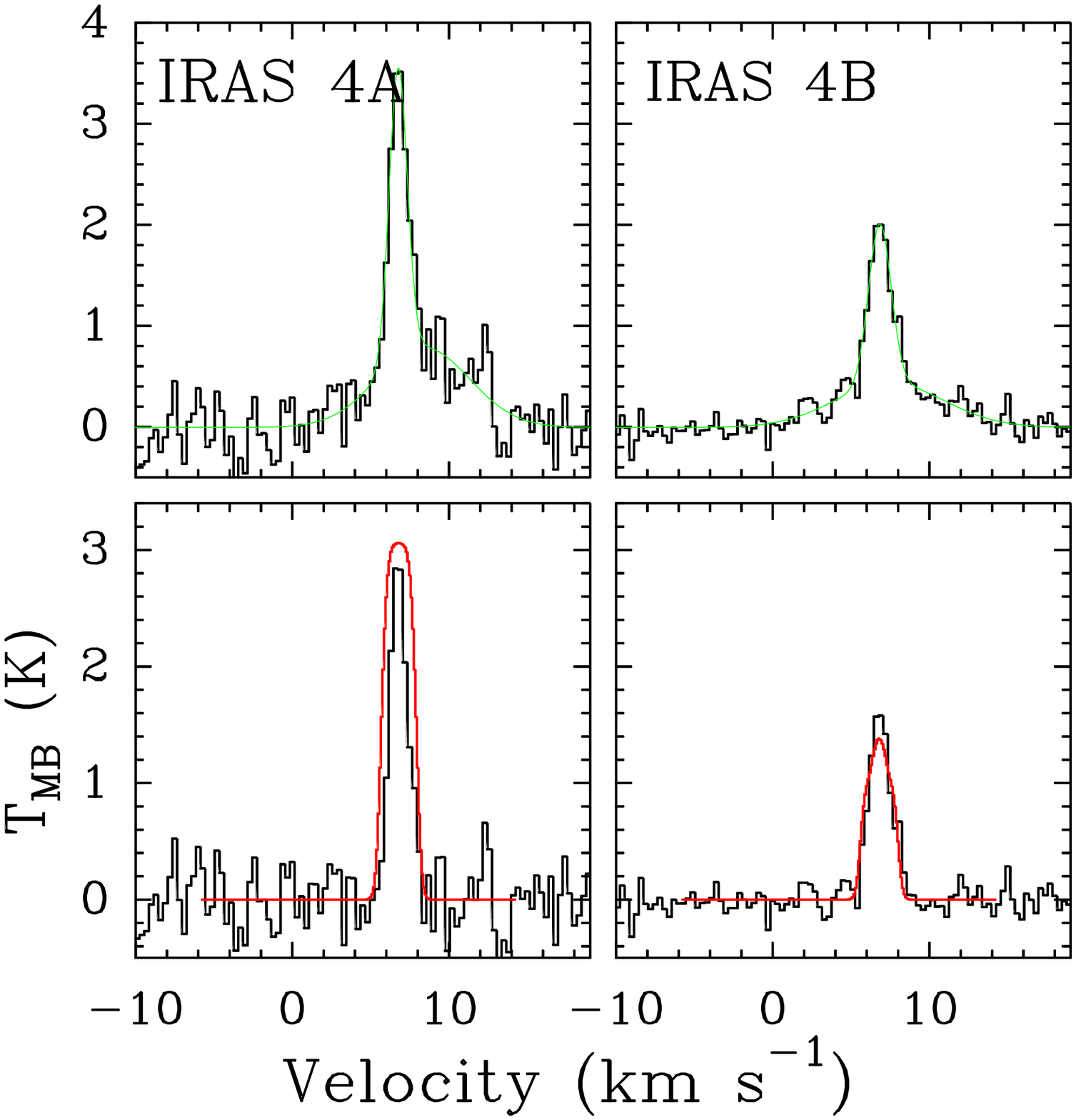}     \caption{\small {\it Top Panels:} $^{13}$CO~6--5 spectra
        of IRAS~4A and 4B at the source positions. The green line is
        the fit to the narrow plus broad components.  {\it Bottom
          panels:} The same lines after subtracting the broad
        component.  The red line indicates the $^{13}$CO envelope
        model emission using the CO drop abundance profile derived
        from the C$^{18}$O data. The figure indicates that a
        substantial fraction of the on-source emission comes from the
        passively-heated envelope. For IRAS 4A, however, there is also
        significant extended emission not due to the envelope (see
        Fig.~\ref{fig:dropThirteenCOUVEnv}). }
    \label{fig:dropThirteenCO}
\end{figure}
}

\def\placeFigureThirteenCOUVEnv{
\begin{figure}[tb]
    \centering
    \includegraphics[scale=0.32]{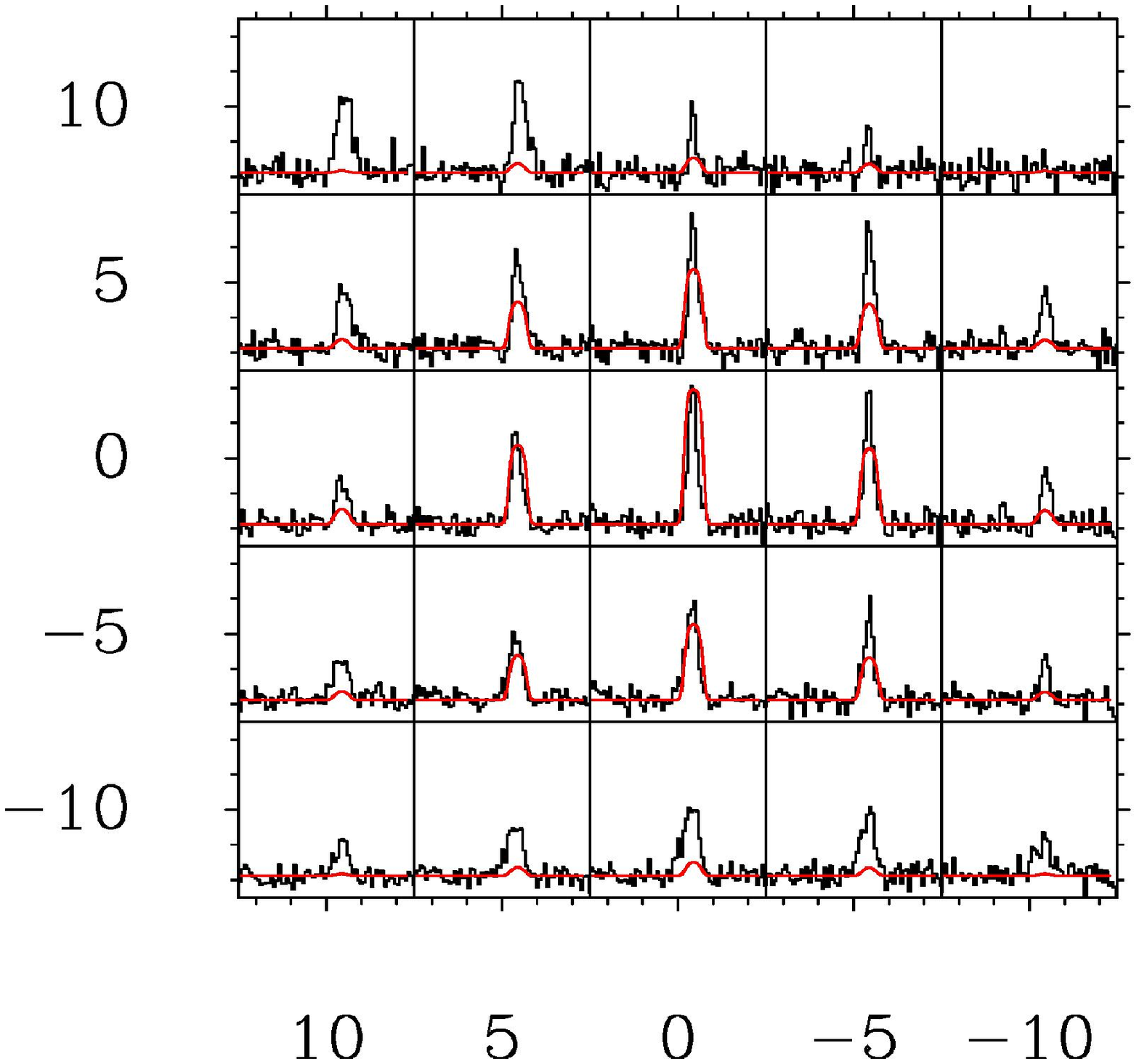}     \includegraphics[scale=0.68]{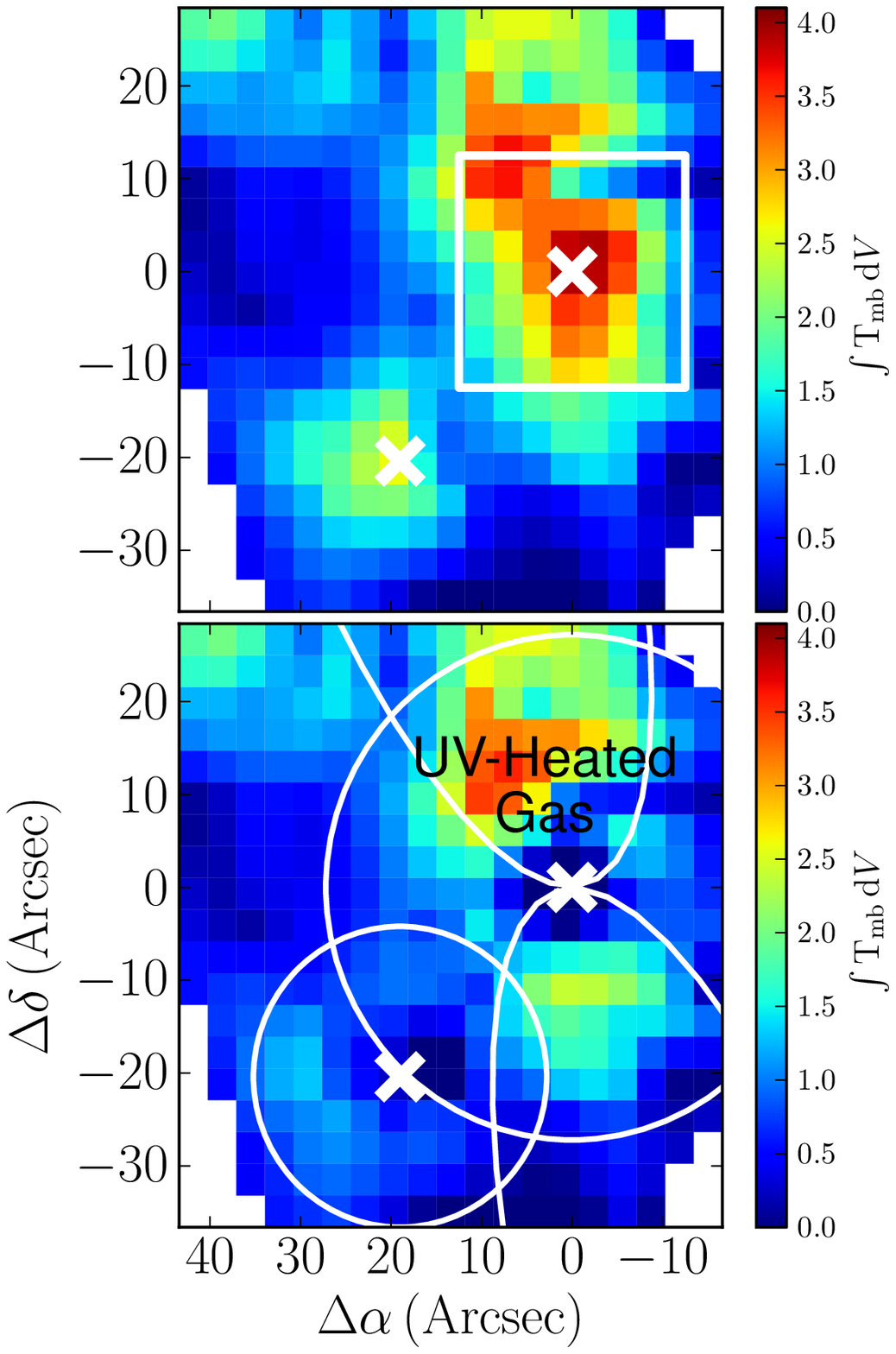}     \caption{\small {\it Top:} Central region of the $^{13}$CO~6--5
      map covering IRAS~4A. The broad component has been removed from
      the entire map. The red lines indicate the envelope model
      emission. {\it Middle:} Integrated intensity map of the narrow
      $^{13}$CO 6--5 emission, obtained by removing the broad
      component. The white square box indicates the region covered in
      the top figure. This map shows both the envelope and UV-heated
      gas. {\it Bottom:} $^{13}$CO map obtained after subtracting the
      model $^{13}$CO 6--5 envelope emission convolved to the APEX
      beam from the above map. White circles show the limits of the
      10~K radius envelope and white cones show the direction of the outflows.
      This map represents the UV-heated gas only.}
    \label{fig:dropThirteenCOUVEnv}
\end{figure}
}

\def\placeFigureIrasFourABCOthreetwospectraandoutflow{
\begin{figure}[!th]
\begin{center}
\includegraphics[scale=0.33]{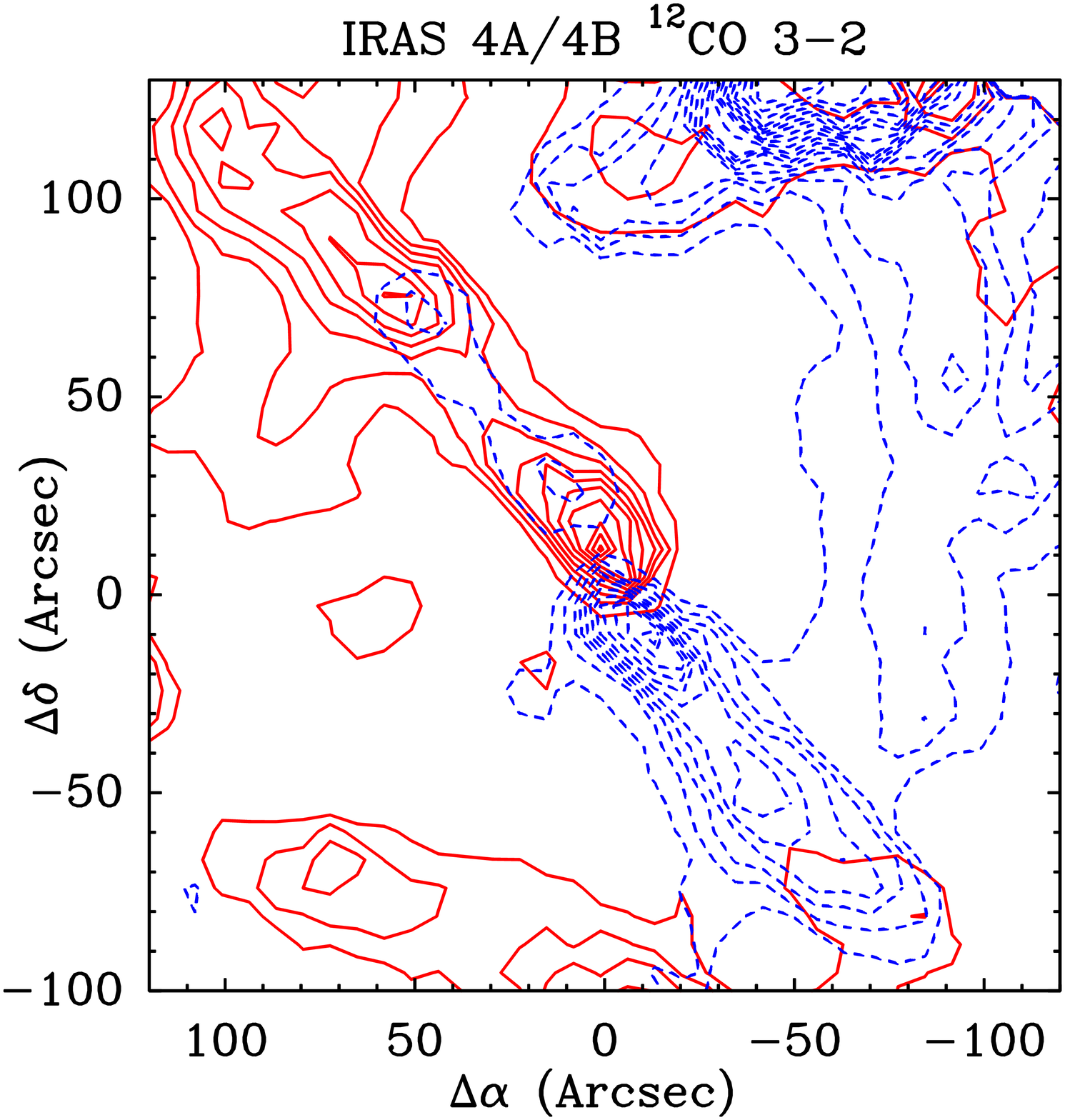} \includegraphics[scale=0.46]{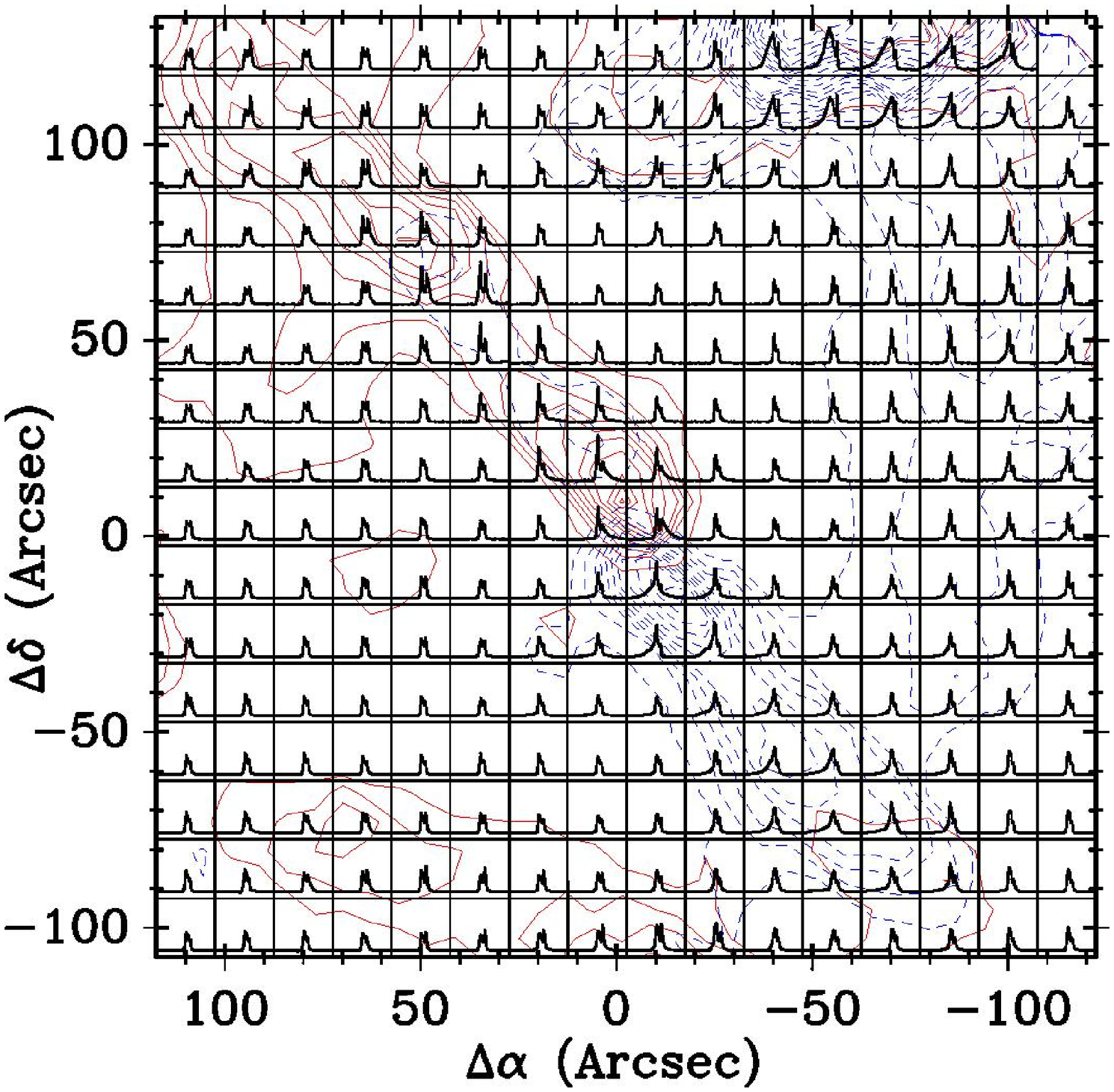} \end{center}
\caption{\mbox{$^{12}$CO~3--2} spectral map of IRAS~4A and 4B over the
  240$\arcsec\times$240$\arcsec$ mapping area. Individual spectra are
  shown on the $T_{\rm MB}$ scale from --2 to 16 K and velocity scale
  from --20 to 30 \mbox{km~s$^{-1}$}. The maps are centered on
  IRAS~4A. The contour levels start from 20$\sigma$ (10~K~km~s$^{-1}$)
  with an increasing step size of 5~K~km~s$^{-1}$.}
\label{fig:co32map}
\end{figure}
}

\def\placeFigureIRASBTwoBullet{
\begin{figure}[tb]
    \centering
    \includegraphics[scale=0.4]{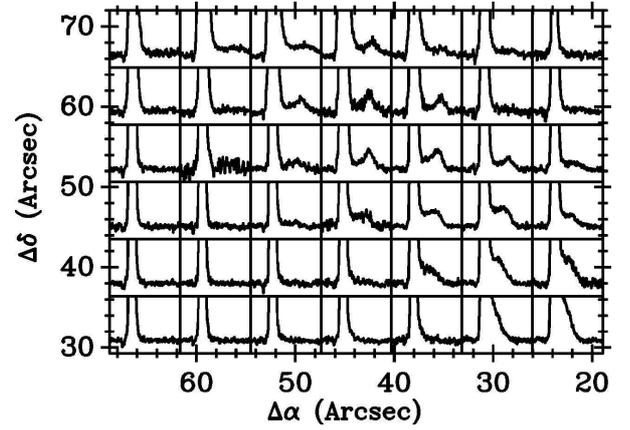}     \caption{\small Blow-up of \mbox{$^{12}$CO~3--2} spectra at the
      IRAS~4A-R2 outflow knot position.  Bullet emission at +35 km
      s$^{-1}$ is visible in the upper left part of the IRAS~4A
      outflow. Individual spectra are shown on the $T_{\rm MB}$ scale
      from --0.7 to 2.5 K and velocity scale from --10 to 45
      \mbox{km~s$^{-1}$}. The coordinates are relative to IRAS~4A.  }
    \label{fig:IRAS4A-B2_bullet}
\end{figure}
}

\def\placeFigureIrasFourABCITwoOneAndTwelveCOSixFiveSpectraAndOutflow{
\begin{figure}[!t]
\begin{center}
\includegraphics[scale=0.32]{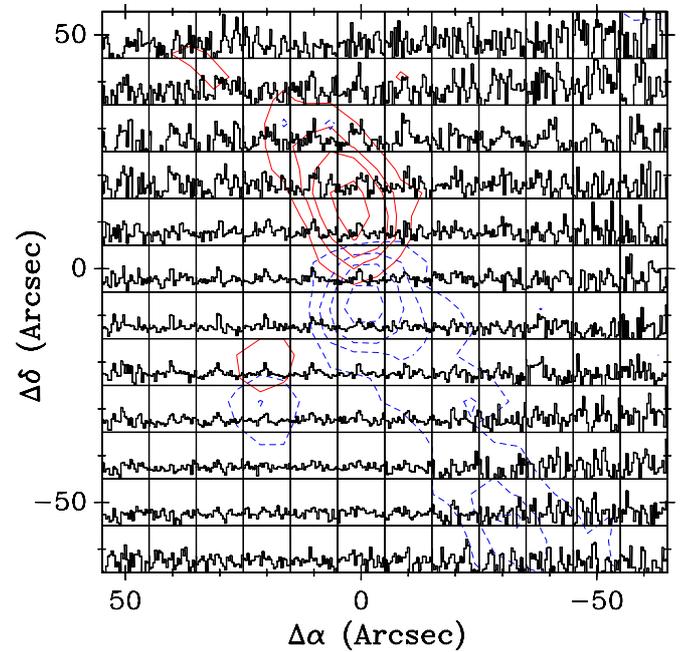} \end{center}
\caption{\mbox{[\ion{C}{i}]~2--1} spectral map (rebinned to
  10$\arcsec$x10$\arcsec$ with a 1~km s$^{-1}$ velocity resolution) is overlaid on a \mbox{$^{12}$CO~6--5}
  outflow contour map. In the \mbox{[\ion{C}{i}]~2--1} map, individual
  spectra are shown on a $T_{\rm MB}$ scale of --1 to 3 K and the
  velocity scale runs from --5 to 20 \mbox{K km s$^{-1}$}. The map is
  centered on the IRAS~4A position.}
\label{fig:cIco65overlaid}
\end{figure}
}

\def\placeTableOverviewOfTheSources{
\begin{table}[!t]
\caption{Source properties}
\tiny
\begin{center}
\begin{tabular}{l l l l l l}
\hline
\hline
 Source & RA(J2000)  & Dec(J2000)  & Distance\tablefootmark{a} & $L_{\rm bol}$\tablefootmark{b}  & $V_{\rm LSR}$\tablefootmark{c} \\
        & [$^{\mathrm{h}}\,^{\mathrm{m}}\,^{\mathrm{s}}$] & \ [$\degr\,\arcmin\,\arcsec$] & [pc] & [$L_{\odot}$] & [km s$^{-1}$]\\
\hline  
IRAS 4A & 03 29 10.5 & +31 13 30.9 & 235 & 9.1 & +7.0\\
IRAS 4B & 03 29 12.0 & +31 13 08.1 & 235 & 4.4 & +7.1\\
\hline
\end{tabular}
\end{center}
\tablefoot{
\tablefoottext{a} Adopted from \citet{Hirota08}.
\tablefoottext{b} Karska et al. (in prep.)
\tablefoottext{c} Obtained from C$^{18}$O and C$^{17}$O lines (this work).
}
\label{tbl:sourceslist}
\end{table}
}

\def\placeTableRotDiag{
\begin{table}[!t]
\caption{Rotational temperatures (in K) for the NGC 1333 sources}
\tiny
\begin{center}
\begin{tabular}{l l l l}
\hline
\hline
 Source & $^{12}$CO  & $^{13}$CO  & C$^{18}$O \\
\hline  
IRAS 2A\tablefootmark{a} & 61$\pm$8 & \dots & 34$\pm$4  \\
IRAS 4A & 69$\pm$7 & 38$\pm$4 & 25$\pm$4  \\
IRAS 4B & 83$\pm$10 & 29$\pm$3 & 26$\pm$4  \\
\hline
\end{tabular}
\end{center}
\tablefoot{
\tablefoottext{a} IRAS~2A rotational temperatures are calculated for comparison
using data from \citet{yildiz10hifi}.}
\label{tbl:rotdiaglist}
\end{table}
}

\def\placeTableOverviewOfTheObservations{
\begin{table*}[!t]
\caption{Overview of the observations of IRAS~4A and IRAS~4B}
\tiny
\begin{center}
\begin{tabular}{l l r r r l r r l r r r}
\hline \hline
Source & Mol. & Trans.& $E_{\rm u}$ & Freq. & Telescope & Beam  & Efficiency & Map & Offset\tablefootmark{a} & Obs. Date & References\\
   &  & $J_{\mathrm{u}}$-$J_{\mathrm{l}}$ & [K] & [GHz]  & & size [$\arcsec$] & $\eta$ & &  [$\Delta\alpha \arcsec, \Delta \delta \arcsec$]\\ 
\hline
IRAS~4A & CO            & 1--0 & 5.5     & 115.271202 & FCRAO                      & 46 & 0.45 & yes  & (0.0,0.0) & 01/01/2000 & (1)\\
 &                               & 2--1 & 16.6   & 230.538000 & JCMT-RxA                 & 22 & 0.69 & no  & (0.0,0.3) & 21/10/1995 & (2)\\
 &                               & 3--2 & 33.2   & 345.795989 & JCMT-HARP-B           & 15 & 0.63 & yes & (0.0,0.0) & 18/03/2010 & (3)\\
 &                               & 4--3 & 55.3   & 461.040768 & JCMT                         & 11 & 0.38 & no  & (0.0,0.2) & 23/12/1994 & (2)\\
 &                               & 6--5 & 116.2  & 691.473076 & APEX-CHAMP$^+$  & 9   & 0.48 & yes & (0.0,0.0) & 11/11/2008 & (3)\\
 &                               & 7--6 & 154.9  & 806.651806 & APEX-CHAMP$^+$  & 8   & 0.45 & yes & (1.5,1.3) & 10/11/2008 & (3)\\
 &                               &10--9 & 304.2 &1151.985452 & {\it Herschel}-HIFI         & 20 & 0.66 & no & (0.0,0.0) & 05/03/2010 & (4)\\
 & $^{13}$CO             & 1--0 & 5.3      & 110.201354 & IRAM 30m                & 23 & 0.77 & yes  & (0.0,0.0) & 21/07/2010 & (5) \\
 &                               & 2--1 & 15.87  & 220.398684 & JCMT-RxA                & 23 & 0.74 & no  & (0.0,0.1) & 11/11/2001 & (2)\\
 &                               & 3--2 & 31.7    & 330.587965 & JCMT-B3                  & 15 & 0.60 & no  & (0.0,0.1) & 16/09/2009 & (2)\\
 &                               & 4--3 & 52.9    & 440.765174 & JCMT                        & 11 & 0.38 & no  & (0.2,0.8) & 25/03/2003 & (2)\\
 &                               & 6--5 &111.05 & 661.067277 & APEX-CHAMP$^+$  & 9   & 0.52 & yes & (0.0,0.0) & 24/08/2009 & (3)\\
 &                               & 8--7 &190.36 & 881.272808 & APEX-CHAMP$^+$  & 7   & 0.42 & no  & (0.0,0.0) & 26/08/2009 & (3)\\
 &                               &10--9 &290.8 & 1101.349597 & {\it Herschel}-HIFI         & 21 & 0.76 & no  & (0.0,0.0) & 04/03/2010 & (4) \\
 & C$^{17}$O             & 1--0 &   5.39  & 112.358777 & Onsala                     & 33 &  0.43 & no & (0.0,0.0) & 13/11/2001 & (6)\\
 &                               & 2--1 & 16.18  & 224.713533 & IRAM~30m              & 17 &  0.43 & no & (0.0,0.0) & 13/11/2001 & (6)\\
 &                               & 3--2 & 32.35  & 337.061513 & JCMT-B3                  & 15 & 0.60 & no  & (1.5,1.2) & 25/08/2001 & (2)\\
 & C$^{18}$O             & 1--0 &   5.27  & 109.782173 & Onsala                     & 34 & 0.43  & no & (0.0,0.0) & 11/03/2002 & (6) \\
 &                               & 2--1 & 15.81  & 219.560354 & JCMT-RxA                & 23 & 0.69 & no & (0.0,0.6) & 03/12/1993 & (2)\\
 &                               & 3--2 & 31.61  & 329.330553 & JCMT-B3                  & 15 & 0.60 & no  & (0.2,1.2) & 26/08/2001 & (2)\\
 &                               & 5--4 & 79.0    & 548.831010 &  {\it Herschel}-HIFI          & 42 & 0.76 & no  & (0.0,0.0) & 15/03/2010 & (4) \\
 &                               & 6--5 & 110.63& 658.553278 & APEX-CHAMP$^+$  & 10 & 0.48 & no  & (0.0,0.0) & 26/08/2009 & (3)\\
 &                               & 9--8 & 237.0 & 987.560382  &  {\it Herschel}-HIFI          & 23 & 0.76 & no  & (0.0,0.0) & 03/03/2010 & (4) \\
 &                               &10--9& 289.7 & 1097.162875 & {\it Herschel}-HIFI         & 21 & 0.76 & no  & (0.0,0.0) & 31/07/2010 & (4) \\
 & [\ion{C}{i}]              & 2--1 &    62.3 & 809.341970 & APEX-CHAMP$^+$  &   8 & 0.43 & yes & (0.0,0.0) & 24/08/2009 & (3)\\
I4A B1\tablefootmark{b} & CO     & 3--2 & 33.2   & 345.795989 & JCMT-HARP-B           & 15 & 0.63 & yes & ($-$12.0,$-$12.0) & 18/03/2010 & (3)\\
           &   CO             & 6--5 & 116.2  & 691.473076 & APEX-CHAMP$^+$  &   9  & 0.48 & yes & ($-$12.0,$-$12.0) & 11/11/2008 & (3)\\
I4A B2\tablefootmark{b} & CO     & 2--1 & 16.6 & 230.538000 & JCMT-RxA  & 22 & 0.69 & no  & ($-$34.5,$-$61.9) & 02/07/2009 & (3)\\
                       &        & 3--2 & 33.2   & 345.795989 & JCMT-HARP-B           & 15 & 0.63 & yes & ($-$34.5,$-$61.9) & 18/03/2010 & (3)\\
             &                  & 6--5 & 116.2  & 691.473076 & APEX-CHAMP$^+$  &   9  & 0.48 & yes & ($-$34.5,$-$61.9) & 11/11/2008 & (3)\\
           & $^{13}$CO   & 2--1 & 15.9 & 220.398684 & JCMT-RxA                   & 22 & 0.69 & no  & ($-$34.5,$-$61.9) & 02/07/2009 & (3)\\
I4A R1\tablefootmark{c} & CO     & 3--2 & 33.2   & 345.795989 & JCMT-HARP-B   & 15 & 0.63 & yes & (12.0,12.0) & 18/03/2010 & (3)\\
           &   CO             & 6--5 & 116.2  & 691.473076 & APEX-CHAMP$^+$  &   9  & 0.48 & yes & (12.0,12.0) & 11/11/2008 & (3)\\
I4A R2\tablefootmark{c} & CO     & 2--1 & 16.6 & 230.538000 & JCMT-RxA  & 22 & 0.69 & no  & (52.5,68.1)     & 02/07/2009 & (3)\\
                        &        & 3--2 & 33.2   & 345.795989 & JCMT-HARP-B   & 15 & 0.63 & yes & (52.5,68.1) & 18/03/2010 & (3)\\
                 &               & 6--5 & 116.2  & 691.473076 & APEX-CHAMP$^+$  &   9  & 0.48 & yes & (52.5,68.1) & 11/11/2008 & (3)\\
           & $^{13}$CO   & 2--1 & 15.9 & 220.398684 & JCMT-RxA                   & 22 & 0.69 & no  & (52.5,68.1)     & 02/07/2009 & (3)\\
\hline
IRAS~4B & CO            & 2--1 & 16.6     & 230.538000 & JCMT-RxA              & 22 & 0.69 & no  & (0.0,0.0) & 21/10/1995 & (2)\\
 &                               & 3--2 & 33.2    & 345.795989 & JCMT-HARP-B         & 15 & 0.60 & yes  & (0.0,0.0) & 18/03/2010 & (3)\\
 &                               & 4--3 & 55.3    & 461.040768 & JCMT                       & 11 & 0.38 & no  & (0.0,0.2) & 22/01/2010 & (2)\\
 &                               & 6--5 & 116.2  & 691.473076 & APEX-CHAMP$^+$ & 9  & 0.48 & yes & (0.0,0.0) & 11/11/2008 & (3) \\
 &                               & 7--6 & 154.9  & 806.651806 & APEX-CHAMP$^+$ & 8  & 0.45  & no & (0.0,0.0) & 10/11/2008 & (3)\\
 &                               &10--9 & 304.2 &1151.985452 & {\it Herschel}-HIFI        & 20 & 0.66 & yes & (0.0,0.0) & 06/03/2010 & (4)\\
 & $^{13}$CO             & 1--0 & 5.3      & 110.201354 & IRAM 30m                & 23 & 0.77 & yes  & (0.0,0.0) & 21/07/2010 & (5) \\
 &                               & 3--2 & 31.7    & 330.587965 & JCMT-B3                 & 15 & 0.60 & no  & (0.0,0.0) & 25/08/2001 & (2)\\
 &                               & 6--5 & 111.1 & 661.067276 & APEX-CHAMP$^+$  & 10 & 0.52 & yes & (0.0,0.0) & 24/08/2009 & (3)\\
 &                               & 8--7 & 190.4 & 881.272808 & APEX-CHAMP$^+$  & 7  & 0.42 & no  & (0.0,0.0) & 30/08/2009 & (3)\\
 &                               &10--9& 290.8 & 1101.349597 & {\it Herschel}-HIFI        & 21 & 0.76 & no  & (0.0,0.0) & 04/03/2010 & (4)\\
 & C$^{17}$O             & 1--0 &   5.39  & 112.358777 & IRAM~30m               & 33 &  0.43 & no  & (0.0,0.0) & 13/11/2001 & (6) \\
 &                               & 2--1 & 16.18  & 224.713533 & Onsala                     & 17 &  0.43 & no  & (0.0,0.0) & 13/11/2001 & (6) \\
 &                               & 3--2 & 32.35  & 337.060513 & JCMT-B3                  & 15 & 0.60 & no  & (0.0,0.0) & 25/08/2001 & (2)\\ 
 & C$^{18}$O             & 1--0 &   5.27  & 109.782173 &  Onsala                    & 34 & 0.43  & no  & (0.0,0.0) & 11/03/2002  & (6)\\
 &                               & 2--1 & 15.81  & 219.560354 & JCMT-RxA               & 23 & 0.69 & no  & (0.0,0.6) & 08/02/1992 & (2)\\
 &                               & 3--2 & 31.61   & 329.330552 & JCMT-B3                 & 15 & 0.60 & no  & (0.2,1.2) & 29/08/2001 & (2)\\
 &                               & 5--4 & 79.0    & 548.831006 & {\it Herschel}-HIFI          & 42 & 0.76 & no  & (0.0,0.0) & 15/03/2010 & (4)\\
 &                               & 6--5 & 110.63 & 658.553278 & APEX-CHAMP$^+$ & 10  & 0.48 & no  & (0.0,0.0) & 30/08/2009 & (3)\\
 &                               & 9--8 & 237.0 & 987.560382  & {\it Herschel}-HIFI          & 23  & 0.76 & no  & (0.0,0.0) & 03/03/2010 & (4)\\
 &                               &10--9& 289.7 & 1097.162875 & {\it Herschel}-HIFI         & 21  & 0.76 & no  & (0.0,0.0) & 31/07/2010 & (4)\\
 & [\ion{C}{i}]              & 2--1 & 62.3     & 809.341970 & APEX-CHAMP$^+$ & 8  & 0.42 & yes & (0.0,0.0) & 24/08/2009 & (3)\\
\hline 
\end{tabular}
\end{center}
\tablefoot{
\tablefoottext{a} Offset from the IRAS~4A and IRAS~4B source coordinates given in Table \ref{tbl:sourceslist}. \tablefoottext{c} I4A-BX are the blueshifted outflow positions, and \tablefoottext{c} I4A-RX are the redshifted outflow positions. These regions are also depicted in Fig. \ref{fig:co65map}.
\tablebib{
(1) \citet{Ridge06}; (2) Archive; (3) this work; (4) \citet{yildiz10hifi}; (5)~\citet{Jorgensen02}; (6) Pagani et al. in prep. 
}}
\label{tbl:overviewobs}
\end{table*}
}

\def\placeTableObservedLineIntensities{
\begin{table*}[!t]
\caption{Observed line intensities for IRAS~4A and 4B in all observed transitions. }
\tiny
\begin{center}
\begin{tabular}{l l l r r r r r}
\hline \hline
Source & Mol.  & Transition & $\int T_{\rm MB} \mathrm{d}V$\tablefootmark{a} & $T_{\mathrm{peak}}$ & Blue ($\int T_{\rm MB} \mathrm{d}V$)\tablefootmark{b} & Red ($\int T_{\rm MB} \mathrm{d}V$)\tablefootmark{c} & rms\tablefootmark{d} \\
         &  &  & [K km s$^{-1}$] & [K] &  [K km s$^{-1}$] &  [K km s$^{-1}$] & [K]\\
\hline
IRAS~4A & CO     & 1--0  & 60.1\phantom{0}    & 13.0\phantom{0} &   1.1 & 26.1  & 0.45    \\
                   &           & 2--1  & 117.1\phantom{0} & 18.4\phantom{0} & 12.9 & 37.8  & 0.11     \\
                   &           & 3--2  & 128.0\phantom{0} & 16.8\phantom{0} & 34.4 & 30.6  & 0.07     \\
                   &           & 4--3  & 220.0\phantom{0} & 23.4\phantom{0} & 47.4 & 86.8 & 0.29     \\
                   &           & 6--5  & 110.5\phantom{0} & 11.9\phantom{0} & 31.8 & 37.4  & 0.33     \\
                   &           & 7--6  & 55.0\phantom{0}   & 10.0\phantom{0} & \dots     & \dots      & 4.39     \\
                   &           &10--9 & 40.7\phantom{0}   & 1.9\phantom{0}   & 9.9 & 17.6  & 0.07     \\
&$^{13}$CO          & 1--0  & 26.2\phantom{0}   & 8.5\phantom{0}   & \dots     & \dots      & 0.03     \\
                   &           & 2--1  & 16.0\phantom{0}   & 6.5\phantom{0}   & \dots     & \dots      & 0.23     \\
                   &           & 3--2  & 11.4\phantom{0}   & 4.0\phantom{0}   & 1.2     & 0.2      & 0.04     \\
                   &           & 4--3  & 15.2\phantom{0}   & 5.7\phantom{0}   & \dots     & \dots      & 0.36     \\
                   &           & 6--5  & 11.4\phantom{0}   & 3.6\phantom{0}   & 0.7        & 1.7         & 0.21     \\
                   &           & 8--7  & 2.4\phantom{0}     & 2.2\phantom{0}   & \dots     & \dots      & 0.39     \\
                   &           &10--9 & 1.1\phantom{0}     & 0.2\phantom{0}   & \dots     & \dots      & 0.02     \\
&C$^{17}$O             & 1--0  & 1.8\phantom{0}   & 0.7\phantom{0}   & \dots     & \dots      & 0.05     \\
                   &           & 2--1  & 3.8\phantom{0}     & 1.9\phantom{0}   & \dots     & \dots      & 0.04     \\
                   &           & 3--2  & 1.6\phantom{0}     & 1.0\phantom{0}   & \dots     & \dots      & 0.09     \\
&C$^{18}$O             & 1--0  & 4.3\phantom{0}  & 2.7\phantom{0}   & \dots     & \dots      & 0.18     \\
                   &           & 2--1  & 4.9\phantom{0}     & 2.7\phantom{0}   & \dots     & \dots      & 0.13     \\
                   &           & 3--2  & 4.2\phantom{0}     & 2.3\phantom{0}   & \dots     & \dots      & 0.17     \\
                   &           & 5--4  & 0.6\phantom{0}     & 0.4\phantom{0} & \dots     & \dots      & 0.006   \\
                   &           & 6--5  & 3.3\phantom{0}     & 2.0\phantom{0} & \dots     & \dots      & 0.21     \\
                   &           & 9--8  & 0.16    & 0.05                                      & \dots     & \dots      & 0.02     \\
                   &           &10--9 & $<$0.05\tablefootmark{e}                      &  \dots\phantom{0} & \dots     & \dots      & 0.02   \\
     & [\ion{C}{i}]       & 2--1  & 2.3\phantom{0}    &  1.0\phantom{0}  & \dots     & \dots      & 0.21\tablefootmark{f}      \\
I4A-B1  & CO           & 3--2  & 92.8\phantom{0}   & 14.0\phantom{0} & 64.8 & 10.1  & 0.37     \\
                   &           & 6--5  & 96.6\phantom{0}   & 13.7\phantom{0} & 81.6 & 4.5   & 0.32     \\
I4A-B2  & CO           & 2--1  & 49.5\phantom{0}   & 7.1\phantom{0}   & 32.8 & 7.6   & 0.04   \\             &$^{13}$CO & 2--1  & 11.6\phantom{0}   & 3.7\phantom{0}   & \dots     & \dots     & 0.03   \\
I4A-B2  & CO           & 3--2  & 71.2\phantom{0}   & 12.3\phantom{0} & 12.8 & 40.5 & 0.42     \\%
                   &           & 6--5  &109.1\phantom{0}  & 10.8\phantom{0} & 16.4 & 79.2 & 0.86     \\%
I4A-R2 & CO            & 2--1  & 41.1\phantom{0}   & 8.8\phantom{0}   & 13.7  & 20.3 & 0.04   \\%
            &$^{13}$CO & 2--1  & 13.3\phantom{0}   & 4.6\phantom{0}   & \dots       & \dots     & 0.03   \\%
\hline 
IRAS~4B & CO     & 2--1 & 54.7\phantom{0}   & 13.9\phantom{0}  & 1.0     & 4.6   & 0.07    \\
                   &           & 3--2 & 57.0\phantom{0}   & 10.5\phantom{0} & 6.6      & 6.7   & 0.05    \\
                   &           & 4--3 & 114.4\phantom{0} & 14.6\phantom{0} & 20.4   & 35.4   & 0.29    \\
                   &           & 6--5 & 52.3\phantom{0}   & 5.9\phantom{0}   & 10.7   & 16.5   & 0.34    \\
                   &           & 7--6 & $<$39.0\tablefootmark{e}   & \dots\phantom{0}   & \dots       & \dots       & 4.51    \\
                   &           &10--9& 29.7\phantom{0}   & 2.9\phantom{0}   & 5.9     & 8.7 & 0.08  \\ &$^{13}$CO          & 1--0  & 23.9\phantom{0}   & 7.9\phantom{0}   & \dots      & \dots       & 0.09     \\
                   &           & 3--2 & 5.9\phantom{0}      & 2.3\phantom{0}   & \dots      & \dots       & 0.03   \\
                   &           & 6--5 & 6.8\phantom{0}     & 2.0\phantom{0}   & 0.4          & 0.8           & 0.16    \\
                   &           & 8--7 & 2.7\phantom{0}     & 2.3\phantom{0}   & \dots       & \dots       & 0.39    \\
                   &           &10--9& 0.6\phantom{0}     & 0.15   & \dots       & \dots       & 0.02  \\
&C$^{17}$O          & 1--0 & 1.3\phantom{0}     & 0.6\phantom{0}   & \dots       & \dots       & 0.05    \\
                   &           & 2--1 & 2.3\phantom{0}     & 1.3\phantom{0}   & \dots       & \dots       & 0.06    \\
                   &           & 3--2 & 0.5\phantom{0}     & 0.4\phantom{0}   & \dots       & \dots       & 0.07    \\
&C$^{18}$O          & 1--0 & 4.4\phantom{0}     & 2.9\phantom{0}   & \dots       & \dots       & 0.18    \\
                   &           & 2--1 & 5.3\phantom{0}     & 2.5\phantom{0}    & \dots       & \dots       & 0.16    \\
                   &           & 3--2 & 1.9\phantom{0}     & 1.7\phantom{0}    & \dots       & \dots       & 0.23    \\
                   &           & 5--4 & 0.3\phantom{0}     & 0.2\phantom{0}   & \dots       & \dots       & 0.007  \\
                   &           & 6--5 & 0.9\phantom{0}     & 0.4\phantom{0}    & \dots       & \dots       & 0.07    \\
                   &           & 9--8 & $<$0.06\tablefootmark{e}  & \dots                     & \dots & \dots       & 0.02  \\
                   &           &10--9& $<$0.06\tablefootmark{e}  & \dots                    & \dots       & \dots       & 0.02    \\
     & [\ion{C}{i}]  & 2--1 & 1.8\phantom{0} &  0.9                & \dots       & \dots       & 0.17\tablefootmark{f}     \\
\hline
\end{tabular}
\end{center}
\tablefoot{
\tablefoottext{a}{Velocity range used for integration: --20 km s$^{-1}$ to 30 km s$^{-1}$.}
\tablefoottext{b}{Blue emission is calculated by selecting a velocity range of \mbox{--20 to 2.7 km~s$^{-1}$}.}
\tablefoottext{c}{Red emission is calculated by selecting a velocity range of \mbox{10.5 to 30 km~s$^{-1}$}.}
\tablefoottext{d}{In 0.5 km~s$^{-1}$ bins.}
\tablefoottext{e}{Upper limits are 3$\sigma$.}
\tablefoottext{f}{In 1.0 km~s$^{-1}$ bins.}
}
\label{tbl:lineintensities}
\end{table*}
}

\def\placeTableOutflowProperties{
\begin{table*}[!t]
\caption{Outflow properties of the red and blue outflow lobes of IRAS~4A and IRAS~4B.}
\small  
\begin{center}
\begin{tabular}{l l l l l l l l l l l l}
\hline \hline
\multicolumn{10}{c}{Outflow properties} \\ \hline
Source & Trans. & Lobe & ${V_{\rm max}}$\tablefootmark{a} & $R_{\rm CO}$\tablefootmark{a} & $t_{\rm{dyn}}$\tablefootmark{a,b} & $N_{\rm H_{2}}$ & M$_{\rm outflow}$\tablefootmark{c} & $\dot{M}$\tablefootmark{a,e} & $F_{\rm CO}$\tablefootmark{a,f} & $L_{\rm{kin}}$\tablefootmark{a,g}  \\ 
& & & [km s$^{-1}$] & [AU] & [yr] & [cm$^{-2}$] & [M$_\odot$] & [M$_\odot$ yr$^{-1}$]  &[M$_\odot$ yr$^{-1}$km s$^{-1}$] & [L$_\odot$]  \\ \hline
IRAS~4A & CO 3--2 & Blue & 22 & 2.8$\times10^{4}$ & 5.9$\times 10^{3}$ & 8.6$\times 10^{21}$ &  1.3$\times 10^{-2}$ & 2.1$\times 10^{-6}$ & 5.0$\times 10^{-5}$ & 2.9$\times 10^{-4}$\\
              &                    & Red  & 18 & 3.9$\times10^{4}$ & 1.0$\times 10^{4}$ & 1.2$\times 10^{22}$ &  1.8$\times 10^{-2}$  & 1.7$\times 10^{-6}$ & 3.9$\times 10^{-5}$ & 2.0$\times 10^{-4}$\\
              & CO 6--5    & Blue & 20 & 2.5$\times10^{4}$ & 5.9$\times 10^{3}$ & 1.0$\times 10^{22}$ &  6.1$\times 10^{-3}$ & 1.0$\times 10^{-6}$ & 3.1$\times 10^{-5}$ & 2.2$\times 10^{-4}$\\
              &                    & Red  & 18 & 3.5$\times10^{4}$ & 9.2$\times 10^{3}$ & 1.8$\times 10^{22}$ &  1.0$\times 10^{-2}$ & 1.1$\times 10^{-6}$ & 3.7$\times 10^{-5}$ & 2.5$\times 10^{-4}$\\
\hline
IRAS~4B & CO 3--2 & Blue & 18 & 3.5$\times 10^{3}$ & \dots & 4.6$\times 10^{20}$ &  6.7$\times 10^{-4}$ & \dots &  1.9$\times 10^{-5}$ &  1.1$\times 10^{-4}$ \\
              &                    & Red & 15  & 4.7$\times 10^{3}$ & \dots & 7.5$\times 10^{20}$ &  1.1$\times 10^{-3}$ & \dots &  2.0$\times 10^{-5}$ &  9.7$\times 10^{-5}$ \\
              & CO 6--5     & Blue & 20 & 1.9$\times 10^{3}$ & \dots & 1.0$\times 10^{21}$ &  6.0$\times 10^{-4}$ & \dots &  3.4$\times 10^{-5}$ &  2.5$\times 10^{-4}$ \\
              &                    & Red & 12  & 7.5$\times 10^{2}$ & \dots & 9.1$\times 10^{20}$ &  5.3$\times 10^{-4}$ & \dots  &  7.7$\times 10^{-5}$ &  4.5$\times 10^{-4}$ \\
\hline
\end{tabular}\\
\end{center}
\tablefoot{
\tablefoottext{a} Not corrected for inclination. 
\tablefoottext{b} Dynamical time scale. \tablefoottext{c} Constant temperature of 75~K assumed for the \mbox{CO~6--5} calculations, and 50~K assumed for the \mbox{CO~3--2} calculations. Not corrected for inclination as explained in Sect. \ref{subsec:mass}.
\tablefoottext{e} Mass outflow rate, \tablefoottext{f} outflow force and \tablefoottext{g} kinetic luminosity.
}
\label{tbl:outflowparam}
\end{table*}
}

\def\placeTableDustyResultsEnv{
\begin{table*}[h]
\caption{\small Spherical envelope models derived from dust continuum radiative transfer calculations of \citet{Kristensen12subm}.}
\small
\begin{center}
\begin{tabular}{l l l l l l l l l l l l}
\hline \hline
Source  & $Y$\tablefootmark{a} & $p$\tablefootmark{b} & $\tau_{100}$\tablefootmark{c}  &  $r_{\rm in}$\tablefootmark{d} & $r_{\rm out}$\tablefootmark{e} & $r_{\rm 10K}$\tablefootmark{f} & $n_{\rm r_{\rm in}}$\tablefootmark{g} &  $n_{r_{\rm out}}$\tablefootmark{h}  & $n_{\rm 10K}$\tablefootmark{i} & $N_{\rm H_{2},\rm 10K}$\tablefootmark{j}     & $M_{\rm 10K}$\tablefootmark{k}     \\
        & & &   & [AU]           & [AU]           & [AU]            &  [cm$^{-3}$] &         [cm$^{-3}$]        & [cm$^{-3}$]   & [cm$^{-2}$] & [$\mathrm{M_{\odot}}$] \\ 
\hline
IRAS~4A & 1000 & 1.8 & 7.7 & 33.5 & 3.3$\times 10^{4}$ & 6.4$\times 10^{3}$ & 3.1$\times 10^{9}$ & 1.2$\times 10^{4}$& 2.4$\times 10^{5}$ & 1.9$\times 10^{24}$   & 5.1\\
IRAS~4B & 800 & 1.4 & 4.3 & 15.0 & 1.2$\times 10^{4}$ & 3.8$\times 10^{3}$ & 2.0$\times 10^{9}$ & 1.8$\times 10^{5}$& 8.7$\times 10^{5}$ & 1.0$\times 10^{24}$   & 3.0\\
\hline 
\end{tabular}
\end{center}
\tablefoot{
\tablefoottext{a} $r_{\rm in}/r_{\rm out}$. 
\tablefoottext{b} Power law index.
\tablefoottext{c} Opacity at 100~$\mu$m. 
\tablefoottext{d} Inner radius of the envelope. 
\tablefoottext{e} Outer radius of the envelope, reaching out to 8~K. 
\tablefoottext{f} 10 K radius. 
\tablefoottext{g} Number density at $r_{\rm in}$. 
\tablefoottext{h} Number density at $r_{\rm out}$. 
\tablefoottext{i} Number density at 10 K radius.
\tablefoottext{j} H$_{2}$ column density. 
\tablefoottext{k} Total mass of the envelope in 10~K radius. 
}
\label{tbl:dustyresultsenv}
\end{table*}
}

\def\placeMassComparison{
\begin{table}[!t]
\caption{\small Comparison of photon-heated and outflow masses over the area mapped by \mbox{$^{13}$CO~6--5}.}
\begin{center}
\begin{tabular}{l l l l l l l l}
\hline \hline
Source & $M_{\rm total}$\tablefootmark{a}  & $M_{\rm cones}$\tablefootmark{b} & $M_{\rm outflow}$\tablefootmark{c}   & $M_{\rm UV}$\tablefootmark{d}    \\
            & Envelope & Envelope & $^{12}$CO~6--5    & $^{13}$CO~6--5\\ 
\hline
IRAS~4A &  5.0 & 1.5 & 3.7$\times 10^{-3}$ & 1.7$\times 10^{-2}$ \\
IRAS~4B &  3.1 & 0.9 & 1.0$\times 10^{-3}$  & \dots \\
\hline 
\end{tabular}
\end{center}
\tablefoot{
  All masses are given in $M_{\odot}$ and the area taken in the
  calculations is shown in Fig. \ref{fig:dropThirteenCOUVEnv}.
  \tablefoottext{a} Total mass of the spherical envelope inferred from
  the continuum radiative transfer modelling using DUSTY.
  \tablefoottext{b} Envelope mass in both of the outflow cones
  assuming that each cone contains $\sim$15$\%$ of the total envelope
  mass out to the mapped radius.  \tablefoottext{c} Outflow
  mass calculated from the $^{12}$CO~6--5 outflow wings over the mapped
  $^{13}$CO area.  \tablefoottext{d} UV photon-heated gas mass calculated
  from the narrow $^{13}$CO~6--5 spectra over the
  mapped area after subtracting the modeled envelope emission.
}
\label{tbl:masscomparison}
\end{table}
}

\def\placeAbundanceTable{
\begin{table}
\caption{Summary of C$^{18}$O abundance profiles for IRAS~4A and 4B.}
\scriptsize
\begin{center}
\begin{tabular}{l l l l l l l }
\hline \hline
Profile      & $X_{\rm in}$ & $T_{\rm ev}$ & $X_{\rm D}$ & $n_{\rm de}$ & $X_{0}$ \\
        &                      & [K]          &                     & [cm$^{-3}$]   &              \\
\hline
IRAS~4A \\
\hline
Constant   & \dots & \dots& \dots& \dots&  1--5$\times$10$^{-8}$ \\
Anti-jump  & \dots & \dots& 1$\times$ 10$^{-8}$ & 7.5$\times$10$^4$ &  5$\times$10$^{-7}$ \\
Drop          &  $\sim$1$\times$10$^{-7}$  & 25 & 5.5$\times$10$^{-9}$ & 7.5$\times$10$^4$ & 5$\times$10$^{-7}$   \\
\hline 
IRAS~4B \\
\hline
Constant  & \dots & \dots& \dots& \dots&  1-6$\times$10$^{-8}$ \\
Jump        & 3$\times$10$^{-7}$ & 25 & 1$\times$ 10$^{-8}$ & \dots &  1$\times$10$^{-8}$ \\
\hline
IRAS~2A\tablefootmark{a} \\
\hline
Constant    & \dots & \dots& \dots& \dots&  1.4$\times$10$^{-7}$ \\
Anti-jump  & \dots & \dots& 3$\times$ 10$^{-8}$ & 7$\times$10$^4$ &  5$\times$10$^{-7}$ \\
Drop          &  1.5$\times$10$^{-7}$  & 25 & $\sim$ 4$\times$10$^{-8}$ & 7$\times$10$^4$ & 5$\times$10$^{-7}$   \\
\hline
\end{tabular}
\end{center}
\tablefoot{
\tablefoottext{a} Results from \citet{yildiz10hifi}.
}
\label{tbl:abundancetable}
\end{table}
}

  \abstract
    {The NGC~1333~IRAS~4A and IRAS~4B
  sources are among the best studied Stage~0 low-mass protostars which
  are driving prominent bipolar outflows. Spectrally resolved molecular emission lines
  provide crucial information regarding the physical and chemical
  structure of the circumstellar material as well as the dynamics of the different components. Most studies have
  so far concentrated on the colder parts ($T$ $\leq$30~K) of these
  regions.}
  {The aim is to characterize the warmer parts of the protostellar 
envelope using the new generation of submillimeter instruments.
This will allow us to quantify the feedback of the protostars on
their surroundings in terms of shocks, UV heating, photodissociation
and outflow dispersal.}
  {The dual frequency 2$\times$7 pixel 650/850~GHz array receiver
  CHAMP$^{+}$ mounted on APEX was used to obtain a fully sampled,
  large-scale $\sim$4$\arcmin \times 4\arcmin$ map at 9$\arcsec$
  resolution of the IRAS~4A/4B region in the \mbox{$^{12}$CO~$J$=6--5}
  line. Smaller maps are observed in the \mbox{$^{13}$CO~6--5} and
  \mbox{[\ion{C}{i}]~$J$=2--1} lines. In addition, a fully sampled
  $^{12}$CO $J$=3--2 map made with HARP-B on the JCMT is presented and
  deep isotopolog observations are obtained at selected outflow
  positions to constrain the optical depth.  Complementary {\it
    Herschel}-HIFI and ground-based lines of CO and its isotopologs,
  from $J$=1--0 up to 10--9 ($E_{\rm u}/k\approx $300 K), are
  collected at the source positions and used to construct velocity
  resolved CO ladders and rotational diagrams.  Radiative-transfer
  models of the dust and lines are used to determine temperatures and
  masses of the outflowing and photon-heated gas and infer the CO
  abundance structure.}
  {{Broad CO emission line profiles trace entrained shocked gas along
    the outflow walls, with typical temperatures of $\sim$100 K. At
    other positions surrounding the outflow and the protostar, the
    6--5 line profiles are narrow indicating UV excitation. The narrow
    $^{13}$CO 6-5 data directly reveal the UV heated gas distribution
    for the first time.  The amount of UV-photon-heated gas and
    outflowing gas are quantified from the combined $^{12}$CO and
    $^{13}$CO 6--5 maps and found to be comparable within a
    20$\arcsec$ radius around IRAS 4A, which implies that UV photons
    can affect the gas as much as the outflows. Weak [C I] emission
    throughout the region indicates a lack of CO dissociating
    photons. Modeling of the C$^{18}$O lines indicates the necessity
    of a ``drop'' abundance profile throughout the envelopes where the
    CO freezes out and is reloaded back into the gas phase through
    grain heating, thus providing quantitative evidence for the CO ice
    evaporation zone around the protostars. The inner abundances are
    less than the canonical value of CO/H$_2$=2.7$\times10^{-4}$,
    however, indicating some processing of CO into other species on
    the grains. The implications of our results for the
      analysis of spectrally unresolved {\it Herschel} data are
      discussed.}}
     {}

   \keywords{Astrochemistry --- stars: formation --- stars: pre-main sequence --- ISM: individual objects: NGC~1333~IRAS~4A, IRAS~4B --- ISM: jets and outflows --- ISM: molecules}
   \maketitle

\section{Introduction}

In the very early stages of star formation, newly forming protostars
are mainly characterized by their large envelopes (\mbox{$\sim$10$^4$
 AU} in diameter) and bipolar outflows \citep{Lada87,
 Greene94}. As gas and dust from the collapsing core
accrete onto the central source, the protostar  drives out
material along both poles at supersonic speeds to distances up to a
parsec or more. 
Outflows have a significant impact on their
surroundings, by creating shock waves which increase the temperature
and change the chemical composition \citep{Snell80,Bachiller99,Arce07}. By
sweeping up material, they carry off envelope mass and limit the
growth of the protostar. They also create a cavity through which
ultraviolet photons from the protostar can escape and impact the
cloud \citep{Spaans95}.  Quantifying these active `feedback' processes and
distinguishing them from the passive heating of the inner envelope by
the protostellar luminosity is important for a complete understanding
of the physics and chemistry during protostellar evolution.

Most studies of low-mass protostars to date have used low-excitation
lines of CO and isotopologs ($J_{\rm u}$ $\leq$ 3) combined with dust
continuum mapping to characterize the cold gas in envelopes and
outflows
\citep[e.g.,][]{Blake95,Bontemps96,Shirley02,Robitaille06}. A
wealth of other molecules has also been observed at mm wavelengths,
but their use as temperature probes is complicated by the fact that
they also have large abundance gradients through the envelope driven
by release of ice mantles
\citep[e.g.,][]{vanDishoeck98,Ceccarelli07,Bottinelli07}.  Moreover,
molecules with large dipole moments such as CH$_3$OH are often highly
subthermally excited unless densities are very high
\citep[e.g.][]{Bachiller95,Johnstone03}. With the opening up of
high-frequency observations from the ground and in space, higher
excitation lines of CO can now be routinely observed so that their
diagnostic potential as temperature and column density probes can now
be fully exploited.

Tracing warm gas with CO up to $J$=7--6 from the ground requires the best 
atmospheric conditions, as well as state-of-art detectors. The combination is offered by the
CHAMP$^+$ 650/850 GHz 2$\times$7 pixel array receiver \citep{Kasemann06} currently mounted at the Atacama
Pathfinder EXperiment (APEX) Telescope at 5100~m altitude on Cerro Chajnantor
\citep[e.g.][]{Guesten08}.  Moreover, the spectroscopic instruments on
the {\it Herschel Space Observatory} have the sensitivity to observe
CO lines up to $J$=44--43 unhindered by the Earth's atmosphere, even
for low-mass young stellar objects
\citep[e.g.,][]{vanKempen10dkcha,vanKempen10hh46,Lefloch10,yildiz10hifi}.
Together, these data allow to address questions such as (i) How is CO
excited, is it due to shock or UV heating?  (ii) How much warm gas is
present in the inner regions of the protostellar envelopes and from
which location does it originate?  What is the swept up mass and how
warm is it? (iii) What is the CO abundance structure throughout the
envelope: where is CO frozen out and where is it processed?

Over the past several years, our group has conducted a survey of
APEX-CHAMP$^+$ mapping of high$-J$ lines of CO and isotopologs of
embedded low-mass Stage 0 and 1 \citep[cf.\ nomenclature
by][]{Robitaille06} young stellar objects (YSOs)
\citep[Paper I and II in this series]{vanKempen09champ2,vanKempen09champ,vanKempen09_southc+}. 
These data complement our earlier surveys at
lower frequency of CO and other molecules with the James Clerk Maxwell
Telescope (JCMT), IRAM~30m, APEX and Onsala telescopes
\citep[e.g.,][]{Jorgensen02,Jorgensen04,vanKempen09_southc+}. More
recently, the same sources are being observed with the {\it Herschel}
Space Observatory in the context of the `Water in star-forming regions
with Herschel' (WISH) key program \citep{vanDishoeck11}. The
\mbox{$^{12}$CO $J$=6--5} ($E_{\rm u}/k$=115~K) line is particularly useful in
tracing the outflows through broad line wings, complementing
recent mapping in the \mbox{$^{12}$CO $J$=3--2} line with the HARP-B
array on the JCMT \citep[e.g.,][]{curtis10_2outflows}. 
The availability of lines up to \mbox{CO $J$=7--6} 
gives much better constraints on the excitation temperature of
the gas, which together with the higher angular resolution of the
high frequency data should result in more accurate determination of
outflow properties such as the force and momentum.

In addition to broad line wings, \citet{vanKempen09champ} also
found {\it narrow} extended $^{12}$CO 6--5 emission along the cavity
walls. Combined with narrow $^{13}$CO 6--5 emission, this was
interpreted as evidence for UV photon-heated gas, following earlier
work by \citet{Spaans95}. The mini-survey by
\citet{vanKempen09_southc+} found this narrow extended emission to be
ubiquitous in low-mass protostars. Further evidence for UV photon heating was provided by
far-infrared CO lines with $J_{\rm u}$=10 to 20 observed with {\it
  Herschel}-PACS \citep{vanKempen10hh46,Visser11}. However, {\it
  Herschel} has only limited mapping capabilities; PACS lacks velocity
resolution and HIFI has a quite large beam (20\arcsec -- 40\arcsec).
Thus, large scale velocity resolved maps at $<10''$ resolution as
offered by APEX-CHAMP$^{+}$ form an important complement to the {\it
  Herschel} data. In this paper, we present fully sampled high-$J$
CHAMP$^{+}$ maps of one of the largest and most prominent low-mass
outflow regions, NGC~1333~IRAS~4.

NGC~1333~IRAS~4A and IRAS~4B (hereafter only IRAS~4A and IRAS~4B) are
two low-mass protostars in the southeast corner of the NGC~1333
region \citep[see][for review]{Walawender08}. They have attracted significant attention due to their strong
continuum emission, powerful outflows and rich chemistry
\citep{Andre94,Blake95,Bottinelli07}.
They were first identified as water maser spots by
\citet{Ho80} and later confirmed as protostellar
candidates by IRAS observations \citep{Jennings87} and
resolved individually in JCMT-SCUBA submm continuum maps
\citep{Sandell91,SandellKnee01,diFrancesco08}.  Using mm
interferometry, it was subsequently found that both protostars are in
proto-multiple systems \citep{Lay95,Looney00}. The projected
separation between IRAS~4A and IRAS~4B is 31$\arcsec$
($\sim$7500 AU). 
The companion to IRAS~4B is clearly detected at a separation of
11$\arcsec$ whereas that of IRAS 4A has a separation of only
2$\arcsec$ \citep{Jorgensen07prosac}. 
The distance to the NGC~1333 nebula is 
still in debate \citep[see][for more
thorough discussions]{curtis10_1data}. 
In this paper, we adopt the distance of 235$\pm$18~pc based on VLBI parallax
measurements of water masers in SVS~13 in the same cluster
  \citep{Hirota08}.

We here present an APEX-CHAMP$^{+}$ $^{12}$CO 6--5 map over a
$4'\times4'$ area at 9$''$ resolution, together with $^{13}$CO~6--5
and [\ion{C}{i}] $J$=2--1 maps over a smaller region ($1'\times1'$).
Moreover, $^{13}$CO~8--7 and C$^{18}$O~6--5 lines are obtained at the
central source positions.  These data are analyzed together with the
higher-$J$ \textit{Herschel}-HIFI observations of CO and isotopologs
recently published by \citet{yildiz10hifi} as well as lower-$J$ JCMT,
IRAM 30m and Onsala archival data so that spectrally resolved
information on nearly the entire CO ladder up to 10--9 ($E_{\rm
  u}/k$=300 K) is obtained for all three isotopologs.  The spectrally
resolved data allow the temperatures in different components to be
determined, and thus provide an important complement to spectrally
unresolved {\it Herschel} PACS and SPIRE data of the CO ladder of such
sources.  In addition, a new JCMT HARP-B map of $^{12}$CO 3--2 has
been obtained over the same area, as well as deep $^{13}$CO spectra at
selected outflow positions to constrain the optical depth.  The
APEX-CHAMP$^{+}$ and JCMT maps over a large area can test the
interpretation of the different velocity components seen in HIFI data
which has so far been based only on single position data.

The outline of the paper is as follows. In
Section~\ref{sec:observations}, the observations and the telescopes
where the data have been obtained are described. In
Section~\ref{sec:results}, the inventory of complementary lines and maps
are presented. In Section~\ref{sec:analysis}, the data are analyzed
to constrain the temperature and mass of the molecular outflows. In Section~\ref{sec:envelopeproperties}, the
envelope abundance structure of these protostars is discussed. 
In Section~\ref{sec:analysisUV}, the amount of  
 shocked gas is compared quantitatively to that of photon-heated gas.
In Section~\ref{sec:conclusions}, the conclusions from this work are
summarized.

\section{Observations}
\label{sec:observations}
\mbox{Table~\ref{tbl:sourceslist}} gives a brief overview of the IRAS
4A and 4B sources.
Spectral line data have been obtained primarily from the \mbox{12-m} \mbox{sub-mm}
Atacama Pathfinder Experiment Telescope, APEX\footnote{This
  publication is based on data acquired with the Atacama Pathfinder
  Experiment (APEX). APEX is a collaboration between the
  Max-Planck-Institut f\"ur Radioastronomie, the European Southern
  Observatory, and the Onsala Space Observatory.}
\citep{Guesten08} at Llano de Chajnantor in Chile. In
addition, we present new and archival results from the 15-m James
Clerk Maxwell Telescope, JCMT\footnote{The JCMT is operated by The
  Joint Astronomy Centre on behalf of the Science and Technology
  Facilities Council of the United Kingdom, the Netherlands
  Organisation for Scientific Research, and the National Research
  Council of Canada.} at Mauna Kea, Hawaii; 
the 3.5-m \textit{Herschel} Space Observatory\footnote{Herschel is an ESA space
  observatory with science instruments provided by European-led
  Principal Investigator consortia and with important participation
  from NASA.} \citep{Pilbratt10} and IRAM 30m telescope. Finally, we use published
data from  the Onsala 20-m and 14-m Five College Radio Astronomy Observatory, FCRAO telescopes.

\placeTableOverviewOfTheSources

{\it APEX:} The main focus of this paper is the high-$J$
\mbox{CO~6--5} and [\ion{C}{i}] 2--1 maps of IRAS~4A and 4B,
obtained with APEX-CHAMP$^{+}$ 
in November 2008 and August 2009. 
The protostellar envelopes and their
complete outflowing regions have been mapped
in \mbox{CO~6--5} emission using the  
on-the-fly mapping mode acquiring more than
100\,000 spectra in 1.5 hours covering a Nyquist sampled 240$\arcsec\times$240$\arcsec$
region. The instrument consists of 2 heterodyne receiver arrays, each
with 7 pixel detector elements for simultaneous operations in the \mbox{620--720
  GHz} and \mbox{780--950} GHz frequency ranges
\citep{Kasemann06, Guesten08}. 
The following two lines were observed
simultaneously:
\mbox{$^{12}$CO~6--5} and \mbox{[\ion{C}{i}]~2--1} (large map);
\mbox{$^{13}$CO 6--5} and \mbox{[\ion{C}{i}]~2--1} (smaller map);
\mbox{C$^{18}$O~6--5} and \mbox{$^{13}$CO~8--7} (staring at source
positions); and \mbox{$^{12}$CO~6--5} and \mbox{$^{12}$CO~7--6}
(staring at source positions).

The APEX beam sizes correspond to 8$\arcsec$ ($\sim$1900 AU at a
distance of \mbox{235 pc}) at \mbox{809 GHz} and 9$\arcsec$
($\sim$2100 AU) at \mbox{691 GHz}. The observations were completed
under excellent weather conditions (Precipitable Water Vapor, PWV
$\sim$0.5 mm) with typical system temperatures of 1900 K for
CHAMP$^{+}$-I (SSB, 691~GHz), and 5600 K for CHAMP$^{+}$-II (SSB,
809~GHz). The relatively high system temperatures are due to the high
atmospheric pathlength at the low elevation of the sources of
$\sim$25$\degr$. Also, for CHAMP$^{+}$-II, there is a significant contribution 
from the receiver temperature. The observations 
were done using position-switching
toward an emission free reference position in settings
\mbox{$^{12}$CO~6--5} + \mbox{[\ion{C}{i}]~2--1} or CO 7--6, and
\mbox{$^{13}$CO~6--5} + \mbox{[\ion{C}{i}]~2--1}. However, in the
setting \mbox{C$^{18}$O~6--5} and \mbox{$^{13}$CO~8--7}, a
beam-switching of +/--90$\arcsec$ is used in staring mode in order to
increase the $S/N$ on the central pixel \citep{vanDishoeck10champ+}.
The CHAMP$^{+}$ array uses the Fast Fourier Transform Spectrometer
(FFTS) backend \citep{Klein06} for all seven pixels with a resolution 
of 0.12 MHz (0.045~km~s$^{-1}$ at 800~GHz).

{\it JCMT:} A \mbox{CO~3--2} fully sampled map is obtained from the
JCMT with the HARP-B instrument in March 2010. HARP-B consists of 16
SIS detectors with 4$\times$4 pixel elements of 15$\arcsec$ each at
30$\arcsec$ separation. The opacity at the time of observations was
excellent ($\tau_{\rm 225 GHz}<$0.04) and the on-the-fly method was
used to fully cover the entire outflow. Apart from the maps, line data
of CO and isotopologs (e.g., 2--1 and 3--2 lines) were obtained from
the JCMT and its public archive\footnote{This research used the
  facilities of the Canadian Astronomy Data Centre operated by the
  National Research Council of Canada with the support of the Canadian
  Space Agency.}.  In Table \ref{tbl:overviewobs}, the offset values
of the archival data from the protostellar source coordinates are
provided.  In addition, we observed four distinct outflow knots of
IRAS~4A in deep $^{12}$CO and \mbox{$^{13}$CO~2--1} integrations to
constrain the optical depth (see Table \ref{tbl:overviewobs} for
coordinates). The B1 and R1 positions are the blue and red-shifted
outflow knots closest to IRAS~4A and B2 and R2 are the two prominent
dense outflow knots further removed.

{\it Herschel:} Spectral lines of \mbox{$^{12}$CO~10--9},
\mbox{$^{13}$CO~10--9}, \mbox{C$^{18}$O~5--4, 9--8 and 10--9} were
observed with the \textit{Herschel} Space Observatory using the
Heterodyne Instrument for Far-Infrared (HIFI); \citep{degraauw10}. All
observations were done in dual-beam-switch (DBS) mode with a chop
reference position located 3$\arcmin$ from the source
positions. Except for the \mbox{C$^{18}$O~10--9} spectra, these data
were presented in \citet{yildiz10hifi} and observational details can
be found there.

{\it IRAM-30m:} The lower-$J$ \mbox{$^{13}$CO~1--0} and \mbox{C$^{17}$O~2--1} transitions 
were observed with the IRAM~30-m telescope \footnote{Based on observations 
carried out with the IRAM 30m Telescope. IRAM is supported by INSU/CNRS (France), 
MPG (Germany) and IGN (Spain).} by \citet{Jorgensen02} and Pagani et al. (in prep).

{\it Onsala:} The lowest-$J$ \mbox{C$^{17}$O} and  
\mbox{C$^{18}$O~1--0} transitions were observed with
the Onsala 20-m radiotelescope by \citet{Jorgensen02}, and the
spectra are used here. 

{\it FCRAO:} \mbox{$^{12}$CO~1--0} spectrum of IRAS~4A is extracted 
from COMPLETE survey map \citep{Arce10} observed with FCRAO. 

\placeTableOverviewOfTheObservations

Table \ref{tbl:overviewobs} summarizes the list of observed lines for
each instrument. Information about the corresponding rest frequencies
and upper-level energies of the transitions are included, together
with the beam sizes and efficiencies of the instruments. The data have
been acquired on the $T_{\rm A}^{*}$ antenna temperature scale, and
have then been converted to main-beam brightness temperatures $T_{\rm
  MB} = T_{\rm A}^{*}/ \eta_{\rm MB}$ by using the stated beam
efficiencies ($\eta_{\rm MB}$). The CHAMP$^+$ beam efficiencies are
taken from the CHAMP$^+$
website\footnote{http:www.mpifr.de/div/submmtech/heterodyne/champplus/
  \\champ\_efficiencies.15-10-09.html} and forward efficiencies are 0.95 in all observations. JCMT beam efficiencies are
taken from the JCMT Efficiencies
Database\footnote{http://www.jach.hawaii.edu/JCMT/spectral\_line/Standards/\\eff\_web.html},
and the \textit{Herschel}-HIFI efficiencies are taken as 0.76 in all
bands except band 5 where it is 0.64 \citep{Roelfsema11}. The Onsala
efficiencies are taken from \citet{Jorgensen02}. 
Calibration errors are estimated as $\sim$20\% for the
ground-based telescopes, and $\sim$10\% for the HIFI lines. 
For the data reduction and analysis, the
``Continuum and Line Analysis Single Dish Software'', \verb1CLASS1
program which is part of the
GILDAS software\footnote{http:www.iram.fr/IRAMFR/GILDAS}, is used. The routines
in GILDAS convolve the irregularly gridded on-the-fly data with a Gaussian kernel of size one third of the beam, yielding a Nyquist-sampled map.

\section{Results}
\label{sec:results}

\placeFigureIrasFourABCOsixfivespectraandoutflowINSET

\subsection{CO line gallery}

\begin{figure*}
\begin{center}
\begin{minipage}{8.5cm}
\includegraphics[width=8.6cm]{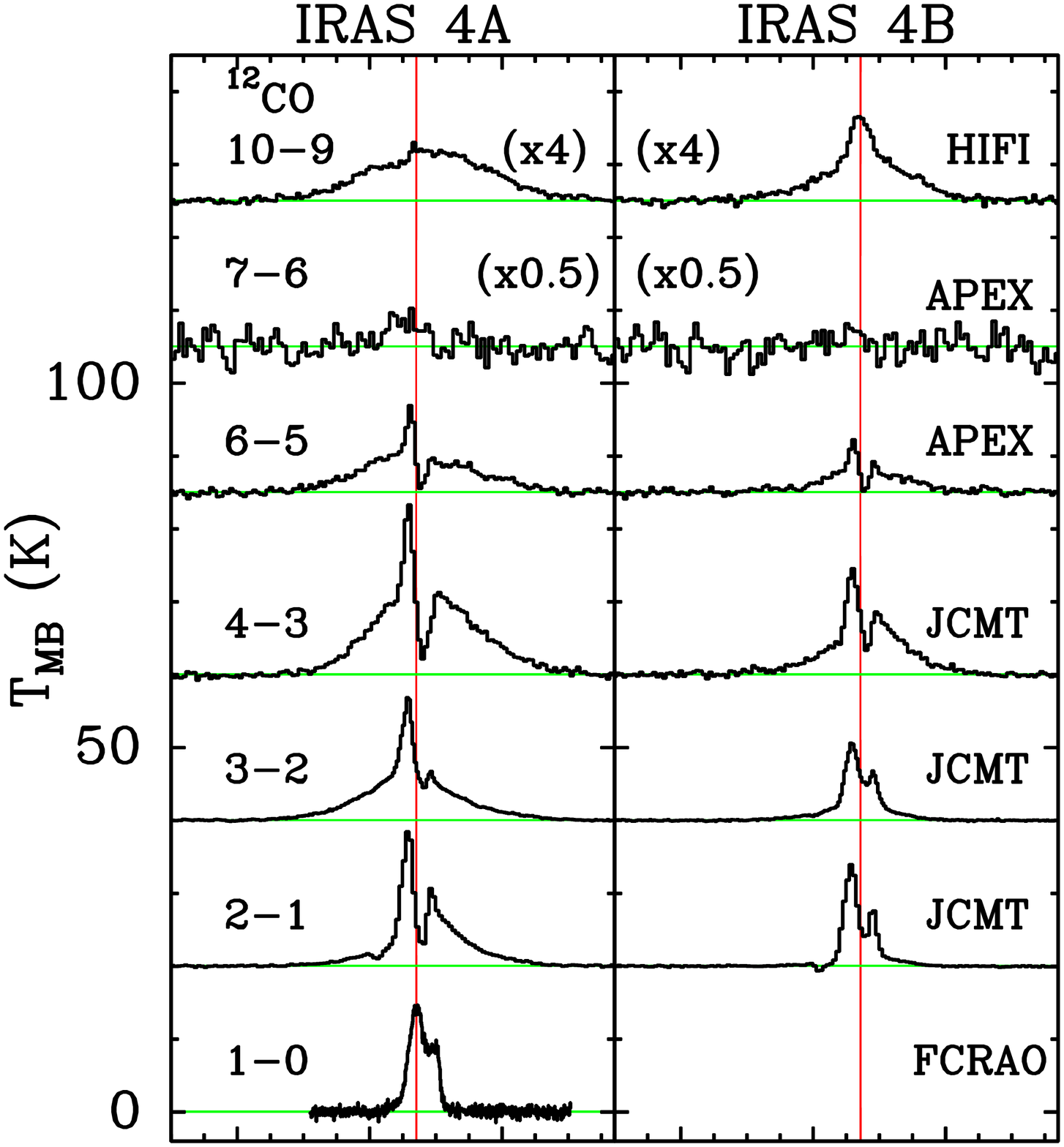} \includegraphics[width=8.6cm]{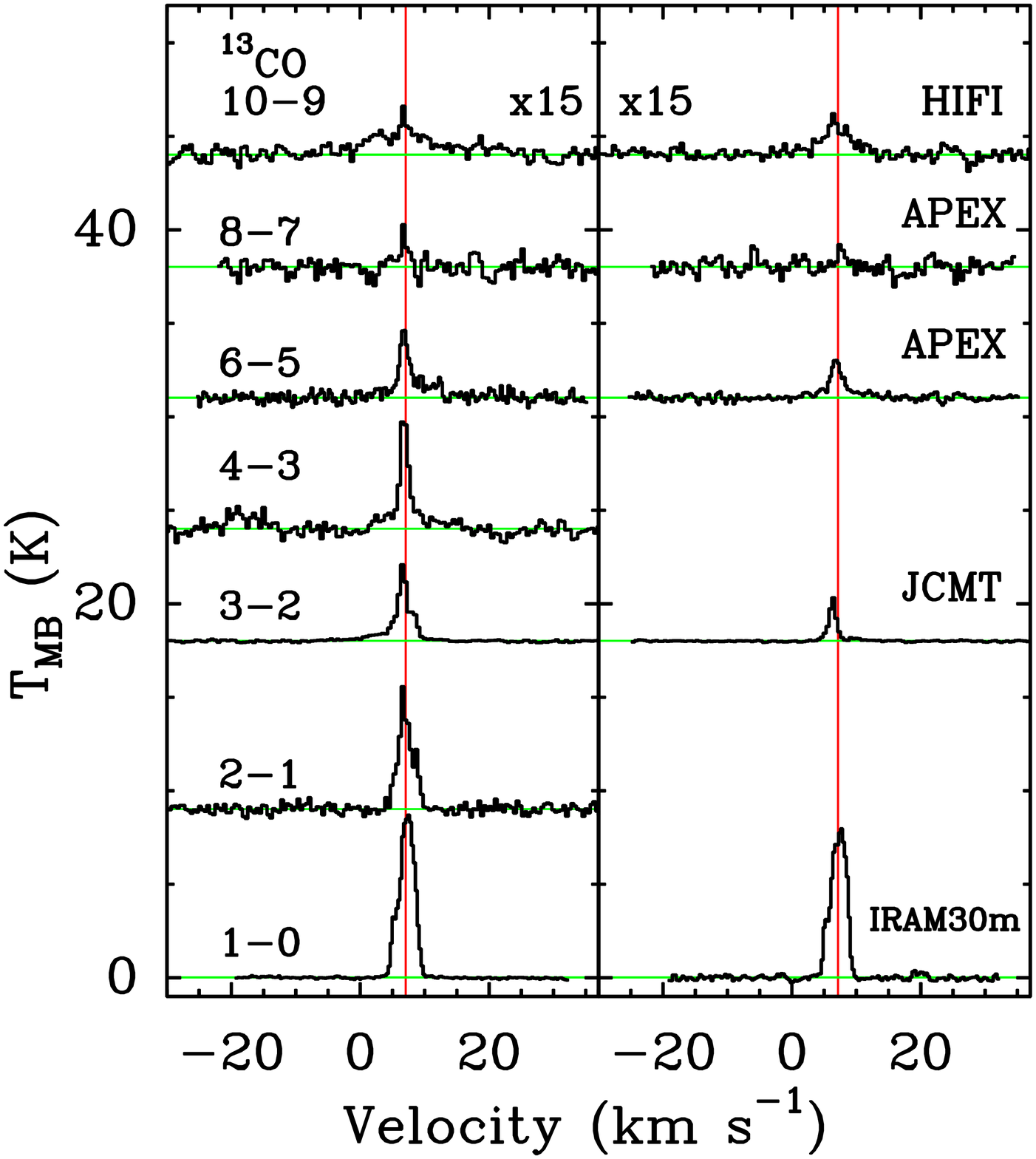} \end{minipage}
\begin{minipage}{8.5cm}
    \hspace*{0.1cm}
\includegraphics[height=10cm]{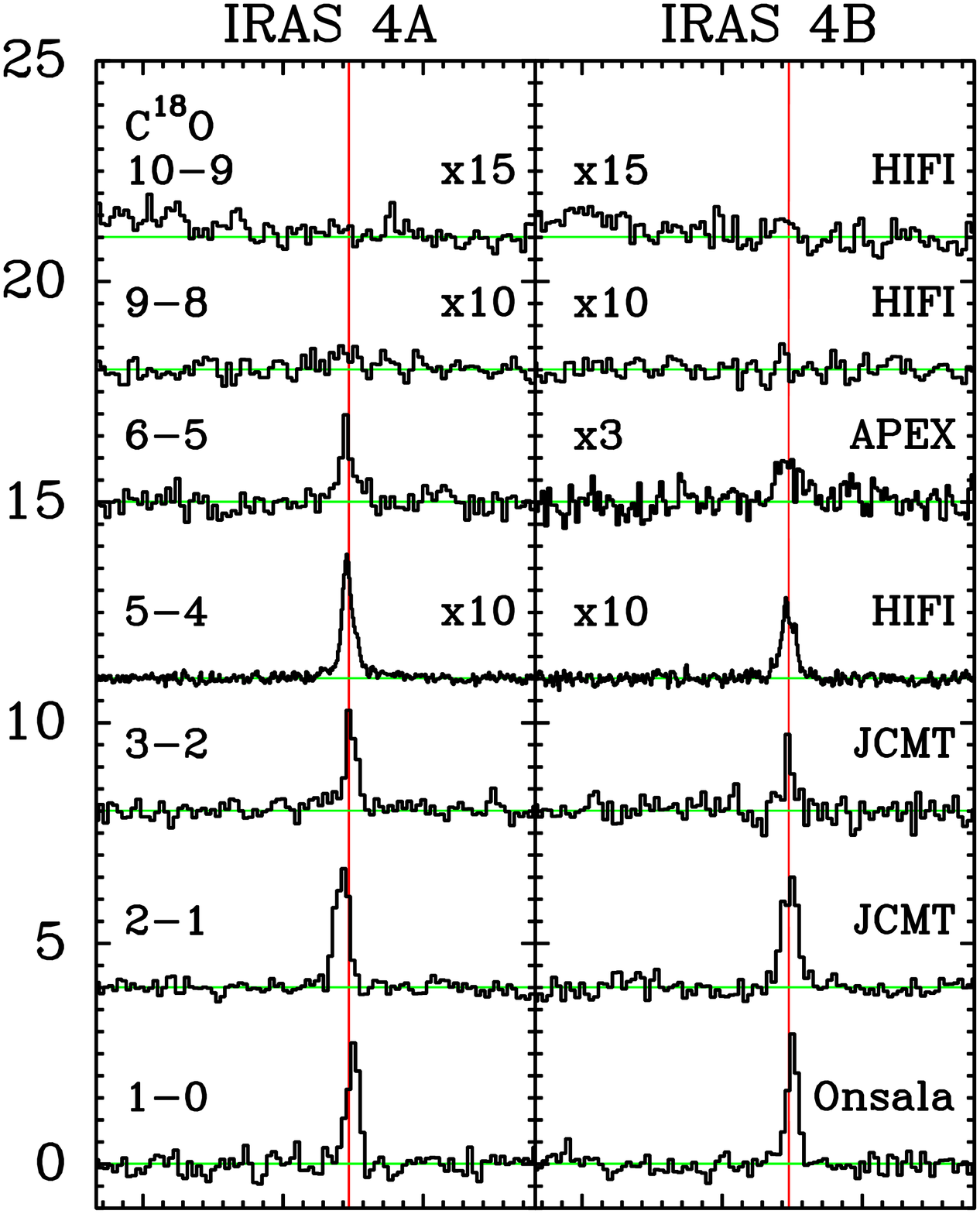}     \hspace*{0.47cm}
\includegraphics[height=9.0cm]{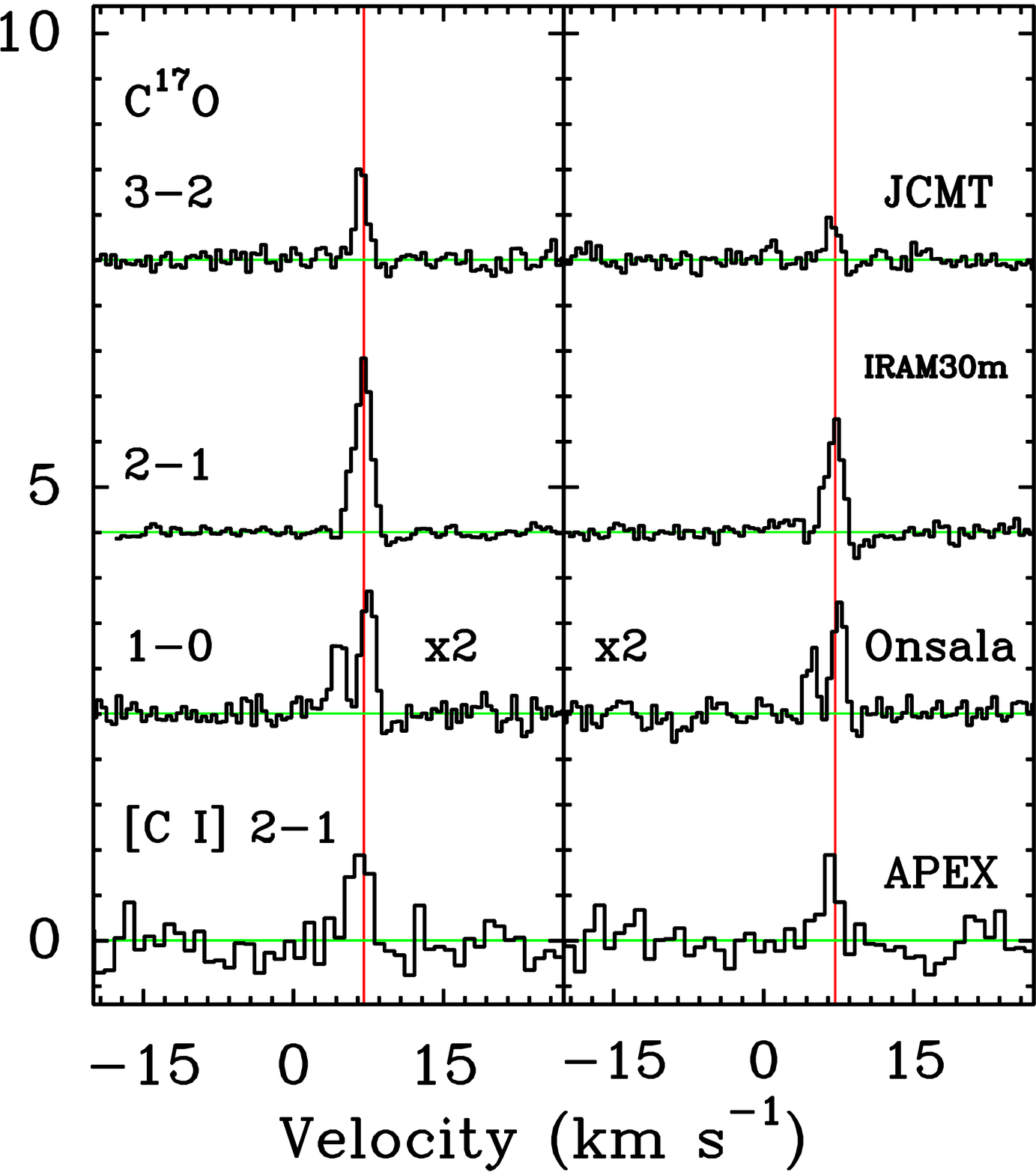} \end{minipage}
\end{center}
\caption{Single spectra obtained from the central positions of IRAS~4A and 4B presented on a $T_{\rm MB}$ scale. 
From bottom to top, \textit{Left:} \mbox{$^{13}$CO~1--0}, \mbox{$^{13}$CO~2--1}, \mbox{$^{13}$CO~3--2}, \mbox{$^{13}$CO~4--3}, \mbox{$^{13}$CO 6--5}, \mbox{$^{13}$CO~8--7}, \mbox{$^{13}$CO~10--9}; \mbox{$^{12}$CO~1--0}, \mbox{$^{12}$CO~2--1}, \mbox{$^{12}$CO~3--2}, \mbox{$^{12}$CO~4--3}, \mbox{$^{12}$CO~6--5}, \mbox{$^{12}$CO~7--6}, \mbox{$^{12}$CO~10--9}; \textit{Right:} \mbox{[\ion{C}{i}] 2--1}, \mbox{C$^{17}$O~1--0}, \mbox{C$^{17}$O~2--1}, \mbox{C$^{17}$O~3--2}; \mbox{C$^{18}$O~1--0}, \mbox{C$^{18}$O~2--1}, \mbox{C$^{18}$O~3--2}, \mbox{C$^{18}$O~5--4}, \mbox{C$^{18}$O~6--5}, \mbox{C$^{18}$O~9--8}, \mbox{C$^{18}$O~10--9}. The spectra have been shifted vertically for viewing purposes and refer to the observing beams presented in Table \ref{tbl:overviewobs}. The red vertical line corresponds to the source velocity, $V_{\rm {LSR}}$, as measured from the C$^{18}$O and C$^{17}$O lines.}
\label{fig:iras4abspectra}
\end{figure*}

Figure~\ref{fig:co65insets} illustrates the quality of the APEX
spectra as well as the variation in line profiles across the
map. Several different velocity components can be identified, which
are best seen at the central source positions.  Figure
\ref{fig:iras4abspectra} presents the gallery of CO lines at IRAS~4A
and 4B using the APEX, JCMT, \textit{Herschel}, IRAM~30m, Onsala and
FCRAO telescopes. Available spectra of $^{12}$CO, $^{13}$CO,
C$^{18}$O, C$^{17}$O and [\ion{C}{i}] ranging from 1--0 up to 10--9
are shown.  Integrated intensities and peak temperatures are
  summarized in \mbox{Table \ref{tbl:lineintensities}} which includes
  the rms of each spectrum after resampling all spectra to the same velocity
  resolution of 0.5 km~s$^{-1}$.
The $S/N$ and dynamic range in the spectra is generally excellent with
peak temperatures ranging from 30 mK to $>$20~K compared with the rms values of 0.006 to 0.4~K.  Note in particular
the very high $S/N$ obtained at the \mbox{C$^{18}$O~5--4} line with
{\it Herschel} ($\sim$6~mK in 0.5 km~s$^{-1}$ bins).  Even C$^{18}$O appears detected up to $J$=10--9 in
IRAS~4B, albeit only tentatively (1.5 $\sigma$) in the 10--9 line
itself.  Together with the IRAS~2A data of \citet{yildiz10hifi}, this
is the first time that the complete CO ladder up to 10--9 is presented
for low-mass protostars, not just for $^{12}$CO but also for the
isotopologs, and with spectrally resolved data.

As discussed in \citet{Kristensen10hifi} based on H$_2$O spectra, the
central line profiles can be decomposed into three components.  A {\it
  narrow} profile with a FWHM of 2--3~km ~s$^{-1}$ can mainly be found
in the optically thin C$^{18}$O and C$^{17}$O isotologue lines at the
source velocity.  This profile traces the quiescent envelope
material. Many $^{12}$CO and $^{13}$CO line profiles show a {\it
  medium} component with a FWHM of 5--10~km~s$^{-1}$ indicative
of small-scale shocks in the inner dense protostellar envelope
($<$1000~AU). The latter assignment is based largely on interferometry
maps of this component toward IRAS~2A \citep{Jorgensen07prosac}. The
$^{12}$CO lines are mainly dominated by the {\it broad} component with
a FWHM 25--30~km~s$^{-1}$ on $>$1000~AU scales
representative of the swept-up outflow gas
(Fig. \ref{fig:iras4abspectra}).

\placeTableObservedLineIntensities

\subsection{Maps}
The observations presented here are large scale $240''\times 240''$
maps in \mbox{$^{12}$CO~6--5} and \mbox{$^{12}$CO~3--2} covering the
entire IRAS 4A/B region, together with smaller scale $80''\times 80''$
maps of \mbox{$^{13}$CO~6--5} and \mbox{[\ion{C}{i}]~2--1} around the
protostellar sources. 

\subsubsection{\mbox{$^{12}$CO~6--5} map}
\label{sec:CO65maps}

The large \mbox{$^{12}$CO~6--5} map over an area of
240$\arcsec\times$240$\arcsec$ \mbox{($\sim$56\,500 $\times$ 56\,500
  AU)} includes all the physical components of both
protostars. Figure~\ref{fig:co65map} (left) shows a
\mbox{$^{12}$CO~6--5} contour map of the blue and red outflow lobes,
whereas Fig.~\ref{fig:co65map} (right) includes the map of individual
spectra overplotted on a contour map. This spectral map has been resampled to
10$\arcsec \times$10$\arcsec$ pixels for visual convenience, however
the contours are calculated through the Nyquist sampling rate of
4$\arcsec$.5 $\times$ 4$\arcsec$.5 pixel size. All 
spectra are binned to \mbox{0.3 km~s$^{-1}$} velocity resolution. The red and blue outflow
contours are obtained by integrating the blue and red wings of each
spectrum separately. 
The selected ranges are  \mbox{--20 to 2.7 km~s$^{-1}$} for the blue
and \mbox{10.5 to 30 km~s$^{-1}$} for the red emission. 
These ranges are free of cloud and envelope emission and are 
determined by averaging spectra from outflow-free regions.

The \mbox{$^{12}$CO~6--5} map shows a well-collimated outflow to the
NE and SW directions centered at IRAS~4A with two knots on each side
like a mirror image. Close to the protostar itself, the outflow appears
to be directed in a pure N-S direction, with the position angle on the
sky rotating by about 45$\degr$ at 10$\arcsec$ (2350~AU)
distance. This N-S direction has been seen in interferometer data of
\citet{Jorgensen07prosac} and \citet{Choi11}, and the high angular
resolution of APEX-CHAMP$^+$ now allows this component also to be
revealed in single dish data. The morphology could be indicative of a
rotating/wandering jet emanating from IRAS 4A or two flows
 from each of the binary components of IRAS4A.
The outflow from IRAS~4B is much more spatially compact moving in the
N-S direction. Overall, the CO~6--5 CHAMP$^{+}$ maps are similar to
the \mbox{CO~3--2} map shown in Fig.~\ref{fig:co32map} and in
\citet{Blake95}. However, because of the $\sim$2 times larger beam,
the N-S extension around IRAS 4A is not obvious in the 3--2 map and
the knots are less `sharp'. Also, the compact IRAS~4B outflow is
revealed clearly in single-dish data here for the first time. In the
north-western part of the map, the southern tip of the SVS 13 flow is seen
\citep[HH~7--11;][]{curtis10_2outflows}.

\subsubsection{\mbox{$^{12}$CO~3--2} map}
\label{sec:CO32maps}

The large and fully sampled \mbox{$^{12}$CO~3--2} JCMT HARP-B map
covers the same area as the \mbox{$^{12}$CO~6--5} map. In
Fig. \ref{fig:co32map}, the \mbox{CO~3--2} contour and spectral maps
presenting blue and red outflow lobes are shown. Here, the spectral map is
resampled to 15$\arcsec \times$15$\arcsec$ pixels and the contours
are calculated in the Nyquist sampling rate of 7.5\arcsec $\times$
7.5\arcsec pixel size. The same velocity ranges as in the
\mbox{CO~6--5} map are used to calculate the blue and red outflow
emission. Overall, the 3--2 map is very similar to those presented by
\citet{Blake95} and \citet{curtis10_2outflows}.  

The line ratio map of \mbox{CO~3--2/CO~6--5} is presented in
Fig.~\ref{fig:12co32-co65mapratio}. The CO~6--5 map is convolved to
the same beam as CO~3--2 and the peak antenna temperatures have been
used to avoid having differences in line widths dominate the
ratios. The distribution of the line ratios is flat at 0.8--1.0 around
the center and outflow knots, with values up to 2.5 in the surrounding
regions. As discussed further in Sect.~\ref{sec:kintemp}, this implies higher
temperatures towards the center and outflow knots than in the envelope
at some distance away from the outflow.

Figure \ref{fig:Vmaxmaps} shows maps of the maximum spectral velocities
$V_{\rm max}$ obtained from the full width at zero intensity (FWZI) at
each position for both the 6--5 and 3--2 maps.  
A 1.5$\sigma$ cutoff is applied in order to determine the FWZIs in both
maps. Because of the lower rms of the data, \mbox{CO~3--2} can trace
higher velocities than \mbox{CO~6--5}.  Overall, the profiles
indicate narrow lines throughout the envelope with broad shocked
profiles along the outflows (see also Fig. \ref{fig:co65insets}).
Similar results have been found by \citet{vanKempen09champ} for
the HH~46 protostar and outflow. The highest velocities with $V_{\rm
  max}=$25--30 km s$^{-1}$ are found at the source positions (where
both red and blue wings contribute) and at the outflow knots. 

Specifically, the IRAS~4A-R2 outflow knot has an extremely high velocity component (EHV or ``bullet'') 
at $V\sim$~20--35~km~s$^{-1}$ as seen clearly 
in the 3--2 map  (Fig. \ref{fig:IRAS4A-B2_bullet} in the Appendix). In 
the CO~6--5 map the ``bullet'' emission is only weakly detected 
($\sim$5$\sigma$, Fig. \ref{fig:co65insets}) and is ignored in the rest of this paper.

\placeFigureIrasFourABCOsixfivespectraandoutflow
\placeFigureTemperatureMaps
\placeFigureDeltaVMaps

\subsubsection{\mbox{$^{13}$CO 6--5} map}
\label{subsec:velrange}

The \mbox{$^{13}$CO 6--5} isotopolog emission was mapped over a
smaller 80$\arcsec~\times$~80$\arcsec$ region presented in
Fig. \ref{fig:iras4ab13co65}. This map only covers the immediate
environment of the protostellar envelopes of both protostars and the
outflow of IRAS~4B. Figure \ref{fig:iras4ab13co65} (left) shows the map
of total integrated intensity whereas Fig. \ref{fig:iras4ab13co65}
(right) shows the spectral map
with the outflow contours obtained using the same velocity range as in the
CO~6--5 map. The $^{13}$CO 6--5 lines are not simple
narrow gaussians, but show clearly the medium outflow component
centered on the protostars. 
The medium component has a FWHM of \mbox{$\Delta\varv$ = 8--10 km
  s$^{-1}$} while the narrow component is again \mbox{$\Delta\varv$ =
  1.5-2 km s$^{-1}$}.

\placeFigureIrasFourABThirteenCOSpectraAndOutflow

\subsubsection{\mbox{[\ion{C}{i}]~2--1} map}
\label{sec:CI21maps}

Figure \ref{fig:cIco65overlaid} (in the Appendix) shows the weak
detection of atomic carbon emission in and around the envelope and the
outflow cavities, with the \mbox{$^{12}$CO~6--5} red and blue contours
overlaid (see also Fig.~\ref{fig:co65map} right panel).  This figure
is the combination of three different observations, with one map
covering only the central region (obtained in parallel with the
\mbox{$^{13}$CO~6--5} map).  Thus, the noise level is higher at the
edges of the figure. The spectra have been resampled to 1 km s$^{-1}$
velocity resolution in order to reduce the noise significantly; still,
the [\ion{C}{i}] line is barely detected with a peak temperature of at
most 1 K.  The weak emission indicates that CO is not substantially
dissociated throughout the region, i.e., the UV field cannot contain
many photons with wavelengths $<$1100 $\AA$
\citep{vanDishoeckBlack88}, as also concluded in
\citet{vanKempen09champ2}.  The low $S/N$ of the [\ion{C}{i}]
data precludes detection of any broad outflow component. Note that
in HH~46, stronger [\ion{C}{i}] emission is found at the bow shock 
position, but this line is still narrow \citep[$\Delta V\sim$1~km~s$^{-1}$;][]{vanKempen09champ}.

\subsection{Morphology}
\label{subsec:morphology}

\placeFigureIRASThereD

By examining the morphology of the outflows from the \mbox{CO~3--2}
and 6--5 maps, it is possible to quantify the width and length of the
outflows. The \mbox{CO~6--5} map is used to calculate these quantities
because of the two times higher spatial resolution. The length of the
outflow, $R_{\rm CO}$, is defined as the total outflow extension
assuming the outflows are fully covered in the map. By taking into
account the distance to the source, the projected $R_{\rm CO}$ is
measured as 105$\arcsec$ \mbox{($\sim$25\,000~AU)} and 150$\arcsec$
\mbox{($\sim$35\,000~AU)} for IRAS~4A for the blue and red outflow
lobes, respectively.  The difference in extent could be a result of
denser gas deflecting or blocking the blue outflow lobe \citep{Choi11}.
For IRAS~4B, the extents are 12$\arcsec$ ($\sim$1\,900~AU) and
9$\arcsec$ ($\sim$750~AU), respectively, but these should be regarded as upper limits since the IRAS~4B outflow is not resolved.
The width of the IRAS~4A outflow is $\sim$20\arcsec
($\sim$4\,700~AU), after deconvolution with the beam size. 
These values do not include corrections for inclination.

The ``collimation factor'', $R_{\rm
  coll}$  for quantifying the outflow bipolarity 
is basically defined as the ratio between the major and minor
axes of the outflow. This quantity has been used to distinguish Stage
0 objects from the more evolved Stage I objects, in which the outflow
angle has widened \citep{Bachiller99,ArceSargent06}. $R_{\rm coll}$
for IRAS~4A is found to be 5.3$\pm$0.5 for the blue outflow lobe and
7.5$\pm$0.5 for the red outflow lobe. 
For IRAS~4B, no collimation factor can be determined
since the outflow is unresolved.
Nevertheless, the much smaller extent of the IRAS~4B outflow raises the question whether IRAS~4B is much younger than IRAS~4A or whether this is simply an
effect of inclination. The inclination of an outflow, which is defined as the
angle between the outflow direction and the line of sight
\citep{CabritBertout90}, can in principle be estimated from the
morphology in the contour maps.

The IRAS~4 system is part of a clustered star-forming region so that
the formation timescales for any of the YSOs in this region are
expected to be similar. Also, the bolometric luminosities of IRAS~4A
and 4B are comparable.  For IRAS~4B, {\it Herschel}-PACS observations
by \citet{Herczeg11} detect only line emission from the blue outflow
lobe, with the red outflow lobe hidden by $>$1000 mag of
extinction. These new data support a close to face-on orientation with
the blue lobe punching out of the cloud with little extinction and the
red lobe buried deep inside the cloud.
High resolution millimeter interferometer data of \citet{Jorgensen07prosac}
as well as our data, however, do not show overlap between the IRAS~4B
blue and red outflow lobes which would imply that they are not
completely, but close to face-on with an inclination close to the line
of sight of $\sim$15--30$\degr$. 
This range is consistent with 10--35$\degr$ suggested for IRAS~4B
based on VLBI H$_2$O water maser observations \citep{Desmurs09}.
The large extent of the collimated outflow of IRAS~4A with, at the same
time, high line-of-sight velocities suggests an inclination of
$\sim$45--60$\degr$ to the line of sight.  It is unlikely to be as
high as 80--85$\degr$ as claimed for L1527 ($i$=85$\degr$) and L483
\citep[80$\degr$;][]{Tobin08}. Karska et al.\ (in prep.) find much
lower velocities ($\sim$6--10 km~s$^{-1}$) in CO 6--5 maps for these
sources than in IRAS~4A/4B ($\sim$20--30 km~s$^{-1}$). 

Under the assumption that the intrinsic lengths of the flows are similar,
Fig. \ref{fig:iras4ab3D} presents various options for the relative
orientation of the two outflows viewed from different angles, all
three of which can lead to the observed projected situation as
seen in Fig. \ref{fig:iras4ab3D}a.  In the first scenario, the
envelopes are very close to each other and interact accordingly
(Fig. \ref{fig:iras4ab3D}b). 
In the second scenario, the
envelopes may be sufficiently separated in distance so that they do
not interact with each other. In this case, IRAS~4A is either in front
of IRAS~4B (Fig. \ref{fig:iras4ab3D}c) or IRAS~4B is in front of
IRAS~4A (\ref{fig:iras4ab3D}d).

The dynamical age of the outflows can be determined by
$t_{\rm dyn} = {R_{\rm CO}} / \overline{V}_{\rm max}$,
where $\overline{V}_{\rm max}$ is the average total velocity extent as measured relative to
the source velocity \citep{CabritBertout92}. $\overline{V}_{\rm max}$ for
IRAS~4A and IRAS~4B are found to be $\sim$20 and \mbox{$\sim$15~km s$^{-1}$}, 
respectively, representative of the outflow tips (Fig.~\ref{fig:Vmaxmaps}).
Using these velocities, the $t_{\rm dyn}$ is 5900 and 9200
years for IRAS~4A for the blue and red outflow lobes,
respectively. \citet{KneeSandell00} found 8900~yr (blue) and 16\,000~yr (red) 
for the IRAS~4A outflow lobes, whereas \citet{Lefloch98_SiO} found
11\,000 years for both of the outflow lobes in IRAS~4A from an
SiO~2--1 map. All these analyses assume a steady flow whereas
the knots have clearly larger widths
than the rest of the flow (Fig. \ref{fig:Vmaxmaps}), indicative of
episodic accretion and outflow. Indeed, the constant flow assumption
is the main uncertainty in the determination of dynamical ages,
although our approach of taking the maximum velocity combined with the
maximum extent should give more reliable estimates than `global'
methods \citep{DownesCabrit07}.

\section{Analysis: Outflow}
\label{sec:analysis}
\subsection{Rotational temperatures and CO ladder}

The most direct quantity that can be derived from the CO lines at the
source position are the rotational temperatures
(Fig.~\ref{fig:RotationDiagrams}). It is important to note that all
lines are well fitted with a single temperature, indicating that they
are probing the same gas up to $J$=10--9.  
Values of 69$\pm$7 and 83$\pm$10 K are found for
$^{12}$CO, whereas those for $^{13}$CO and C$^{18}$O are up to a factor of two lower
(see Table~\ref{tbl:rotdiaglist}). Since the $^{12}$CO integrated intensities
are dominated by the line wings, this may indicate that the outflowing
gas is somewhat warmer than the bulk of the envelope which dominates the isotopolog emission. On the other
hand, the higher optical depths of the $^{12}$CO lines can also result
in higher rotational temperatures. A quantitative analysis of the
implied kinetic temperatures is given in Sect. \ref{sec:kintemp}.  

\placeTableRotDiag
\placeFigureRotDiags
\placeFigureCOladder 

Another way of representing the CO ladder is provided in
Fig.~\ref{fig:COladder} where \mbox{$^{12}$CO} and \mbox{$^{13}$CO}
line fluxes are normalized relative to the $J$=4--3 and $J$=6--5
lines, respectively.  Such figures have been used in large scale Milky
Way and extragalactic studies to characterize the CO excitation
\citep[e.g.][]{Weiss07}.  Other astronomical sources are overplotted
for comparison, including the weighted average spectrum of diffuse gas
in the Milky Way as measured by COBE-FIRAS from \citet{Wright1991};
the dense Orion Bar PDR from {\it Herschel}-SPIRE spectra from
\citet{Habart10}; SPIRE spectra of the ultraluminous infrared galaxy
Mrk231 from \citet{vanderWerf10}; and broad absoption line quasar
observations of APM08279+5255 from ground-based data of
\citet{Weiss2007}. For IRAS~4A and 4B, the $^{12}$CO and $^{13}$CO
maps are convolved to 20$\arcsec$, where available, in order to
compare similar spatial regions.
It is seen that the low-mass YSOs studied here have very similar CO
excitation up to $J_u=10$ to the Orion Bar PDR and even to
ultraluminous galaxies; in contrast, the excitation of CO of the
diffuse Milky Way and Mkr 231 appears to turn over at lower $J$.  Our
conclusion that the $^{13}$CO high-$J$ lines trace UV heated gas (\S
6) is consistent with its similar excitation to the Orion Bar.

\subsection{Observed outflow parameters}
The CO emission traces the envelope gas swept up by the outflow over
its entire lifetime, and thus provides a picture of the overall outflow
activity. The outflow properties can be derived by converting the CO
line observations to physical parameters. Specifically, kinetic
temperatures, column densities, outflow masses, outflow forces and kinetic
luminosities can be derived from the molecular lines. In the following
sections, the derivation of these parameters will be discussed.

\subsubsection{Kinetic temperature}
\label{sec:kintemp}

The gas kinetic temperature is obtained from CO line ratios.
Figure~\ref{fig:i4a_co3-2_6-5ratios} presents the observed line wing
ratios of \mbox{CO~3--2/6--5} at the source positions of IRAS~4A and 4B
as well as the four outflow knots identified in Fig. \ref{fig:co65map}.
The CO~6-5 map is resampled to a
15$\arcsec$ beam so that the lines are compared for the same beam. The
ratios are then analyzed using the \verb1RADEX1 non-LTE excitation and
radiative transfer program \citep{vanderTak07}, as shown in
Fig. \ref{fig:radexmodels} (right). The density within the beam is
taken from the modelling results of \citet{Kristensen12subm}
based on spherically symmetric envelope models assuming a power-law
density structure \citep[see][Sect. \ref{sec:envelopeproperties}]{Jorgensen02}. The analysis assumes that the
lines are close to optically thin, which is justified in Section
\ref{sec:opticaldepth}.
For the \mbox{CO~6--5} transition, the critical density is $n_{\rm
cr}$=1$\times$10$^{5}$ cm$^{-3}$ whereas for \mbox{CO~3--2}, $n_{\rm
cr}$=2$\times$10$^{4}$ cm$^{-3}$ based on the collisional rate
coefficients of \citet{Yang10}. 
For densities higher than $n_{\rm cr}$, the levels are close to being
thermalized and are thus a clean temperature diagnostic; for lower
densities the precise value of the density plays a role in the analysis.

From the adopted envelope model, the density inside $\sim$1750~AU
(7.5$\arcsec$) is $>$10$^{6}$ cm$^{-3}$ for both sources, i.e., well
above the critical densities. The inferred temperatures from the
\mbox{CO~3--2/6--5} line wings are $T_{\rm kin}$ $\sim$60--90 and
$\sim$90--150 K at the source centers of IRAS~4A and 4B.  These values
are somewhat lower than, but consistent with, the temperatures of
90--120~K and 140--180~K found in \citet{yildiz10hifi} using the
\mbox{CO~6-5/10-9} line ratios.  For the outflow positions B1 and R1,
the density is \mbox{$\sim$3$\times$10$^{5}$ cm$^{-3}$} which results
in temperatures of 100--150 K. The B2 and R2 positions are beyond the
range of the envelope model, however assuming typical cloud densities
of $\sim$10$^{4-5}$ cm$^{-3}$, the ratios indicate a higher
temperature range of 140--200~K. Note that the line ratios in
Fig.~\ref{fig:i4a_co3-2_6-5ratios} are remarkably constant with
velocity showing little to no evidence for a temperature change with
velocity.

\placeFigureIrasFourACOThreetoTwoandSixtoFiveRatios

\smallskip
\subsubsection{Optical depths}
\label{sec:opticaldepth}

The optical depth $\tau$ is obtained from the line ratio of two
different isotopologs of the same transition.  In
Fig.~\ref{fig:i4a_sou_coratios}, spectra of \mbox{$^{12}$CO~6--5} and
\mbox{$^{13}$CO~6--5} at the IRAS~4A and 4B and \mbox{$^{12}$CO~3--2}
and \mbox{$^{13}$CO~3--2} at the IRAS~4B source centers are shown. For
presentation purposes, only the wing with the highest $S/N$ ratio is shown,
but the same trend holds for the other wing.
Figure~\ref{fig:i4a_b2r2_coratios} includes the spectra and line wing
ratios of two dense outflow knots in \mbox{$^{12}$CO~2--1} and
\mbox{$^{13}$CO~2--1} at positions labeled I4A-R2 (northern red
outflow knot) and I4A-B2 (southern blue outflow knot).  Line ratios
are taken only from the line wings excluding the central narrow
emission or self-absorption. The optical depths are then derived
assuming that the two species have the same excitation temperature and
the $^{13}$CO lines are optically thin.
The abundance ratio of $^{12}$CO/$^{13}$CO is taken 
as 65 \citep{LangerPenzias90}.  The resulting optical
depths of $^{12}$CO as a function of velocity are shown on the
right-hand axes of Figs.~\ref{fig:i4a_sou_coratios} 
and \ref{fig:i4a_b2r2_coratios}.
High optical depths $>$2 are found at velocities which are very close to the central 
emission implying that the central velocities are optically thick and getting 
optically thinner away from the center in the line wings of the outflowing gas.

\placeFigureIrasFourABlueRedOutflowsOpticalDepthRatioSou
\placeFigureIrasFourABlueRedOutflowsOpticalDepthRatioOutflows

\subsubsection{Outflow Mass}
\label{subsec:mass}

The gas mass in a particular region can be calculated from 
the product of the column densities at each position and the surface area:

\begin{equation}
M_{\rm outflow} = \mu_{\mathrm{H_{2}}} \, m_{\rm H} \times \sum_{i} (N_{{\rm H_{2}},i} \times A)  
\end{equation}
where the factor $\mu_{\mathrm{H_{2}}}$=2.8 includes the contribution
  of Helium \citep{Kauffmann08}, $m_{\rm H}$ is the mass of the
hydrogen atom, $A$ is the surface area in one pixel (4\farcs5$\times$4\farcs5), $\sum_{i} N_{\rm
  H_{2},i}$ is the pixel averaged $\rm H_{2}$ column density over the selected
velocity range, and the sum is over all pixels.
In order to calculate the mass of the outflowing material, the \mbox{CO~3--2} and
6--5 maps are resampled to a Nyquist sampling rate and calculated
separately for each 15$''$ and 9$''$ pixel, respectively. 
As found in Sect.~\ref{sec:opticaldepth}, the bulk of the 
emission in the line wings has low optical depth.
The CO column density is then obtained from

\begin{equation}
\frac{N_{\rm u}}{g_{\rm u}} = \beta \frac{(\nu [\rm{GHz])^{2}}\, W [\mathrm{K\, km\, s^{-1}}]}{A_{\rm ul} [\rm s^{-1}]\, g_{\rm u}}
\end{equation}
where $\beta$ = 1937~cm$^{-2}$; $g_{\rm u}$=2$J$+1 and $W = \int T_{\rm mb} {\rm d}V$
is the integrated intensity over the line wing. This intensity is
calculated separately for the blue and red line wings with the
velocity ranges as defined in Sect. \ref{subsec:velrange}.
The total CO column density, $N_{\rm t}$, can then be found by
\begin{equation}
N_{\rm t} = N_{\rm u} Q_{\rm T}  \frac{1}{g_{\rm u} \, e^{-E_{\rm u}/kT_{\rm ex}}}
\end{equation}
where $Q_{\rm T}$ is the partition function corresponding to a
specific excitation temperature, $T_{\rm ex}$. The assumed $T_{\rm
ex}$ is 75~K based on Sect. \ref{sec:kintemp}, but using $T_{\rm ex}$=100~K 
results in only $\sim$10$\%$ less mass.  The column density
$N_{\rm H_{2}}$ is obtained 
assuming an $^{12}$CO/H$_{2}$ abundance ratio of $10^{-4}$ which is lower than the canonical value of 2.7$\times$10$^{-4}$ \citep{Lacy94}. The precise value of the CO abundance in the outflow is uncertain because some of the CO may be frozen out onto dust grains.
The total H$_{2}$ column densities in the outflows derived from the CO
6--5 data are 1.0 and 1.8$\times10^{22}$ cm$^{-2}$ for IRAS~4A, and
1.0 and 0.9$\times10^{21}$ cm$^{-2}$ for IRAS~4B, summed over the
entire blue and red outflow lobes, respectively (see Table
\ref{tbl:outflowparam}). 

The masses of the outflowing material in the
IRAS~4A blue and red lobes are then 6.1$\times10^{-3}$ and
1.0$\times10^{-2}$ M$_{\odot}$, and for IRAS~4B, 6.0$\times10^{-4}$
and 5.3$\times10^{-4}$ M$_{\odot}$, respectively. 
The masses have also been calculated from the \mbox{CO~3--2} map, and
the resulting values are $\sim$2 times larger, which is partly due to
the fact that this line traces the colder gas with assumed $T_{\rm
  ex}$=50~K.  \citet{curtis10_2outflows} used the JCMT \mbox{CO~3--2}
map of the entire Perseus molecular cloud to calculate the masses of
the outflows from many sources in the region. They obtained around
a factor of two higher mass for the total outflow in IRAS~4A
(\mbox{7.1$\times10^{-2}$} vs.\ our measurement of
\mbox{3.0$\times10^{-2}$ $M_{\odot}$} from the 3--2 data) and around a
factor of three higher value for IRAS~4B outflow
(\mbox{1.1$\times10^{-2}$} vs.\ our measurement
\mbox{1.8$\times10^{-3}$ $M_{\odot}$}).  These differences are well
within the expected uncertainties, i.e., caused by choosing slightly
different velocity ranges.

\placeTableOutflowProperties

\subsection{Outflow energetics}

Theories of the origin of jets and winds and models of the `feedback'
of young stars on their surroundings require constraints on the
characteristic force and energetics of the flow to infer the
underlying physical processes. Specifically, the outflow force,
kinetic luminosity and mass ouflow rate can be measured from our data. 
The outflow force is defined as
\begin{equation}
F_{\rm CO}=\frac{1}{R_{\rm CO}} \,\mu_{\mathrm{H_{2}}}\, m_{\rm H} \times \sum_{i} (N_{{\rm H_{2}},i} \times A \times \Delta V_{{\rm max},i}^{2}.) 
\end{equation}
So far, this
parameter has been determined by using lower-$J$ lines for several
young stellar objects \citep{CabritBertout92, Bontemps96, Hogerheijde98,
vanKempen09champ}. 
The kinetic luminosity can be obtained from
\begin{equation}
L_{\rm{kin}}=\frac{1}{2 R_{\rm CO}} \,\mu_{\mathrm{H_{2}}} \, m_{\rm H} \times \sum_{i} (N_{{\rm H_{2}},i} \times A \times \Delta V_{{\rm max},i}^{3}) 
\end{equation}
and the mass outflow rate,
\begin{equation}
\dot{M}=\frac{M_{\rm outflow}}{t_{\rm{dyn}}}.
\end{equation}

The outflow parameters derived from the observations are presented in Table
\ref{tbl:outflowparam}. 
No corrections factors cf.\ \citet{CabritBertout90} have been applied.

\section{Analysis: Envelope properties and CO abundance}
\label{sec:envelopeproperties}

\subsection{Envelope model}

In order to quantify the density and temperature structure of each
envelope, the continuum emission is modelled using the 1D spherically
symmetric dust radiative transfer code \verb1DUSTY1
\citep{IvezicElitzur97}. The method follows closely that of
\citet{Schoier02} and \citet{Jorgensen02,Jorgensen05},
discussed further in \citet{Kristensen12subm}. The inner boundary of the envelope is
set to be where the dust temperature has dropped to
250 K ($=r_{\rm in}$). The density structure of the envelope is
assumed to follow a power law with index $p$, i.e., $n \propto
r^{-p}$, with $p$ being a free parameter. The other free parameters
are the size of the envelope, $Y=r_{\rm out}/r_{\rm in}$ and the
opacity at 100 $\mu$m, $\tau_{100}$. A grid of \verb1DUSTY1 models was
run and compared to the SEDs as obtained from the literature and
radial emission profiles at 450 $\mu$m and 850 $\mu$m
\citep{diFrancesco08}. The best-fit solutions were obtained using a
$\chi^2$ method and are listed in Table \ref{tbl:dustyresultsenv},
where also the derived physical parameters for the envelopes are
listed.

A complication for the IRAS~4A/4B system is that they are so close to each
other that their envelopes could overlap.  Figs. \ref{fig:co65map} and \ref{fig:envelope_plot}
compare the envelopes at the 10~K radius together with the observed
beam sizes.
The model envelopes show that they start to overlap
almost immediately from the central protostars if the two sources are at
the same distance. In that case, the summed density of the two envelopes
does not drop below 1.5$\times$10$^{6}$ cm$^{-3}$
(Fig.~\ref{fig:envelope_plot}, {\it left}). Another scenario discussed in
Sect.~\ref{subsec:velrange} is that
the two sources are sufficiently separated in distance so that they do
not interact and therefore have separate envelopes
(Fig. \ref{fig:iras4ab3D}c and \ref{fig:iras4ab3D}d).  The density and
temperature profiles as function of radius for such a scenario are
shown in Fig. \ref{fig:envelope_plot} ({\it right}).  Since the overlap area is
small even in the case that the sources are at exactly the same
distance, the subsequent analysis is done adopting this latter
scenario.

\placeFigureEnvelopePlot

\placeTableDustyResultsEnv

The resulting envelope structure is used as input for the
\verb1Ratran1 line radiative transfer modeling code
\citep{Hogerheijde00}. In Table \ref{tbl:dustyresultsenv}, the
inferred values from \verb1DUSTY1 that are used in \verb1Ratran1 are
given.
In IRAS~4A, the outer radius is taken to be the radius where the density
$n$ drops to 1.2$\times$10$^{4}$~cm$^{-3}$, and the
temperature is considered to be constant after it reaches  8~K.
The total masses of the envelopes are 5.1~M$_{\odot}$ (out to a radius of 6.4$\times$10$^{3}$~AU) and 3.0~M$_{\odot}$  (3.8$\times$10$^{3}$~AU) at the
10~K radius and 37.0~M$_{\odot}$ (3.3$\times$10$^{4}$~AU) and 18.0~M$_{\odot}$ (1.2$\times$10$^{4}$~AU) at the 8~K radius for IRAS~4A
and 4B, respectively. The turbulent velocity is set to 0.8~km~s$^{-1}$
which is representative of the observed C$^{18}$O line widths. However, the narrow component of the $^{13}$CO lines is best fit with 0.5 and 0.6~km~s$^{-1}$ for IRAS~4A and 4B, respectively. The
model emission is convolved with the beam in which the line has been
observed.

\subsection{CO Abundance Profile}

\citet{yildiz10hifi} present \textit{Herschel}-HIFI single pointing
observations of CO and isotopologs up to 10--9 ($E_{\rm u}/k$=300 K)
for NGC 1333 IRAS 4A, 4B and 2A.
They used the C$^{18}$O and C$^{17}$O isotopolog data from 1--0 up to
9--8 to infer the abundance structure of CO through the envelope of
the IRAS~2A protostellar envelope. For that source, the inclusion of
the higher-$J$ lines demonstrates that CO must evaporate back into the
gas phase in the inner envelope. In contrast, the low-$J$ lines trace
primarily the freeze-out in the outer envelope
\citep{Jorgensen02,Jorgensen05}. The maximum possible abundance of CO with respect to H$_2$ is 2.7$\times$10$^{-4}$ as measured in warm dense gas. Interestingly, the inner abundance in the warm
gas was found to be less for IRAS~2A by a factor of a few. One goal of this study is
to investigate whether this conclusion holds more commonly, in
particular for the CO abundance profiles in IRAS 4A and 4B.

The CO abundance profile models were constructed for both IRAS~4A and
4B in the isotopolog lines of \mbox{C$^{18}$O} and \mbox{C$^{17}$O}
using the methods outlined above. The lines are optically thin and
have narrow line widths characteristic of the quiescent surrounding
envelope.
The CO-H$_{2}$ collision parameters from \citet{Yang10} are used. The
calibration errors are taken into account in the modelling. Following
the recipe from \citet{yildiz10hifi} for IRAS~2A,
``constant'', ``anti-jump'', ``drop'' and ``jump'' abundance
profiles are investigated (see Fig.~\ref{fig:envelope_plot}, right).
The abundance ratio of C$^{18}$O/C$^{17}$O is taken as 3.65 \citep{WilsonRood94}.

\subsubsection{Constant abundance profile}
\label{sec:constant}
As a first iteration, a constant abundance  was used
to model the C$^{18}$O and C$^{17}$O lines,
but it was not possible to reproduce all line intensities with this
profile.  In IRAS~4A, higher-$J$ C$^{18}$O lines converge well around
an abundance of \mbox{$X$$\sim$6$\times$10$^{-8}$}, however it is
necessary to have lower abundances to produce lower-$J$ lines, around 
\mbox{$X$$\sim$1--2$\times$10$^{-8}$}. In IRAS~4B, higher-$J$ lines
fit well $\sim$1$\times$10$^{-7}$ and lower-$J$ lines again with
1--2$\times$10$^{-8}$. Here, low-$J$ refers to the $J$ $\leq$3 lines and high-$J$ to the $J$ $\geq$5 lines.

\subsubsection{Anti-jump abundance profile}
\label{sec:antijump}
For IRAS~4A, an anti-jump profile was run for the C$^{18}$O and C$^{17}$O
lines. In an anti-jump profile, the evaporation jump in the
   inner envelope is lacking, that is, the inner abundance, $X_{\rm in}$=$X_{\rm D}$, and the
depletion density, $n_{\rm de}$, are varied keeping the outer
abundance high at $X_0$=5$\times$10$^{-7}$ corresponding to a
$^{12}$CO abundance of 2.7$\times$10$^{-4}$ for
$^{16}$O/$^{18}$O=550 \citep[see][for motivation of keeping $X_0$ at this value]{yildiz10hifi}.
Reduced-$\chi^{2}$ plots are shown in
Fig. \ref{fig:antijumpallchi4ab} where lower and higher-$J$ lines are
shown separately in order to illustrate their different constraints.
\placeFigureAntiJumpAllChi
Lower-$J$ lines indicate an $n_{\rm de}$ of
7.5$\times$10$^{4}$~cm$^{-3}$ and $X_{\rm D}$ of
1$\times$10$^{-8}$.  The higher-$J$ lines do not constrain $n_{\rm de}$ 
but an upper limit of 2.5$\times$10$^{5}$~cm$^{-3}$ and a well-determined
$X_{\rm D}$ value of 5$\times$10$^{-8}$ are obtained.
Because the density at the outer edge of the IRAS~4B envelope does not 
drop below 1.8$\times$10$^{5}$~cm$^{-3}$, applying an anti-jump profile is not 
possible. CO remains frozen-out throughout the outer parts of the envelope.

\subsubsection{Drop and Jump abundance profile}
\label{sec:drop}
In order to fit all observed lines, a
``drop'' abundance profile is needed in which the inner abundance $X_{\rm
  in}$ increases above the ice evaporation temperature, $T_{\rm ev}$
\citep{Jorgensen05freeze}, as was found for the IRAS~2A
envelope \citep{yildiz10hifi}. The $n_{\rm de}$ and $X_{\rm D}$
parameters inferred from the anti-jump profile from the low-$J$ lines are used to determine the
evaporation temperature and inner abundance ($X_{\rm in}$). As for
IRAS~2A, the reduced $\chi ^{2}$ plots (not shown) 
indicate that the evaporation temperature is not well determined, thus a
laboratory lower limit of $\sim$25~K is taken.
Figure~\ref{fig:dropallXDXin} {\it left} shows the $\chi^2$ plots
in which the inner abundance $X_{\rm in}$ and $X_{\rm D}$ are varied.
The models are run for a desorption density of 7.5$\times$10$^{4}$~cm$^{-3}$ in
IRAS~4A.
Best fit values for the lower- and higher-$J$ lines are $X_{\rm in} \sim$1$\times$10$^{-7}$ and  $X_{\rm D}$=5.5$\times$10$^{-8}$.  
For IRAS~4B, a jump abundance profile was applied in which the CO abundance stays low in the outer part (see Sect. \ref{sec:antijump}). With this model, again, $X_{\rm  in}$ and  $X_{\rm D}$ values are varied (Fig.~\ref{fig:dropallXDXin} {\it right}). The best fit gives $X_{\rm in}~\sim$3$\times$10$^{-7}$ and  $X_{\rm D}$=1$\times$10$^{-8}$. 

\placeFigureDropAllChiXDXin
\placeFigureBestfitXR

Best fit values obtained with the above mentioned models 
are summarized in Table~\ref{tbl:abundancetable} and a simple 
cartoon is shown in Fig. \ref{fig:bestfit_X_vs_R}.   
Modeled lines are overplotted
on the observed C$^{18}$O lines in Fig. \ref{fig:dropoverplot4ab}
convolving each line to the beam in which they have been observed. In
the models, the \mbox{C$^{18}$O~1-0} and 5-4 lines are underproduced
due to the fact that their much larger beam sizes pick up emission
from the extended surroundings not included in the model.

\placeAbundanceTable

\placeFigureDropOverplots

Table \ref{tbl:abundancetable} includes the IRAS~2A results from
\citet{yildiz10hifi}. 
It is found that $X_{\rm in}$ is a factor of $\sim$3--5 lower than
$X_{0}$ in IRAS~2A and a factor of 5 lower in IRAS~4A. 
\citet{Fuente2012} find a similar factor for the envelope of the intermediate
mass protostar NGC 7129 IRS.
Thus, the conclusion of
\citet{yildiz10hifi} for IRAS 2A that $X_{\rm in} < X_0$ holds more
generally and is not linked to a specific source. This, in
turn, may imply that a fraction of the CO is processed into other
molecules in the cold phase when the CO is on the grains. The lack of
strong centrally-peaked [C I] emission in the [C I] map indicates that
CO is not significantly (photo)dissociated in the inner envelope.

\section{Analysis: UV-heated gas}
\label{sec:analysisUV}

In addition to shocks, UV photons can also heat the
gas. Qualitatively, the presence of UV-heated gas is
demonstrated by the detection of extended narrow $^{12}$CO
and $^{13}$CO 6--5 emission in our spectrally resolved data
\citep{Hogerheijde98,vanKempen09champ,vanDishoeck10champ+}. The
fact that this emission is observed to surround the outflow walls
(Sect. \ref{sec:CO65maps}) suggests a scenario in which UV photons
escape through the outflow cavities and either impact directly the
envelope or are scattered into the envelope on scales of a few
thousand AU \citep{Spaans95}.
Our map also shows narrow $^{12}$CO 6--5 
emission at larger scales as well as in and
around the bow-shock regions (Fig. \ref{fig:co65insets}). 
At these locations, the UV photons are most likely produced by the
bow- and jet-shocks themselves, with the UV photons directly impacting
the cavity walls and quiescent envelope. At velocities of 80 km
s$^{-1}$ or more, these shocks produce photons with high enough
energies that they can even photodissociate CO \citep{Neufeld89}.

Quantitatively, the best constraints on the UV heated gas come from
the narrow component of the $^{13}$CO emission. However, at
the source positions, the passively-heated envelope also contributes
to the intensity. To model this component, the best fitting
C$^{18}$O abundance profile for each source are taken and multiplied the
abundance by the $^{13}$C/$^{18}$O abundance ratio of 8.5.
Figure \ref{fig:dropThirteenCO} presents the resulting \verb1Ratran1
\mbox{$^{13}$CO~6--5} line profiles at the central positions.  The observed
spectra for IRAS~4A and 4B are overplotted; here the (weak) broad
component has been removed by fitting two gaussians to the
spectra. The model spectra obtained with this profile are found
to fit the $^{13}$CO~6--5 narrow emission profiles very well, implying
that the contribution from the envelope is indeed significant.

In Fig. \ref{fig:dropThirteenCOUVEnv}, the same method is applied to
the entire \mbox{$^{13}$CO~6--5} map to probe the extent of the
  envelope emission. In the {\it middle} panel, the observed
integrated intensity map of only the narrow component is
plotted. In the {\it bottom} panel, the $^{13}$CO map from the
envelope model is convolved with the APEX beam and subsequently
subtracted from the corresponding observed spectra. 
For both sources, the envelope model reproduces $^{13}$CO~6--5 
emission at the central position. For IRAS~4B, no significant emission remains at off-source positions. 
For IRAS~4A, however, there is clearly narrow and extended emission visible 
beyond the envelope. Figure~\ref{fig:dropThirteenCOUVEnv} {\it top} panel overplots the
model envelope profiles on top of the observed profiles, showing that
only the central positions are well reproduced by this model.  The
excess emission has a width of just a few km s$^{-1}$ so that it is not
related to the outflow. Heating by UV photons is the only
other plausible explanation. This interpretation is strengthened by 
the fact that the excess emission occurs precisely along the cavity
walls, as shown in the {\it bottom} panel.
The \mbox{$^{13}$CO~6--5} transition requires a temperature of $T$$\approx$50~K to be excited which is
consistent with model predictions of \citet{Visser11} showing a
plateau around this temperature on scales of a few 1000~AU from the
protostar. Hence these observations constitute the first direct
observational evidence for the presence of UV-heated cavity walls.

In order to compare the amount of gas heated by UV photons and gas
swept up by the outflows, the masses in each of these components are
calculated using the \mbox{CO~6--5} and \mbox{$^{13}$CO~6-5} data
(narrow component only) over the same region.  The mass of the 
UV-heated gas is calculated assuming $T_\mathrm{ex}$=75~K and CO/H$_2$=10$^{-4}$. Since the
\mbox{$^{13}$CO~6--5} map is smaller than that of \mbox{CO~6--5}, the
outflow masses cannot simply be taken from Table~5, but are recomputed
over the smaller area covered by the $^{13}$CO data.  Both numbers are
compared to the total gas mass in this area, obtained from the
spherical model envelope based on the \verb1DUSTY1 results 
(see~Sect. \ref{sec:envelopeproperties}). To compare the same area as
covered by the outflows, only the mass in an elliptical biconical
shape is considered, with each cone taking $\sim$15\% volume of the
entire envelope out to the 10~K radius that would be present if the
area had not been evacuated (see Fig.~\ref{fig:iras4ab13co65}). The
10~K limit is still within the borders of the $^{13}$CO map
(Fig.~\ref{fig:co65map}).  The UV photon-heated gas mass is derived
from the \mbox{$^{13}$CO 6--5} narrow emission map where the envelope
emission has been subtracted ({\it bottom} panel of
Fig. \ref{fig:dropThirteenCOUVEnv}).
The inferred masses --- total,
UV-heated and outflow --- are summarized in Table~\ref{tbl:masscomparison}.

Interestingly, for IRAS~4A, the mass of UV-photon-heated gas is
somewhat larger than that of the outflowing gas, demonstrating that UV
photons can have at least an equally large impact on their surroundings as the
outflows. Although the uncertainties in derived values are a factor of
2-3 (largely due to uncertainties in CO/H$_2$), both masses are only a few \% of the total quiescent envelope
mass in the same area, however. For IRAS~4B, UV photons are apparently
unable to escape the immediate protostellar environment \citep[see also
discussion in][]{Herczeg11}. In addition, the near pole-on
geometry makes the detection of extended emission along the outflow
cavity more difficult in this case.

\placeMassComparison

\placeFigureDropThirteenCO

\placeFigureThirteenCOUVEnv

\section{Conclusions}
\label{sec:conclusions}

The two nearby Stage 0 low-mass YSOs, NGC1333~IRAS~4A and IRAS~4B,
have been mapped in $^{12}$CO 6--5 using APEX-CHAMP$^+$, with the map
covering the large-scale molecular outflow from IRAS 4A. $^{12}$CO
6--5 emission is detected everywhere in the map. Velocity-resolved
line profiles appear mainly in two categories: broad lines with
$\Delta\varv>$10 km s$^{-1}$ and narrow lines with $\Delta\varv <$2
km s$^{-1}$. The broad lines originate in the molecular outflow
whereas the narrow lines are interpreted as coming from UV heating of
the gas. This interpretation is supported by the location of the
narrow profiles: they ``encapsulate'' the broad outflow lines.

Comparing the CO~6--5 map with a CO~3--2 map obtained at the JCMT
allows for a determination of the kinetic temperature in the outflow
gas as a function of position through the outflow. The temperature
peaks up at the outflow knots and exceeds 100~K. The temperature is
found to be constant with velocity, and there is no indication of
higher temperatures being reached at higher velocities.
Our high $S/N$ multi-line data of $^{12}$CO and isotopologs have
allowed us to derive excitation temperatures, line widths and optical
depths, and thus the outflow properties, more accurately than before.

Smaller $^{13}$CO~6--5 maps centered on the source positions have also
been obtained with APEX-CHAMP$^+$. The \mbox{$^{13}$CO~6--5} emission
is detected within a 20\arcsec\ radius of each source, and the line
profiles are narrower than observed for the outflowing gas. The
narrow $^{13}$CO emission traces gas with a temperature of $\sim$50~K at
these densities, with the gas being heated by the UV photons. The mass
of the outflowing gas is measured from the $^{12}$CO data, whereas the
mass of the UV-heated gas is measured from the narrow
$^{13}$CO spectra after subtracting the spherical envelope and
  outflow contributions. 
For IRAS~4A, the mass of the UV-heated gas is at least comparable to the mass of the
outflow.
This result shows that close to the source position on
scales of a few thousand AU, UV heating is just as important as shock
heating in terms of exciting CO to the $J$=6 level.  Outflow- and
envelope-subtracted $^{13}$CO 6--5 maps clearly reveal the first
direct observational images of these UV-heated cavity walls.

Single-pointing C$^{18}$O data have been obtained at the JCMT,
APEX-CHAMP$^+$ and most recently with \textit{Herschel}-HIFI, the
latter observing lines up to $J$=10--9.  The data are used to
constrain the CO abundance throughout the envelopes of the two
sources. To reproduce the high-$J$ C$^{18}$O emission, a ``drop'' in
the abundance profile is required. This ``drop'' corresponds to the
zone where CO is frozen out onto dust grains, and thus provides
quantitative evidence for the physical characteristics of this
zone. The CO abundance rises in the inner part where $T$ $>$ 25 K, but
not to its expected canonical value of 2.7$\times$10$^{-4}$
\citep{Lacy94}, indicating that some further processing of the
molecule is taking place.

The combination of low-$J$ CO lines (up to $J$=3--2) and higher-$J$ CO
lines such as $J$=6--5 opens up a new window for quantifying the warm
($T$ $\sim$100 K) gas surrounding protostars and along their
outflows. These spectrally resolved data form an important complement
to spectrally unresolved data of the same lines such as being observed
of similar sources with {\it Herschel}-SPIRE. 
From our data
  it is clear that the $^{12}$CO lines covered by SPIRE are dominated
  by the entrained outflow gas with an excitation temperature of
  $\sim$100 K. For $^{13}$CO, lines centered on the protostar are
  dominated by emission from the warm envelope which is passively
  heated by the protostellar luminosity. Off source on scales of a few
  thousand AU, however, UV-photon heated gas along the cavity walls
  dominates the emission. The UV-heated component becomes visible in
  $^{12}$CO lines higher than 10--9, but it is likely that for Stage 0
  sources this component will be overwhelmed by shocks for all
  lines in spectrally unresolved data \citep{Visser11}. Thus, the
  $^{12}$CO and $^{13}$CO data provide complementary information on
  the physical processes in the protostellar environment: $^{12}$CO
  traces swept-up outflow (lower-$J$) and currently shocked
  (higher-$J$) gas, $^{13}$CO traces warm envelope and photon-heated
  gas. Our results imply that spectrally unresolved
  $^{12}$CO/$^{13}$CO line ratios have only a limited meaning.

Understanding the excitation of chemically simple molecules such as CO
is a prerequisite for interpreting other molecules, in particular
H$_2$O data from \textit{Herschel}-HIFI. Furthermore, understanding
the distribution of warm CO on large spatial scales ($>$ 1000 AU) is
necessary for interpreting future high spatial resolution data from
ALMA.

\begin{acknowledgements}
The authors would like to thank the anonymous referee for suggestions and comments which improved this paper.
This work is supported by Leiden Observatory. UAY is grateful to the
APEX, JCMT and {\it Herschel} staff for carrying out the observations. Also
thanks to NL and MPIfR observers for all APEX observations, 
Remo Tilanus for the observation of \mbox{CO~3--2} in JCMT
with the HARP-B instrument, Laurent Pagani for the \mbox{$^{13}$CO~1-0}
observations at IRAM 30m, and Hector Arce for the CO~1-0 data from FCRAO.  
Special thanks to Daniel Harsono for the help with scripting issues.
TvK is grateful to the JAO for supporting his research 
during his involvement in ALMA commissioning.
Astrochemistry in Leiden is supported by the Netherlands Research
School for Astronomy (NOVA), by a Spinoza grant and grant 614.001.008
from the Netherlands Organisation for Scientific Research (NWO), and
by the European Community's Seventh Framework Programme FP7/2007-2013
under grant agreement 238258 (LASSIE).
Construction of CHAMP+ is a collaboration between the
Max-Planck-Institut für Radioastronomie Bonn, Germany; SRON
Netherlands Institute for Space Research, Groningen, the Netherlands;
the Netherlands Research School for Astronomy (NOVA); and the Kavli
Institute of Nanoscience at Delft University of Technology, the
Netherlands; with support from the Netherlands Organization for
Scientific Research (NWO) grant 600.063.310.10.  The authors are
grateful to many funding agencies and the HIFI-ICC staff who has been
contributing for the construction of \textit{Herschel} and HIFI for
many years. HIFI has been designed and built by a consortium of
institutes and university departments from across Europe, Canada and
the United States under the leadership of SRON Netherlands Institute
for Space Research, Groningen, The Netherlands and with major
contributions from Germany, France and the US.  Consortium members
are: Canada: CSA, U.Waterloo; France: CESR, LAB, LERMA, IRAM; Germany:
KOSMA, MPIfR, MPS; Ireland, NUI Maynooth; Italy: ASI, IFSI-INAF,
Osservatorio Astrofisico di Arcetri- INAF; Netherlands: SRON, TUD;
Poland: CAMK, CBK; Spain: Observatorio Astron{\'o}mico Nacional (IGN),
Centro de Astrobiolog{\'i}a (CSIC-INTA). Sweden: Chalmers University
of Technology - MC2, RSS $\&$ GARD; Onsala Space Observatory; Swedish
National Space Board, Stockholm University - Stockholm Observatory;
Switzerland: ETH Zurich, FHNW; USA: Caltech, JPL, NHSC.

\end{acknowledgements}

\bibliographystyle{aa}
\bibliography{bibdata}

\Online
\newpage
\newpage

\appendix

\section{Auxillary Figures}
We here present the CO~3-2 map obtained from JCMT, which is discussed
in Sect. \ref{sec:CO32maps}, and the CHAMP$^{+}$ map of
\mbox{[\ion{C}{i}]~2--1}, which is discussed in Sect.
\ref{sec:CI21maps}.

\placeFigureIrasFourABCOthreetwospectraandoutflow

\placeFigureIRASBTwoBullet

\placeFigureIrasFourABCITwoOneAndTwelveCOSixFiveSpectraAndOutflow

\end{document}